\documentclass[aps,10pt,reprint,groupedaddress,superscriptaddress, prx, nofootinbib]{revtex4-2}

\usepackage{amssymb}
\usepackage{graphicx}
\usepackage{amsmath}
\usepackage{multirow}
\usepackage{blkarray}
\usepackage{nicematrix}
\usepackage{tikz}
\usetikzlibrary{quantikz2}
\usepackage{mathrsfs}
\NiceMatrixOptions{
code-for-first-row = \color{black} ,
code-for-last-row = \color{black} ,
code-for-first-col = \color{black} ,
code-for-last-col = \color{black}
}
\usepackage[bookmarks=true,colorlinks,citecolor=blue,urlcolor=blue]{hyperref}
\usepackage{braket}
\usepackage{nicematrix}
\usepackage{babel}
\usepackage{blindtext}
\usepackage{soul}
\usepackage[compat=1.1.0]{tikz-feynman}
\usepackage[normalem]{ulem}
\usepackage{physics}

\usepackage{mathtools}

\definecolor{mypurp}{rgb}{0.35, 0, 0.7}

\usepackage{amsthm}
\usepackage{txfonts}

\newcommand{\eq}[1]{Eq.\thinspace\eqref{#1}}

\newcommand{\fig}[1]{Fig.\thinspace{}\ref{#1}}

\newcommand{\figc}[2]{Fig.\thinspace{}\ref{#1}\thinspace{}#2}

\theoremstyle{definition}

\begin{document}
\def\papertitle{{Quantum Phase Transitions between Symmetry-Enriched Fracton Phases}}
\title{\papertitle}

\newcommand{\TUM}{\affiliation{Technical University of Munich, TUM School of Natural Sciences, Physics Department, 85748 Garching, Germany}}
\newcommand{\MCQST}{\affiliation{Munich Center for Quantum Science and Technology (MCQST), Schellingstr. 4, 80799 M{\"u}nchen, Germany}}
\newcommand{\MIT}{\affiliation{Center for Theoretical Physics, Massachusetts Institute of Technology, Cambridge, MA 02139, USA}}

\author{Julian Boesl} \TUM \MCQST
\author{Yu-Jie Liu} \TUM \MCQST \MIT
\author{Wen-Tao Xu} \TUM \MCQST
\author{Frank Pollmann}  \TUM \MCQST
\author{Michael Knap}  \TUM \MCQST

\begin{abstract}
    Topologically ordered phases exhibit further complexity in the presence of global symmetries: Their anyonic excitations may exhibit different transformation patterns under these symmetries, leading to a classification in terms of symmetry-enriched topological orders. We develop a generic scheme to study an analogous situation for three-dimensional fracton phases by means of isometric tensor network states (isoTNS) with finite bond dimension, which allow us to tune phase transitions between different symmetry fractionalization patterns. We focus on the X-Cube model, a paradigmatic fracton model hosting two types of excitations: lineons, which are mobile in a single direction only, and fractons that are immobile on their own. By deforming the local tensors of the fixed point ground state, we find a family of exact wavefunctions for which the symmetry fractionalization under an anti-unitary symmetry on both types of excitations is directly visible. These wavefunctions are non-stabilizer states and have non-vanishing correlation lengths. They even exhibit power-law correlations at criticality between two symmetry-enriched topological orders. Furthermore, the isoTNS description allows for the explicit construction of a linear-depth quantum circuit to sequentially realize these exotic 3D states on a quantum processor, including a holographic scheme using only a pair of two-dimensional qubit arrays alongside measurements. Our approach provides a construction to enrich phases with exotic topological or fracton order and to study 3D quantum phase transition with exact wavefunctions, and offers a tractable route to implement and characterize fracton order on quantum devices.
\end{abstract}

\maketitle

\section{Introduction}\label{sec:Introduction}
The study of exotic quantum phases of matter has led to the discovery of topological order as a novel classification paradigm~\cite{dijkgraaf_topological_1990, wen_topological_1990, wen_zoo_2017}. Topologically ordered systems host anyonic excitations with braiding statistics beyond bosons and fermions~\cite{LeinaasMyrheim,Wilczek1982,kitaev_fault-tolerant_2003}. In the presence of additional global symmetries, the interplay between topology and symmetry can lead to interesting features such as symmetry fractionalization~\cite{essin_classifying_2013, chen_symmetry_2017}. The anyons can transform non-trivially under the application of the symmetry operator, in which case the phase is termed to be a non-trivial \textit{symmetry-enriched topological} (SET) phase, analogously to how symmetries may distinguish different symmetry-protected topological (SPT) phases which would appear to belong to the same trivial phase otherwise~\cite{pollmann_entanglement_2010}. Indeed, the most prominent example of symmetry enrichment is the $\nu = 1/3$ fractional quantum Hall effect, which has been observed experimentally, where the Laughlin anyons are enriched by $U(1)$ charge conservation~\cite{tsui_two-dimensional_1982}. 
More generally, a number of theoretical investigations have shed light on SET phases in two dimensions, including their emergence from partial gauging of SPT phases or their construction via decoration procedures of topological phases~\cite{barkeshli_symmetry_2019, mesaros_classification_2013, swingle_interplay_2014, levin_braiding_2012, haller_quantum_2023, ye_classification_2024}.

\begin{figure}[h!]
    \includegraphics[width=.92\columnwidth]{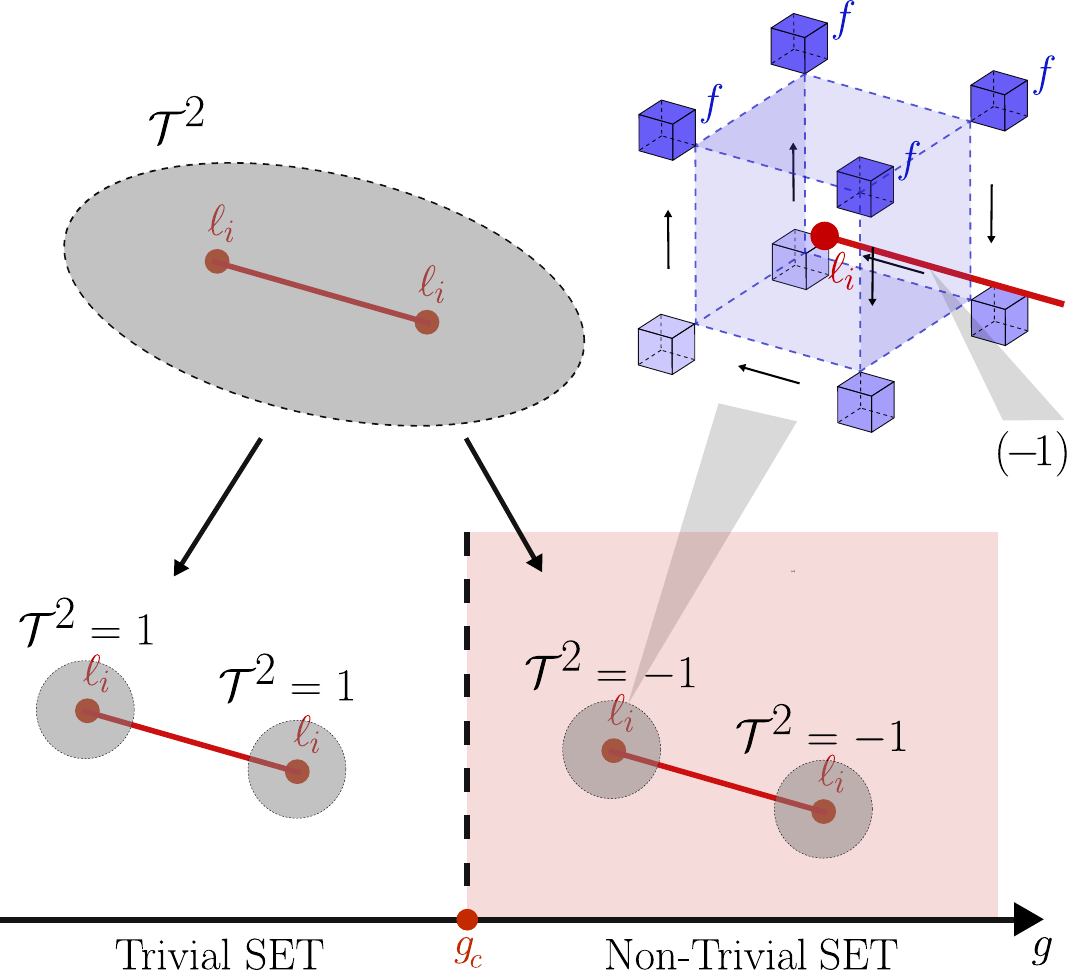}
    \caption{\label{fig:Enrichment}
        \textbf{Quantum phase transition between symmetry-enriched fracton phases.} The excitations of topological or fractonic systems can transform non-trivially under physical symmetries. A global symmetry reduces to local operators acting in the vicinity of quasi-particles. In the case of time-reversal symmetry $\mathcal{T}$, the excitations transform either trivially as $\mathcal{T}^2 = 1$ (left) or exhibit Kramers' degeneracy for $\mathcal{T}^2 = -1$ (right). In the non-trivial case, the effect of $\mathcal{T}^2$ on a subsystem is equivalent to an anyon braiding process around the subsystem, which acts as an interferometer to detect the symmetry-enriched excitation; a possible braiding of a fracton dipole around a lineon is colored in blue in the figure. Using isometric tensor network states, we deform the trivial fixed point to tune across a quantum critical point with power-law correlations at $g_c$.
    }
\end{figure}

Over the last few years, a novel class of quantum phases in three-dimensional systems has attracted increasing interest, which are broadly dubbed \textit{fracton phases}~\cite{chamon_quantum_2005, haah_local_2011, vijay_new_2015, vijay_fracton_2016, ma_topological_2018, pai_fracton_2019, pretko_fracton_2020, zhu_topological_2023, song_fracton_2024}. They are commonly characterized by their excitations which are either completely immobile on their own or can move freely only on a submanifold of the underlying lattice, as well as their subextensive ground state degeneracy on closed manifolds, which renders them distinct from more well-known cases of 3D topological order. Due to their ground state properties and the slow glassy dynamics of their fractonic excitations, these phases of matter have been proposed as useful quantum memory at finite temperatures \cite{bravyi_quantum_2013, prem_glassy_2017}. While there is no rigorously proven classification of these phases, significant process has been achieved in describing fracton order in terms of defect topological quantum field theories~\cite{aasen_topological2020, song_topological2023}. Furthermore, various connections to two-dimensional topological phases have been established: Some of these models can be realized with a construction through coupled layers of two-dimensional systems, and a partial classification can be achieved by a foliation procedure which posits an equivalence between two fracton phases when they can be connected by a finite-depth unitary upon addition of a number of decoupled two-dimensional layers~\cite{ma_fracton_2017,prem_cage-net_2019, shirley_fracton_2018, shirley_foliated_2019, slagle_foliated_2019, shirley_twisted_2020}.

In this work, we initiate a general program that unifies two unique features: First, it allows for the intuitive construction of SET phases and their transitions in three dimensions. Second, our approach shows how these exotic states can be efficiently prepared on quantum processors, even at the critical points between two gapped phases. We illustrate this on fracton models, which are some of the most exotic quantum many-body systems. Regarding the first goal, while some progress has already been made~\cite{you_symmetric_2020, tantivasadakarn_searching_2020, rayhaun_higher2023}, it seems highly pertinent to develop a simple strategy for the following problem: Given some known fracton or topological stabilizer code with a global or subsystem symmetry, how can the ground state wavefunction be modified so to have non-trivial symmetry fractionalization on some excitation? In fact, we should pose this question in more general terms: While normally fixed points with vanishing correlation length are studied, how can one systematically perturb these simple models as to allow for finite correlations and can continuous phase transitions between such two phases be constructed and analyzed. Achieving this is much more involved than it might appear at first glance: A generic perturbation on the virtual level of the tensor network will immediately destroy the fracton order at any strength, as this corresponds to a non-local perturbation on the physical level~\cite{williamson_matrix2016, shukla_boson2018}. However, even if we identify paths which leave the relevant virtual symmetries of the tensor network intact, some excitation may condense along the path and lead to an intermediate phase of different order, for instance a trivial phase~\cite{haller_quantum_2023}. These considerations are not limited to transitions between different SET orders; they will arise generally if we try to continuously connect two phases which are not related by an anyon or fracton condensation process.

One possible way to circumvent these issues lies in the structure of the ground states of stabilizer codes: As they are equal-weight superpositions of physical configurations obeying some local constraints, they can be written as tensor network states (TNS) using the so-called plumbing structure~\cite{liu_simulating_2023, yu_dual2024}. In essence, this property locks physical and virtual degrees of freedom of a local tensor, thereby allowing us to express all information about the state in a local matrix which is obtained by stripping off plumbing tensors between the physical and the virtual level. We refer to this matrix as \emph{$W$-matrix}. This may be interpreted as a mapping from the quantum state norm to the partition function of a classical statistical model~\cite{verstraete_criticality_2006}. From this construction, it is easy to read off which perturbations to the $W$-matrix leave the virtual symmetries of the topological order invariant. At the same time, it may be modified in such a way that a local symmetry operation on the physical legs corresponds to a non-trivial operator on the virtual legs, leading to symmetry fractionalization on excitations which are detected by this virtual operator~\cite{cirac_matrix_2021}.

Assuming the $W$-matrix along the path fulfills a normalization condition further restricts possible deformations and allows for evaluation of diagonal correlation functions by associating the tensor network to a stochastic process. Importantly, this can be used to check whether the correlation length diverges, signaling a second-order phase transition. In fact, this condition implies the states are isometric tensor network states (isoTNS)~\cite{zaletel_isometric_2020, haghshenas_conversion_2019, Kadow2023, malz_computational_2024}. This subclass of general TNS obeys an additional isometry condition which allows for efficient contraction; despite this strong restriction, it has been found to represent a wide array of quantum phases~\cite{soejima_isometric_2020}. Similar to 1D matrix product states (MPS), isoTNS can be prepared efficiently with a linear-depth sequential quantum circuit~\cite{schon_sequential_2005, wolf_quantum_2006, jones_skeleton_2021, smith_crossing_2022, wei_sequential_2022, chen_sequential_2024}. As topologically ordered states have recently been realized on various quantum processing platforms~\cite{satzinger_realizing_2021, liu_methods_2022, cochran_visualizing_2024, semeghini_probing_2021, verresen_prediction_2021, iqbal_non-abelian_2024, tantivasadakarn_shortest_2023}, and first proposals of protocols for stabilizer codes exhibiting fracton order arose~\cite{nevidomskyy_realizing_2024, chen_quantum_2024}, this suggests to further these concepts to studying finite-correlation states away from these fixed points, and even driving phase transitions between different orders.

In summary, the purpose of this paper is to demonstrate how to leverage plumbed isoTNS in the study of fracton phase transitions, with particular focus on symmetry enrichment: They can be used to transparently construct fracton wavefunctions with non-trivial symmetry fractionalization, they allow parametrized paths between the fixed points of the distinct phases which undergo a direct phase transition, and they provide a simple scheme to realize this transition on a quantum simulation platform. Importantly, this can be achieved without increasing the bond dimension of the tensors, even at the phase transition. We apply this program to one of the most paradigmatic models with fracton order, the X-Cube model. Its fractional excitations come in two types: lineons, which can move in a single direction, and completely immobile fractons. In Section \ref{sec:Xcube}, we introduce the model and its different tensor network representation, which feature the plumbing property discussed above. In Sections \ref{sec:Lineon} and \ref{sec:Fracton}, we discuss deformations of the local tensors in each of the two networks, which allow tuning to phases where one of the excitations exhibits non-trivial symmetry fractionalization under an anti-unitary $\mathbb{Z}_2^{T}$ symmetry. In Section \ref{sec:Preparation}, we provide the associated sequential quantum circuits which generate the wavefunctions along the path. We further propose a holographic preparation scheme which uses a pair of two-dimensional lattices, significantly reducing the number of qubits required. Finally, in Section \ref{sec:Outlook}, we point out further phase transitions to which our approach is applicable, within the realm of symmetry enrichment and beyond it.

\begin{figure}
    \includegraphics[width=0.95\columnwidth]{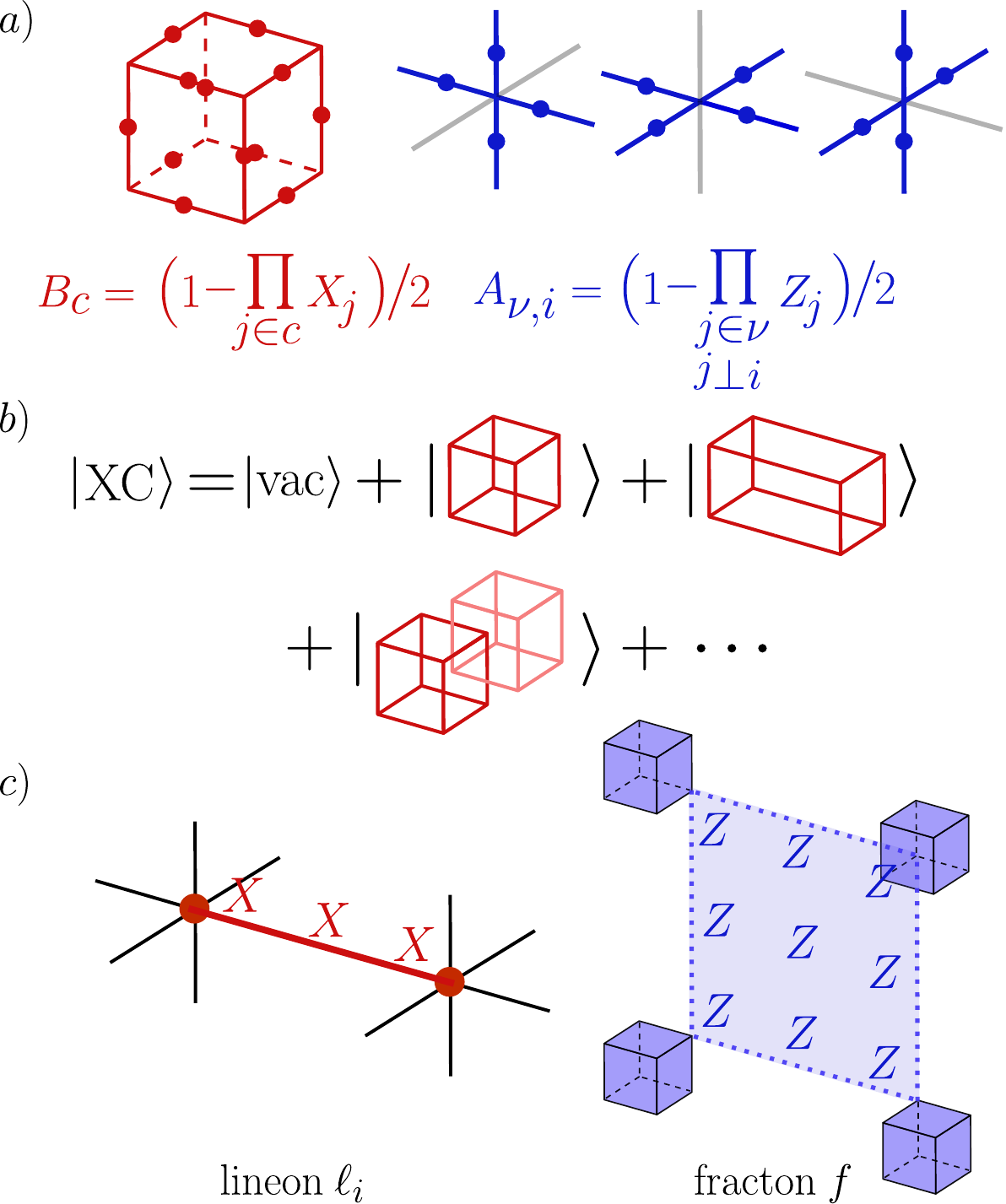}
    \caption{\label{fig:XCube}
        \textbf{The X-Cube model.} a) The local stabilizers of the X-Cube model on a three-dimensional cubic lattice. b) The ground state of the X-Cube model. The ground state in the bulk is an equal-weight cuboid condensate in the $Z$ basis on top of the vacuum. It is built up by sequentially applying the sum of the identity operator and an $X$ stabilizer to the vacuum, thereby realizing all allowed combinations. On a closed manifold, the ground state degeneracy is subextensive. c) The local excitations of the X-Cube model. Strings of $X$ operators create pairs of lineons living on the end vertices, which move freely in one direction, while changing the lineon's direction is possible only by creating further excitations according to the triple fusion rule of three distinct types of lineons fusing to the identity. Membranes of $Z$ operators create fours of fractons in the cubes at the corners of the membrane, which are fully immobile on their own. The stabilizers in a) are associated to the minimal braiding processes of lineon/fracton dipoles around a single cube/vertex.
    }
\end{figure}
\section{X-Cube model}\label{sec:Xcube}
The starting point of our investigation is the X-Cube model~\cite{vijay_fracton_2016, shirley_fracton_2018, slagle_x-cube_2018}. It is defined on a three-dimensional cubic lattice. On each edge we put a qubit. Introducing the Pauli operators $Z$ (with $Z \ket{0} = \ket{0}$ and $Z \ket{1} = -\ket{1}$) and $X$ (with $X \ket{0} = \ket{1}$ and $X \ket{1} = \ket{0}$), the X-Cube Hamiltonian is defined as

\begin{equation}
    H_{\text{XC}} = \sum_c B_c + \sum_{\nu, i = x,y,z} A_{\nu, i},
   \label{eq:XCHamiltonian}
\end{equation}
where the projectors $B_c = (1 -\prod_{j \in c} X_j)/2$ and $A_{\nu,i} = (1-\prod_{j \in \nu, j \perp i} Z_j)/2$ act on the twelve edges around each cube $c$ and the four edges around each vertex $\nu$ perpendicular to direction $i$, respectively (see \figc{fig:XCube}{a} for a visualization). 

This Hamiltonian is exactly solvable due to its stabilizer-code structure; $A_{\nu,i}$ and $B_c$ commute and a ground state $\ket{\text{XC}}$ minimizes every stabilizer $B_c \ket{\text{XC}} = 0 \;\forall c, \; A_{\nu, i} \ket{\text{XC}} = 0 \; \forall \nu, i$. As a stabilizer code, it has vanishing correlation length. The ground state is written explicitly by applying projectors to the vacuum state in the form $\ket{\text{XC}}\propto \prod_c \left( 1 + \prod_{j \in c} X_j \right) \ket{00\cdots0}$; this is a condensate of cuboid configurations in the $Z$ basis, which generalizes the string-net liquid construction in two dimensions~\cite{levin_string-net_2005, lin_generalized_2021} (see \figc{fig:XCube}{b}). Indeed, this connection can be formalized using a coupled layer construction of stacks of toric code ground states which produces the X-Cube state in a strong-coupling limit~\cite{ma_fracton_2017}. As is typical for fractonic systems, in the case of periodic boundary conditions the dimension of the ground state manifold of this Hamiltonian scales exponentially in the linear system size in all three directions due to a subextensive number of non-local loop operators~\cite{shirley_fracton_2018, he_entanglement_2018}. However, as each ground state is locally indistinguishable, the choice of boundary conditions is not relevant in this section.

The excitations of this ground state correspond to violations of the stabilizers. The application of a straight line of $X$ operators in direction $i$ creates two so-called lineons $\ell_i$ living on the vertices $\nu, \nu^\prime$ at the end of the string with $A_{\nu^{(\prime)}, j} = 1$ for $j \neq i$. These particles can only be created in pairs and are mobile on their own only in the specified direction $i$, as the change of direction of a lineon leads to the creation of further lineon excitations. This process can be put succinctly in the triple fusion rule $\ell_x \times \ell_y \times \ell_z = 1$, which implies that three lineons along different directions fuse to the vacuum. On the other hand, $B_c = 1$ violations can be induced by the application of a membrane of $Z$ operators, with a fracton $f$ living on a cube at each corner of this membrane. They can only be created in fours and are completely immobile on their own, as there is no local operator which can move a single fracton without the creation of further excitations (see \figc{fig:XCube}{c} for an illustration of the particle types and the corresponding string/membrane operators). Furthermore, both lineons and fractons are their own anti-particles, $\ell_i \times \ell_i = f \times f = 1$.

Bound states of both lineons and fractons feature enhanced mobility, as they can freely move in a plane perpendicular to their internal direction. Braiding a lineon dipole around a fracton $f$ (equivalent to a wireframe of $X$ operators) or braiding a fracton dipole around a lineon $\ell_i$ where $i$ is different from the dipole's internal direction (equivalent to a tube of $Z$ operators) leads to a sign of $-1$, allowing the braiding statistics of these quasiparticles to be defined~\cite{shirley_fractional_2019}. The stabilizers $B_c$ and $A_{\nu,i}$ thus check whether fractons or lineons are present by applying the smallest possible braiding processes; they are simultaneously minimized by the ground state, i.e. the vacuum state of lineons and fractons.

The system features different types of symmetries: Besides the often-discussed subsystem symmetries which consist of $Z$ or $X$ operators on a single plane in the lattice, it also hosts global $Z$ and $X$ symmetries. In particular, as the ground state wavefunction is real, it remains invariant under complex conjugation $\mathcal{K}$; we can therefore promote the two global symmetries to two anti-unitary time-reversal-like symmetries $\mathcal{T}_X = \left( \prod_i X_i \right) \mathcal{K}$ and $\mathcal{T}_Z = \left( \prod_i Z_i \right) \mathcal{K}$. At the X-Cube fixed point discussed so far, the excitations transform linearly under both $\mathbb{Z}^{T}_2$ symmetries. In the following sections, we will explore how allowing for complex entries in the tensor can instead lead to projective transformations.
\begin{figure}
    \includegraphics[width=0.98\columnwidth]{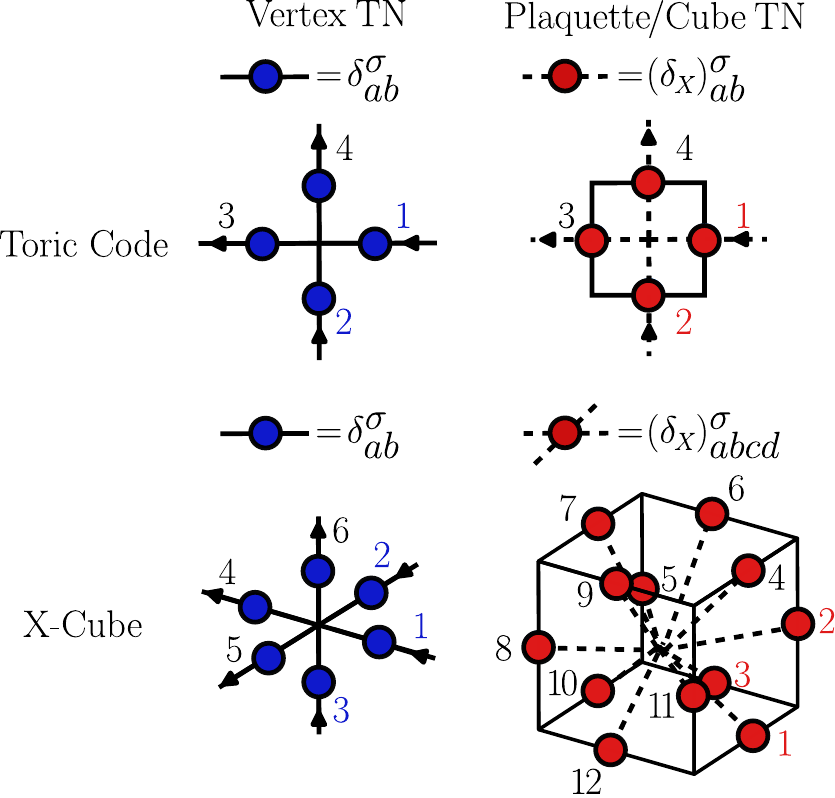}
    \caption{\label{fig:Tensors}
        \textbf{Tensor networks for wavefunctions.} 
        Graphical illustration of the isometric tensor networks (isoTNS) for the ground states of the toric code and the X-Cube model. For each tensor network, an $n$-leg tensor is dressed with ``plumbing tensors" on each leg (blue or red symbols); they determine the physical degree of freedom depending on the virtual degrees of freedom. The local tensor encoding the information on the virtual degrees of freedom can be written in terms of a $W$-matrix, the rows corresponding to configurations on the incoming legs and the columns corresponding to the outgoing legs (indicated with arrows; the labels of the incoming legs are colored in blue and red, respectively). The $W$-matrix for both toric code networks is the same; the X-Cube $W$-matrices are distinct and are given in \eq{eq:WVertex} and \eq{eq:WCube}, respectively.
    }
\end{figure}
To study the symmetry enrichment of the fractonic excitations, it is useful to go to a description of the ground state in terms of local objects. An effective way to do so is to use tensor network states (TNS)~\cite{cirac_matrix_2021}. For a three-dimensional cubic lattice with one spin degree of freedom on each edge, we can define multi-rank tensors $T^{\sigma\rho\tau}_{i_1\cdots,i_n}$ which encode the local information in each unit cell. The three legs $\sigma, \rho, \tau$ with dimension $d =2 $ correspond to the three qubits in the unit cell, while the virtual legs $i_1, \dots, i_n$ with a certain bond dimension $D$ connect the considered tensor to adjacent tensors. The total number $n$ of virtual legs may vary depending on the structure of the tensor network. For a system of $N$ qubits, the full wavefunction is obtained by a contraction over all virtual degrees of freedom
\begin{equation}
    \ket{\Psi} = \sum_{\sigma_1, \cdots, \sigma_N} \text{tTR}\left( \left\{ T^{\sigma_1\sigma_2\sigma_3}, \cdots, T^{\sigma_{N-2}\sigma_{N-1}\sigma_N} \right\} \right) \ket{\sigma_1\cdots\sigma_N},
    \label{eq:TensorWF}
\end{equation}
where $\text{tTR}$ denotes the tensor contraction.

There are multiple ways to represent the X-Cube ground state as a TNS. The simplest choice is to introduce vertex tensors $\left(T_\nu \right)^{\sigma\rho\tau}_{i_1 \cdots i_6}$ with six virtual legs corresponding to the six edges adjacent to the vertex. We can further express this tensor in terms of a $W$-matrix as

\begin{align}
    &\left(T_\nu \right)^{\sigma\rho\tau}_{i_1\cdots i_6} = \sum_{i,j,k} \delta^\sigma_{i_4i} \delta^\rho_{i_5j} \delta^\tau_{i_6k} \left( W_\nu \right)_{(i_1i_2i_3)(ijk)} \nonumber \\
    & \includegraphics[scale=0.5]{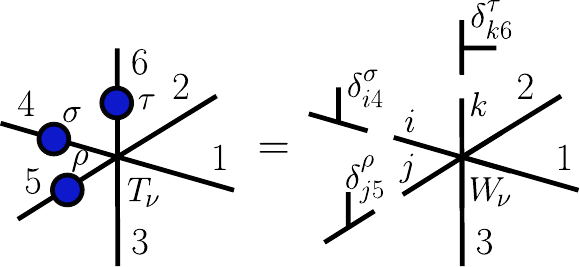},
    \label{eq:TVertex}
\end{align}
where $\delta^\sigma_{ab}$ is a ``plumbing tensor" which takes the value $\delta^\sigma_{ab} = 1$ if $\sigma = a = b$ and $0$ otherwise. The equation is also visualized graphically for clarity. This procedure makes the virtual legs of the tensor equivalent to the physical legs and provides a mapping to classical partition functions (see Appendix~\ref{apx:ClassPart}). An important property of this tensor is that $X$ and $Z$ operators can be pulled from the physical degree of freedom to the virtual legs:

\begin{equation}
    \includegraphics[scale=0.5]{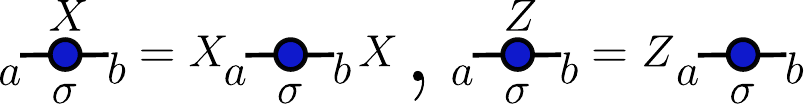}.
    \label{eq:PlumbingZ}
\end{equation}
Due to symmetry, the $Z$ operator can equally be applied to the other virtual leg. In the following, this will allow us to rewrite the effect of a physical symmetry as an operator on the virtual legs.
As we consider qubits, the bond dimension of the vertex tensor $T_\nu$ is $d = D = 2$. Accordingly, $W_\nu$ is an $8 \times 8$ matrix, the columns of which correspond to qubit configurations of the edges dressed with plumbing tensors, while the rows correspond to the configurations of the other legs. As the wavefunction in question is an equal-weight superposition of all closed cuboid configurations, the $W$-matrix for this TNS has a simple structure:

\begin{widetext}
\begin{equation}
    W_\nu = \begin{pNiceMatrix}[first-row,last-col]
 \ket{000} & \ket{001} & \ket{010} & \ket{011} & \ket{100} & \ket{101} & \ket{110} & \ket{111} &    \\
 \frac{1}{\sqrt{2}}  & 0   & 0  & 0  & 0 & 0  & 0 & \frac{1}{\sqrt{2}}  & \;\ket{000} \\
 0   & \frac{1}{\sqrt{2}}   & 0 & 0  & 0 & 0 & \frac{1}{\sqrt{2}} & 0  & \;\ket{001} \\
0   & 0   & \frac{1}{\sqrt{2}}  & 0  & 0 & \frac{1}{\sqrt{2}}  & 0 & 0  & \;\ket{010} \\
0   & 0  & 0  & \frac{1}{\sqrt{2}}  & \frac{1}{\sqrt{2}} & 0  & 0 & 0  & \;\ket{011} \\
0   & 0  & 0  & \frac{1}{\sqrt{2}}  & \frac{1}{\sqrt{2}} & 0  & 0 & 0  & \;\ket{100} \\
0   & 0   & \frac{1}{\sqrt{2}}  & 0  & 0 & \frac{1}{\sqrt{2}}  & 0 & 0  &  \;\ket{101} \\
 0   & \frac{1}{\sqrt{2}}   & 0 & 0  & 0 & 0 & \frac{1}{\sqrt{2}} & 0  & \;\ket{110} \\
\frac{1}{\sqrt{2}}  & 0   & 0  & 0  & 0 & 0  & 0 & \frac{1}{\sqrt{2}}  & \;\ket{111} \\
\end{pNiceMatrix}.
\label{eq:WVertex}
\end{equation}
\end{widetext}
The tensor defined by this $W$-matrix exhibits virtual symmetries: In each plane, the number of $\ket{1}$ states on  the four edges in this plane is even. The presence of these symmetries is equivalent to the closed cuboid constraint; breaking them leads to loss of fracton order as lineons condense~\cite{shukla_boson2018}.

This tensor falls in the class of isometric tensor networks (isoTNS)~\cite{zaletel_isometric_2020} as it exhibits the defining isometric property

\begin{align}
    & \sum_{\sigma, \rho, \tau, i_1, i_2, i_3} \left( \left(T_\nu \right)^{\sigma\rho\tau}_{i_1i_2i_3i_4i_5i_6}\right)^* \left(T_\nu \right)^{\sigma\rho\tau}_{i_1i_2i_3i^\prime_4i^\prime_5i^\prime_6} = \delta_{i_4 i^\prime_4} \delta_{i_5 i^\prime_5} \delta_{i_6 i^\prime_6} \nonumber \\
    & \includegraphics[scale=0.5]{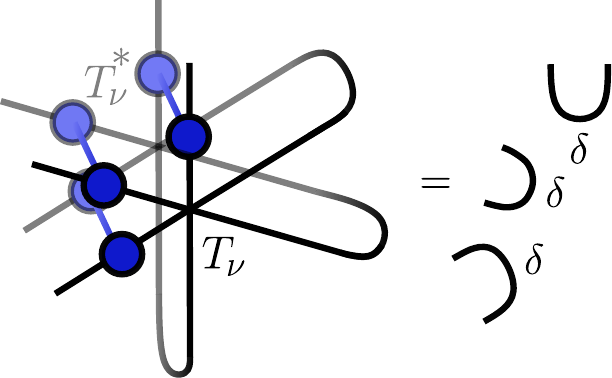}.
    \label{eq:IsoTNS}
\end{align}

On the level of the $W$-matrix, this is equivalent to the fact that the columns of the matrix are normalized, $\sum_{i_1, i_2, i_3} \vert \left(W_\nu\right)_{(i_1i_2i_3)(i_4i_5i_6)}\vert^2 = 1 \forall i_4, i_5, i_6 $. IsoTNS can be seen as higher-dimensional analogues of 1D matrix product states. They correspond to an expressive subclass of tensor networks that can describe a large class of gapped phases~\cite{soejima_isometric_2020} and may also support power-law correlations~\cite{liu_simulating_2023}. The identity (\ref{eq:IsoTNS}) allows for efficient contraction of the tensor network in the direction given by the edges dressed with plumbing tensors; we may label them as outgoing legs of the local tensor, while the remaining legs are incoming (see \fig{fig:Tensors} for a visualization of the tensor in which arrows indicate the direction that satisfies the isometry condition). Furthermore, they can be generated by linear-depth sequential quantum circuits and are therefore promising candidates for the realization on current quantum simulation platforms (see Appendix~\ref{apx:IsoTNS} for a more detailed discussion on isoTNS in three dimensions).

The construction of the vertex tensor $T_\nu$ is a rather straightforward generalization of the isoTNS which describes the 2D toric code fixed point on a square lattice, which is a condensate of loop configurations~\cite{kitaev_fault-tolerant_2003, satzinger_realizing_2021, liu_simulating_2023}. We describe it as a tensor network consisting of vertex tensors, which is decomposed into a $W$-matrix $W^{(TC)}$ and plumbing tensors $\delta^\sigma_{ab}$. However, due to the geometric properties of the square lattice and a duality between the $X$ and $Z$ stabilizers in this system, a second tensor network description is possible, in which we put a matrix $W^{(TC)}$ in the center of each plaquette. The $W$-matrices are connected by modified plumbing tensors $(\delta_X)^\sigma_{ab}$ on each edge shared by two plaquettes, which are defined as $\left(\delta_X\right)^\sigma_{ab} = 1$ for $a = b = 0$, $\left(\delta_X\right)^\sigma_{ab} = (-1)^{\sigma}$ for $a = b = 1$ and $0$ otherwise (see \fig{fig:Tensors} for a visualization of the geometric structure of both tensor networks). While an analogous duality does not exist in the X-Cube model, it inspires another tensor-network representation of the X-Cube ground state. We define a tensor $\left(T_c \right)^{\sigma\rho\tau}_{i_1\cdots i_{18}}$ in the middle of a cube $c$ as

\begin{equation}
    \left(T_c \right)^{\sigma\rho\tau}_{i_1\cdots i_{18}} = \sum_{i,j,k} \left(\delta_X\right)^\sigma_{i_1i_2i_3i} \left(\delta_X\right)^\rho_{i_4i_5i_6j} \left(\delta_X\right)^\tau_{i_7i_8i_9k} \left( W_c \right)_{(ijk)(i_{10}\cdots i_{18})},
    \label{eq:TCube}
\end{equation}
where $\left(\delta_X\right)^\sigma_{abcd}$ is a modified plumbing tensor which takes the value $\left(\delta_X\right)^\sigma_{abcd} = 1$ for $a = b = c = d = 0$, $\left(\delta_X\right)^\sigma_{abcd} = (-1)^{\sigma}$ for $a = b = c = d = 1$ and $0$ otherwise. This plumbing tensor also allows a pulling-through procedure for the operators acting on the physical qubit, in which the operator is transformed between the physical leg and the virtual legs. The relations are

\begin{equation}
    \includegraphics[scale=0.5]{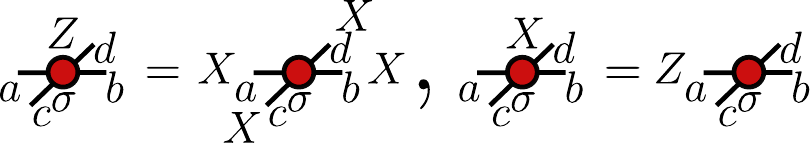}.
    \label{eq:PlumbingX}
\end{equation}
Here again, the $Z$ operator in the second relation can be applied to any single virtual leg. The structure of the tensor $T_c$ can be understood as follows: On three edges which share a corner, a physical leg representing a qubit is put. From each of these edges, three virtual legs connect the cube tensor to the three other cubes which share this edge. Finally, the other nine virtual legs lead to the nine other edges of the cube (see \fig{fig:Tensors} for a visualization of this tensor). This local structure leads to a tensor network which spans the entire lattice. The bond dimension is again $d =D = 2$. Therefore, $W_c$ is a $2^3 \times 2^9$ matrix. It can be checked that this object will return the X-Cube ground state if we define $W_c$ as
\begin{equation}
    \left(W_c \right)_{(i_1i_2i_3),(i_4\cdots i_{12})} = \begin{cases} \frac{1}{2} & \text{if} \;\; \left( \sum_{j = 1, \dots, 12} i_j \right) \mod 2 = 0 \\
    0 & \text{if} \;\; \left( \sum_{j = 1, \dots, 12} i_j \right) \mod 2 = 1.
    \end{cases}
    \label{eq:WCube}
\end{equation}
Here as well, a virtual symmetry which forbids configurations with an odd parity of $\ket{1}$ states on the surrounding edges is respected. This matrix is again normalized as $\sum_{i_1, i_2, i_3} \vert \left(W_c\right)_{(i_1i_2i_3)(i_4\cdots i_{12})}\vert^2 = 1 \forall i_4, \dots, i_{12}$. This construction allows for sequential contraction of the network in a similar way as in the vertex network; in this case, the three legs of the $W$-matrix which connect to the plumbing tensors in the full tensor (\ref{eq:TCube}) are the incoming legs, while the other nine legs are outgoing (see Appendix~\ref{apx:IsoTNS} for a detailed discussion of the contraction properties of this tensor network). While this definition appears unhandily complicated at first glance, it will become useful when considering fractonic excitations, which live on the cubes of the lattice.

\begin{figure}
    \includegraphics[width=0.95\columnwidth]{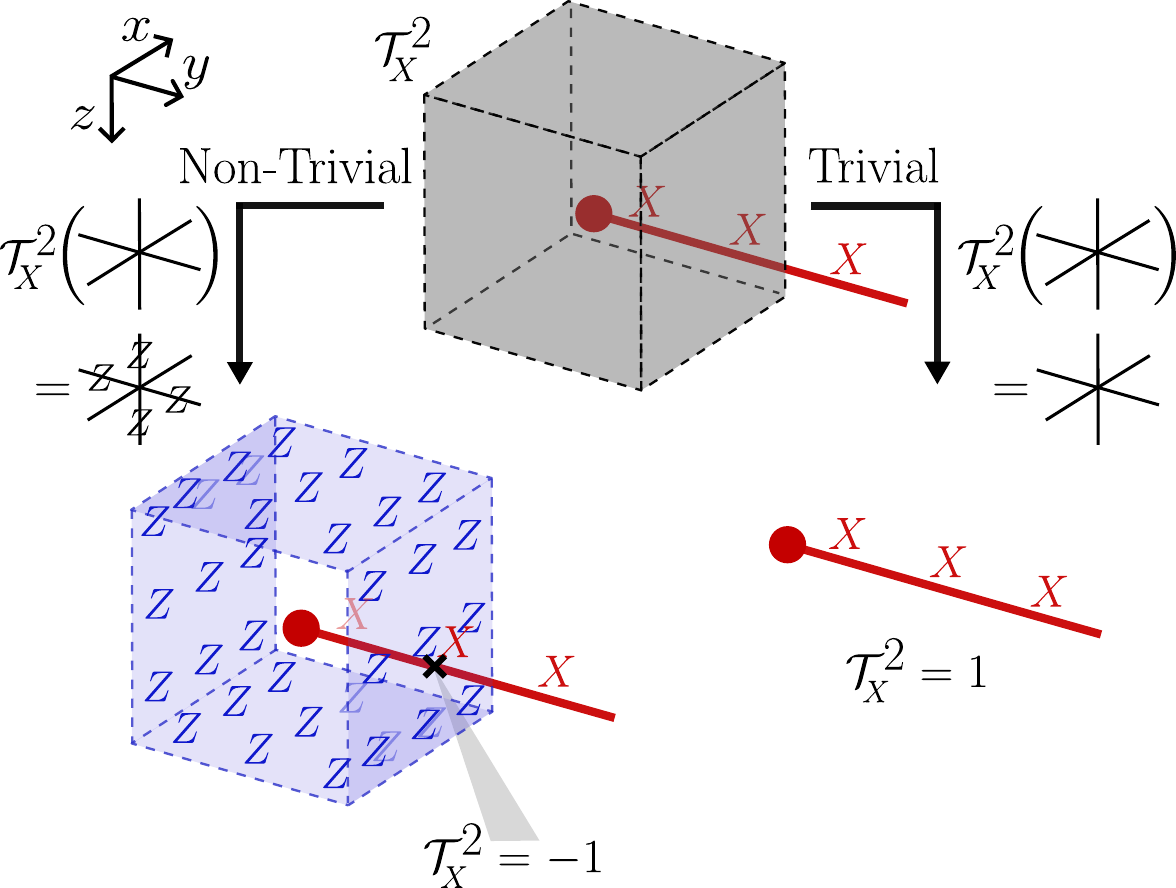}
    \caption{\label{fig:LineonCubeOperator}
        \textbf{Effect of local symmetry operation.}  When excitations are present on top of a fractonic or topological ground state, the effect of a global symmetry can be reduced to local operators in the vicinity of these excitations. This is visualized here for lineon excitations (red semi-infinite line) in the X-Cube model and its behavior under double application of the anti-unitary symmetry $\mathcal{T}_X = \left( \prod_i X_i \right) \mathcal{K}$: In the trivial phase, the local operator is the identity, as can be seen by considering the effect of the symmetry on a single vertex tensor. In a phase with non-trivial symmetry enrichment, by contrast, $\mathcal{T}_X^2$ is equivalent to a boundary operator of $Z$s. This process corresponds to braiding a fracton dipole with a specific directionality around the subsystem. This operator then measures the presence of lineon excitations as the string and the tube operator do not commute at the black cross, implying symmetry fractionalization on these lineons as $\mathcal{T}_X^2 = -1$.
    }
\end{figure}

\section{Lineon enrichment}\label{sec:Lineon}
A topologically ordered system with a global physical symmetry is called a symmetry-enriched topological (SET) phase. The system belongs to a non-trivial SET phase when its excitations transform non-trivially under the physical symmetry~\cite{essin_classifying_2013, chen_symmetry_2017, barkeshli_symmetry_2019}. A physical way to imagine such a scenario is shown in \fig{fig:Enrichment}: Suppose a pair of well-separated quasi-particle excitations has been created on top of the ground state. Since the bulk of the wavefunction stays invariant under the global physical symmetry,  the action of the symmetry can be reduced to certain unitaries applied only in the vicinity of the excitations. In particular, these unitaries may transform under a projective representation of the symmetry group; when this is the case, the symmetry is said to fractionalize non-trivially on the excitations~\cite{essin_classifying_2013, chen_symmetry_2017, barkeshli_symmetry_2019}. 

As long as the symmetry is preserved, different SET phases cannot be adiabatically transformed into each other but instead must be separated by a quantum phase transition. The possible SET orders for a symmetry group and a set of anyons are classified by cohomology theory~\cite{essin_classifying_2013}. In the case of an anti-unitary $\mathbb{Z}_2^T$ symmetry and a $\mathbb{Z}_2$ anyon, the possible fractionalization patterns are given by the second cohomology group $H^{(2)}(\mathbb{Z}_2^T,\mathbb{Z}_2)=\mathbb{Z}_2$. The symmetry either fractionalizes trivially, if the local unitaries on the anyons transform under the linear representation of the $\mathbb{Z}_2^T$ symmetry, or non-trivially if the representation is projective. In the latter case, the excitation behaves under the symmetry like a half-integer object under time-reversal, i.e. it exhibits Kramer's degeneracy~\cite{you_symmetric_2020}. In this section, we want to study these distinct SET orders in the particular example of the lineons in the X-Cube model under the aforementioned anti-unitary symmetry $\mathcal{T}_X$. At the fixed point given by the Hamiltonian in  Eq. \eqref{eq:XCHamiltonian}, the lineons transform under a linear representation of the symmetry group, $\mathcal{T}_X^2 = 1$. In the following, we will construct states where the representation $\mathcal{T}_X^2 = -1$ is projective, implying that the lineons are doubly degenerate under this symmetry. The number of possible fractionalization patterns is further restricted by the triple fusion rule $\ell_x \times \ell_y \times \ell_z = 1$: As the vacuum $1$ has to remain invariant under the symmetry, precisely two types of lineons have to be non-trivially enriched at the same time, while the third transforms trivially. In addition to the regular X-Cube phase without any non-trivial symmetry fractionalization, this implies that there are at least three additional non-trivial SET phases.

To tune between the different possible phases, we introduce a two-parameter vertex $W$-matrix $W_\nu(g_1, g_2)$ with $g_{1/2} \in [-1,1]$ defined as

\begin{widetext}
\begin{equation}
    W_\nu(g_1, g_2) = \begin{pNiceMatrix}[first-row,last-col]
 \ket{000} & \ket{001} & \ket{010} & \ket{011} & \ket{100} & \ket{101} & \ket{110} & \ket{111} &    \\
 \frac{1}{\sqrt{2}}  & 0   & 0  & 0  & 0 & 0  & 0 & \frac{1}{\sqrt{2}}  & \;\ket{000} \\
 0   & \frac{1}{\sqrt{1+\vert g_1\vert}}   & 0 & 0  & 0 & 0 & \sqrt{\frac{\text{sign}(g_1) \vert g_1\vert }{1+\vert g_1\vert }} & 0  & \;\ket{001} \\
0   & 0   & \frac{1}{\sqrt{1+\vert g_2\vert }}  & 0  & 0 & \sqrt{\frac{\text{sign}(g_2) \vert g_2\vert }{1+\vert g_2\vert }}  & 0 & 0  & \;\ket{010} \\
0   & 0  & 0  & \frac{1}{\sqrt{1+\vert g_1 g_2\vert }}  & \sqrt{\frac{ \text{sign}(g_1 g_2) \vert g_1 g_2\vert }{1+\vert g_1 g_2\vert }} & 0  & 0 & 0  & \;\ket{011} \\
0   & 0  & 0  & \sqrt{\frac{ \text{sign}(g_1 g_2) \vert g_1 g_2\vert }{1+\vert g_1 g_2\vert }} & \frac{1}{\sqrt{1+\vert g_1 g_2\vert }} & 0  & 0 & 0  & \;\ket{100} \\
0   & 0   & \sqrt{\frac{\text{sign}(g_2) \vert g_2\vert }{1+\vert g_2\vert }}  & 0  & 0 & \frac{1}{\sqrt{1+\vert g_2\vert }}  & 0 & 0  &  \;\ket{101} \\
 0   & \sqrt{\frac{\text{sign}(g_1) \vert g_1\vert }{1+\vert g_1\vert }}   & 0 & 0  & 0 & 0 & \frac{1}{\sqrt{1+\vert g_1\vert }} & 0  & \;\ket{110} \\
\frac{1}{\sqrt{2}}  & 0   & 0  & 0  & 0 & 0  & 0 & \frac{1}{\sqrt{2}}  & \;\ket{111} \\
\end{pNiceMatrix}.
\label{eq:WSETLineon}
\end{equation}
\end{widetext}
The $W$-matrix remains normalized for all values of $g_{1/2}$, ensuring that the tensor network retains the isoTNS property. As the deformation does not allow for new local configurations, the closed cuboid constraint is fulfilled; as this virtual symmetry is always exactly conserved, we do not expect a sudden first-order transition into a trivial state (as is confirmed numerically). Furthermore, the corresponding wavefunction remains time-reversal symmetric for the entire parametrization. This ensures that there is no gapped adiabatic path between states hosting distinct fractionalization patterns.

For $g_1 = g_2 = 1$, the $W$-matrix is that of the X-Cube fixed point as given in \eq{eq:WVertex}. If at least one of the two parameters is negative, $g_{1/2} <0$, certain off-diagonal entries become complex. The set of configurations with complex entries can be explained as follows: Depending on the sign of $g_1$, $g_2$ and $g_1g_2$, a spatial direction is chosen. The four legs of the tensor in the plane orthogonal to this direction are split into two pairs: a complex unit will be allocated to a configuration if precisely one of these two pairs is in the state $\ket{11}$. This is an example of a more general scheme to modify stabilizer-code isoTNS in such a way as to obtain non-trivial symmetry fractionalization. In the next section, a similar logic will be used to treat the fractonic excitations of the X-Cube model; in Appendix~\ref{apx:3DTC}, we discuss the vertex excitations of the three-dimensional toric code as an additional example.

Before discussing how the different fractionalization patterns are expressed on the level of the $W$-matrix, we want to point out that parent Hamiltonians for these states can easily be constructed~\cite{liu_simulating_2023, yu_dual2024}. For $g_{1/2} > 0$, the wavefunction $\ket{\Psi(g_1, g_2)}$ which is constructed from $W_\nu(g_1, g_2)$ is connected to the X-Cube wavefunction $\ket{\text{XC}}$ by an imaginary time evolution
\begin{equation}
    \ket{\Psi(g_1, g_2)} =\prod_\nu e^{\sum_j \beta_j P_\nu^{(j)}} \ket{\text{XC}}.
\end{equation}
Here,  $P_\nu^{(j)}$ is a projector to a specific configuration of the six qubits around the vertex $\nu$. The imaginary times $\beta_j < \infty$ are functions of $g_1$ and $g_2$; the total transformation thus amounts to locally changing the weights of the configurations. In fact, the parametrization of \eq{eq:WSETLineon} is continuously connected to a path where the wavefunction remains analytic in $g_{1/2}$ for the entire range of values. The definition of $\beta_j$ can therefore be extended to negative $g_{1/2}$~\cite{haller_quantum_2023}.

The deformed state $\ket{\Psi(g_1, g_2)}$ is annihilated by a modified cube operator
\begin{equation}
    \tilde{B}_c(g_1, g_2) = \left( \prod_\nu e^{\sum_j \beta_j P_\nu^{(j)}} \right) B_c \left( \prod_\nu e^{-\sum_j \beta_j P_\nu^{(j)}} \right).
\end{equation}
It should be noted that $\tilde{B}_c(g_1, g_2)$ is a 36-body local operator as $B_c$ commutes with all projectors which are not defined on adjacent vertices. The vertex stabilizers $A_{\nu, i}$ need not be modified as they trivially commute with all projectors. A frustration-free, local parent Hamiltonian of $\ket{\Psi(g_1, g_2)}$ is then given by

\begin{equation}
    H(g_1, g_2) = \sum_c \tilde{B}_c^\dagger(g_1, g_2)\tilde{B}_c(g_1, g_2) \mathcal{P}_c + \sum_{\nu, i = x,y,z} A_{\nu, i},
\end{equation}
where $\mathcal{P}_c = \prod_{\nu \in c, i =x,y,z} (1- A_{\nu,i})$ is a projector onto the closed-cuboid subspace around each cube $c$. ($\nu \in c$ signifies the vertex $\nu$ lies on a corner of the cube $c$.) This additional projector is present to ensure that the Hamiltonian respects the symmetry $\mathcal{T}_X$ also for negative $g_{1/2}$.

\begin{figure}
 \includegraphics[width=\columnwidth]{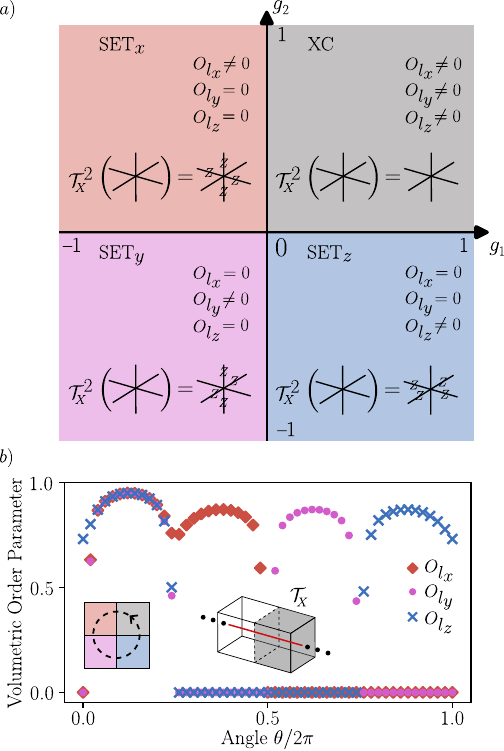}
    \caption{\label{fig:LineonDrawings}
        \textbf{Lineon Enrichment.} a) The phase diagram of the wavefunction given by the matrix $W_\nu (g_1, g_2)$, Eq.~\eqref{eq:WSETLineon}, with different symmetry fractionalization indicated in the insets. As long as $g_{1/2} >0$, the wavefunction is in the same trivial phase as the X-cube fixed point. When at least one parameter changes sign, certain lineon types exhibit non-trivial symmetry fractionalization. We also indicate the associated (non-)vanishing order parameters. b) Non-local volumetric order parameters $O_{l_i}$ with $i = x,y,z$ on a circular path across the phase diagram (illustrated in the inset, left). The order parameter is evaluated by splitting the system into two semi-infinite halves, each of which hosts an excitation, and applying the relevant symmetry to one half (illustrated in the inset, right). When the symmetry fractionalizes non-trivially on a lineon type $l_i$, the corresponding order parameter $O_{l_i}$ vanishes.
    }
\end{figure}

As long as both parameters are positive, $g_{1/2} > 0$, the system is in the same phase as the X-Cube fixed point with regard to the symmetry fractionalization under the time-reversal symmetry. We may consider the effect of applying $\mathcal{T}_X = \left( \prod_{i=1}^6 X_i \right) \mathcal{K}$ on a vertex matrix $W_\nu$ with a plumbing tensor $\delta_{ab}^\sigma$ attached to each of the six virtual legs. As all entries are real, complex conjugation $\mathcal{K}$ has no effect, while \eq{eq:PlumbingZ} allows us to pull the $X$ operators on the virtual legs of the plumbing tensor $\delta_{ab}^\sigma$. $W_\nu$ is invariant under a flip of all entries, leaving only the $X$ on the external virtual legs. Applying $\mathcal{T}_X$ twice is therefore equivalent to the identity,

\begin{equation}
    \includegraphics[scale=0.75]{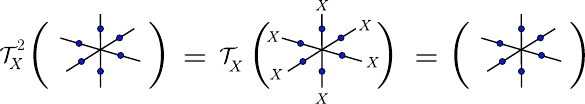},
    \label{eq:LineonTrivial}
\end{equation}
and so $\mathcal{T}_X^2 = 1$ not only on the ground state, but on the excited states as well, as the operators on the virtual legs of the tensor cancel after applying the symmetry twice. Therefore, the symmetry fractionalization is trivial.

This situation changes when $g_{1/2} < 0$. We may consider the case $g_1 < 0, g_2 >0$. In this case, some entries in the $W$-matrix are complex and therefore change signs under complex conjugation $\mathcal{K}$, leading to a modified boundary operator on the external legs,

\begin{equation}
    \includegraphics[scale=0.75]{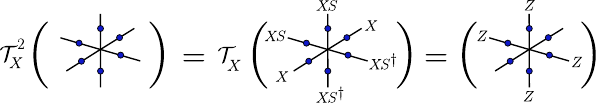},
    \label{eq:LineonEnriched}
\end{equation}
where $S$ is a phase gate with $(S)_{01} = (S)_{10} = 0$, $(S)_{00} = 1$ and $(S)_{11} = i$. If the symmetry is applied to a subsystem consisting of multiple connected vertices, the virtual operators in the bulk will cancel, leaving non-trivial operators only on the boundary of the subsystem. Using the second identity in \eq{eq:PlumbingZ}, the $Z$ operators on the virtual boundary legs can be pulled on the associated physical legs, leading to a physical cage operator of $Z$s around the subsystem which is open in one direction (see \fig{fig:LineonCubeOperator} for a depiction of the cage operator). This operator can be interpreted as the braiding of a fracton dipole around the subsystem. The system picks up a minus sign if this tube operator induced by $\mathcal{T}_X^2$ is pierced by a $X$ operator string of a lineon living inside the subsystem. Arbitrarily choosing the preferred direction in this case to be the $x$ direction, this implies $\mathcal{T}_X^2 = -1$ for $\ell_y$ and $\ell_z$. In this phase, these excitations therefore do indeed host a Kramer's doublet under $\mathcal{T}_X$. Equivalently, for $g_1 > 0, g_2 < 0$ and $g_1 < 0, g_2 < 0$ the tube operators exhibit the other two possible preferred directions, leading to non-trivial symmetry fractionalization for a different pair of lineons. In total, this parametrization thus exhausts all possibilities for lineon symmetry fractionalizations; its phase diagram is shown in \figc{fig:LineonDrawings}{a}.

In two-dimensional topological systems, different symmetry enrichment patterns can be diagnosed by non-local membrane order parameters~\cite{pollmann_detection_2012, zaletel_detecting_2014, huang_detection_2014}. They capture the symmetry fractionalization on certain excitations, as additional superselection rules force them to be zero when non-trivial fractionalization is present. Crucially, they require the construction of so-called minimally entangled states on a torus geometry~\cite{zhang_quasiparticle_2012, huang_detection_2014}. These states form a basis of the ground state space and have the special property of being simultaneous eigenstates of
loop operators in the same direction (Wilson and 't Hooft loops for the toric code); each minimally entangled state can be associated with a type of anyonic excitation including the vacuum. When the system is cut open, each half-cylinder effectively hosts the associated anyon on the respective open boundary.

While the notion of a minimally entangled state is less clear for fractonic systems which feature a system-size-dependent number of Wilson loop operators~\cite{vijay_fracton_2016, he_entanglement_2018, shirley_fracton_2018, slagle_quantum_2017}, it is still possible to develop analogous order parameters $O_{\ell_i}$ which measure symmetry fractionalization on the lineon degrees of freedom $\ell_i$. For an in-depth discussion on the construction of these objects, we refer to Appendix~\ref{apx:MOP}; the basic idea is to build a state $\ket{\Psi_{\ell_i}}$ on an $L_i\times L \times L$-lattice with periodic boundary conditions. It is of infinite extension $L_i \rightarrow \infty$ in direction $i$ and has finite size $L$ in the two other directions. The value of the Wilson loop operators in the finite directions is chosen such that $\ket{\Psi_{\ell_i}}$ hosts one pair of lineons $\ell_i$ which is separated over the third, infinite direction $i$; then, analogously to the two-dimensional case, a non-local volumetric order parameter is introduced as
\begin{equation}
    O_{\ell_i} = \lim_{L \rightarrow \infty} \Big\vert \bra{\Psi_{\ell_i}} \prod_{\alpha \in R} X_\alpha  \ket{\Psi_{\ell_i}} \Big\vert^{1/L^2},
    \label{eq:MOPLineon}
\end{equation}
where $R$ is the set of edges in one semi-infinite half of the system. For this construction, the order parameter is expected to vanish when the symmetry fractionalizes on the excitation $\ell_i$; otherwise, it is 
finite.
For small $L$, the order parameters (\ref{eq:MOPLineon}) can be evaluated numerically, see Appendix~\ref{apx:MOP}. In \figc{fig:LineonDrawings}{b}, we show numerical results for all three order parameters in a system of cross section $L \times L = 4 \times 4$ (in number of vertices) for a path through the phase diagram which we parameterize as $\begin{pmatrix} g_1, g_2 \end{pmatrix} = g_r \begin{pmatrix} \cos \theta, \sin \theta \end{pmatrix}, g_r = 0.8, \theta \in [0, 2\pi]$. In the regular X-Cube phase, $O_{\ell_i} \neq 0$ for $i = x,y,z$ as no fractionalization is present; in a non-trivial symmetry-enriched phase, two of the three quantities vanish as the time-reversal symmetry fractionalizes over the lineons.  

Similar to two-dimensional models, we can associate the $W$-matrix of our system to a three-dimensional classical vertex problem, where the probabilities of local spin configurations are given by $\vert W_\nu \vert^2$~\cite{liu_simulating_2023} (see Appendix \ref{apx:ClassPart}). Moreover, the isometry condition allows for efficient numerical evaluation of $\langle Z_i Z_j \rangle$ correlations: The weight matrix $\vert W_\nu \vert^2$ may be interpreted as a local update rule in a stochastic process, where the qubit configuration on the incoming legs is determined by the normalized columns of the matrix, i.e. the probabilities depend on the configuration on the outgoing legs. Starting from a boundary state $\ket{\psi_0}$, operators in the diagonal basis can thus be evaluated by Monte Carlo sampling of this probabilistic automaton. 

\begin{figure}
 \includegraphics[width=\columnwidth]{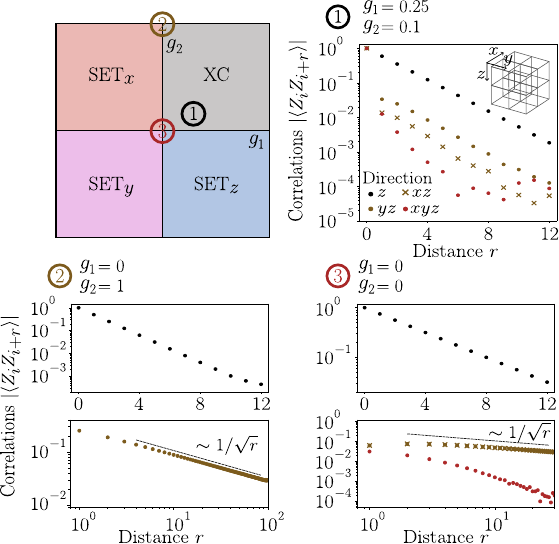}
    \caption{\label{fig:LineonCorrelations}
        \textbf{Correlation Functions.} Spatial correlation functions at three points in the lineon-enrichment phase diagram evaluated using Monte Carlo techniques, starting from a disordered boundary $\ket{\psi_0} = \ket{++\cdots+}$. (1) Away from the critical lines, all correlations decay exponentially. (2) On the critical lines, certain directions exhibit critical $\langle Z_i Z_j \rangle$ correlations which can be mapped to stacks of 2D six-vertex models. (3) At the special multi-critical point $g_1 = 0, g_2 = 0$, all three directions feature planes of this criticality, while other directions also exhibit slowly decaying correlations, the precise nature of which is difficult to determine numerically. We show the absolute value as the correlations in $xyz$ direction are negative for $r \neq 0$.
    }
\end{figure}

We show the correlations at different points in the phase diagram in \fig{fig:LineonCorrelations}. Away from the phase transition lines where neither $g_1$ nor $g_2$ vanishes, all sixteen configurations of the X-Cube fixed point are allowed but in general with different weights. All correlations decay exponentially independent of the boundary state $\ket{\psi_0}$, with the fixed points $\vert g_{1/2} \vert = 1$ having zero correlation length, which increases as $\vert g_{1/2}\vert$ decreases.

We next want to study the correlation functions at the transition lines between two SET orders. This criticality is particularly simple to interpret at the critical point specified by $g_1 = 0$ and  $g_2 = 1$ (or vice versa): as the probability for some of the vertex configurations completely vanishes, each layer orthogonal to the $x$ direction essentially decouples and is equivalent to a two-dimensional six-vertex model up to some local basis transformation. On each of these layers, the stochastic process is equivalent to the motion of hard-core particles through the plane, with $\ket{1}$ implying the presence of a particle. Importantly, the number of particles is conserved throughout the process; this $U(1)$ conservation law allows for an hydrodynamic treatment of the system~\cite{baxter_exactly_1985, friedman_spectral_2019, kos_chaos_2021}. For disordered boundary conditions $\ket{\psi_0} = \ket{++\cdots+}$, this implies critical correlations $\langle Z_i Z_{i+r} \rangle \sim 1/\sqrt{r}$ in the diagonal direction of the layer (the time direction of the stochastic process), with other directions in the plane featuring exponential decay. Correlations in most other directions vanish. If we tune away from the special point $g_2 = 1$, inter-layer correlations gradually increase. Intriguingly, there is a special multi-critical point at $g_1 = g_2 = 0$. Here, diagonal correlations in all three possible planes follow a power law $\sim 1/\sqrt{r}$. Remarkably, correlations in several other directions outside these planes also seem to decay slowly. Confirming the precise functional character of these correlations from a numerical study alone is difficult. It would be interesting to explore this peculiar state using analytical methods from classical statistical physics to confirm whether there are more directions which host power-law correlations. This is an especially interesting task as it has been established recently that 2D isoTNS may support power-law correlations in one spatial dimension alongside the more generic exponential decay of correlation functions~\cite{liu_simulating_2023}. However, it is still unclear whether this is the most general case. In particular, one may study whether 3D isoTNS of low bond dimension $D$ as given by \eq{eq:WSETLineon} can support power-law correlations in a two-dimensional subsystem.

\section{Fracton enrichment}\label{sec:Fracton}
In the case of the two-dimensional toric code, the dual description in terms of vertex and plaquette tensors as shown in \fig{fig:Tensors} extends to its excitations. A parametrized $W$-matrix which tunes between different fractionalization schemes under $\mathcal{T}_X$ for the electric excitations in the vertex network describes the same transition under $\mathcal{T}_Z$ for the magnetic excitations in the plaquette network. An equally simple duality is not present in the X-Cube model, as the $X$ and $Z$ stabilizers in \eq{eq:XCHamiltonian} are manifestly distinct. Nevertheless, one can still adapt the logic of the vertex $W$-matrix (\ref{eq:WSETLineon}) to access different symmetry-enriched phases with the same topological order as the regular X-Cube wavefunction.

Our goal in this section is to construct a symmetry-enriched topological phase transition described by isoTNS, in which the fractons $f$ exhibit trivial or non-trivial symmetry fractionalization under the time-reversal symmetry $\mathcal{T}_Z$. The X-Cube fixed point is again invariant, $\mathcal{T}_Z \ket{\text{XC}} = \ket{\text{XC}}$, as $\mathcal{T}_Z$ commutes with every factor $(1  + \prod_{j \in c} X_j)$, and all excitations transform trivially under this symmetry with $\mathcal{T}_Z^2 = 1$. The tensor network we consider has the same structure as in \eq{eq:WCube}: In each cube, we put a 12-leg cube tensor which we express in terms of a parametrized $2^3 \times 2^9$ matrix $\left(W_c(g)\right)_{(i_1i_2i_3)(i_4\cdots i_{12})}$. Each leg extends to one edge of the cube; the three qubits $i_1, i_2, i_3$ sit on edges which meet at a corner of the cube~(see the graphical summary in \fig{fig:Tensors}). These legs are labeled as the incoming legs of the isoTNS-like structure, while the other nine legs are labeled as outgoing legs. On each edge, we put the modified plumbing tensor $(\delta_X)^\sigma_{abcd}$.

The matrix $W_c(g)$ with $g \in [-1,1]$ is defined as follows: We split the twelve edges of the cube into two sets $P_1 = i_1, \dots, i_6$ and $P_2 = i_7,\dots, i_{12}$ of six apiece, where the incoming qubits are in the same set (see \fig{fig:Tensors} for a possible choice of the two sets).\footnote{There are other valid prescriptions for the sets $P_1$ and $P_2$ which exhibit the same phase transition. They share the property that, under global application of $\mathcal{T}_Z$ on the ground state wavefunction, all $S^{(\dagger)}$ cancel, leaving the state invariant.} For each possible configuration, we can calculate the number $N_{P1/P2}$ of pairs of $\ket{1}$ states in both sets, evaluated for a given configuration as

\begin{align}
    N_{P1} = & \frac{1}{2}\left(\sum_{j=1}^6 i_j \right)\left(\sum_{j=1}^6 i_j -1\right) \nonumber\\
    N_{P2} = & \frac{1}{2}\left(\sum_{j=7}^{12} i_j\right) \left(\sum_{j=7}^{12} i_j -1\right).
\end{align}
Then, defining the parity $P$ of both numbers, $P = (N_{P1} + N_{P2}) \mod 2$, together with the number $Q$ of $\ket{1}$ states for the qubits on the incoming legs, $Q = i_1 + i_2 + i_3$, the $W$-matrix is given as

\begin{equation}
    \left(W_c (g) \right)_{(i_1i_2i_3),(i_4\cdots i_{12})} = \begin{cases}
    \;\;\;\;\; 0 & \text{if} \; \left( \sum_{j = 1, \dots, 12} i_j \right) \; \text{mod} \; 2 = 1 \\
    & \text{else:}\\
    \frac{1}{\sqrt{1 + 3 \vert g\vert}} & \text{if} \; P = 0 \; \text{and} \; Q = 0 \; \text{or} \; 3  \\
    \frac{1}{\sqrt{3 +\vert g\vert}} & \text{if} \; P = 0 \; \text{and} \; Q = 1 \; \text{or} \; 2 \\
    \sqrt{\frac{\text{sign}(g) \vert g \vert}{3 + \vert g\vert}} & \text{if} \; P = 1 \; \text{and} \; Q = 0 \; \text{or} \; 3  \\
    \sqrt{\frac{\text{sign}(g) \vert g \vert}{1 + 3 \vert g\vert}} & \text{if} \; P = 1 \; \text{and} \; Q = 1 \; \text{or} \; 2.
    \end{cases}
    \label{eq:WSETFracton}
\end{equation}
At the fixed point $g= 1$, this is the X-Cube tensor (\ref{eq:WCube}), while for $g = -1$, the matrix differs only by complex factors in some of its entries. Indeed, the logic of this construction can be seen as a generalization of the $W$-matrix (\ref{eq:WSETLineon}) we have introduced to enrich the lineon excitations. There, only a single pair of legs was present in each set $P_{1/2}$. As no new configurations are introduced, the virtual symmetry of the X-Cube fixed point is not violated; just as in the lineon case, this should preclude a sudden first-order transition into a trivial state, as the virtual perturbation can be pulled on the physical level.

\begin{figure}
 \includegraphics[width=0.98\columnwidth]{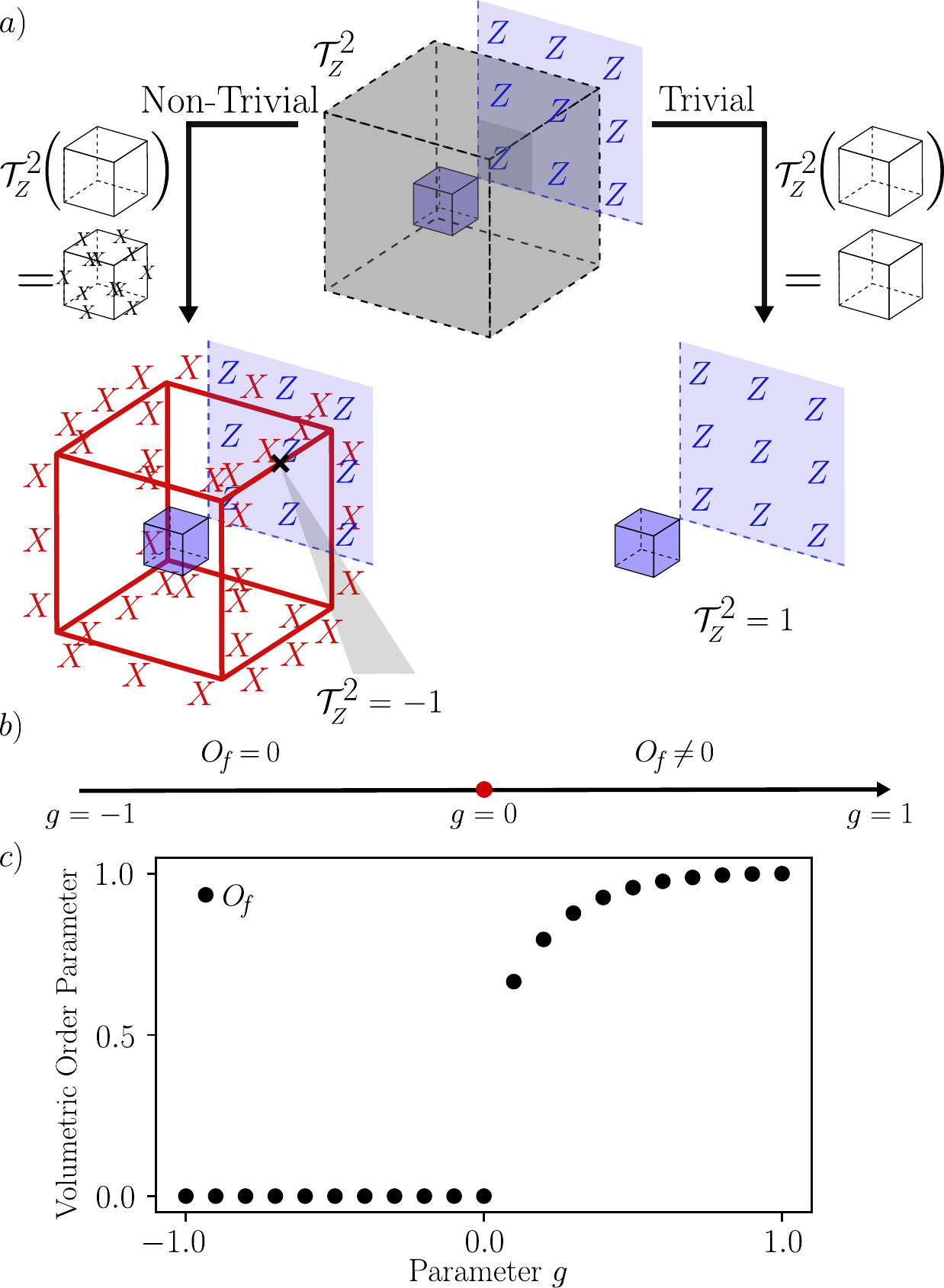}
    \caption{\label{fig:FractonDrawings}
        \textbf{Fracton Enrichment.} a) Possible fractionalization schemes in the phase diagram spanned by the $W$-matrix given in \eq{eq:WSETFracton}. In the trivial phase, the double application of the anti-unitary symmetry $\mathcal{T}_Z = \left( \prod_i Z_i \right) \mathcal{K}$ on a subsystem is equal to the identity; in the phase with non-trivial symmetry fractionalization, its effect is equivalent to a wireframe operator of $X$ operators, which can be seen as the braiding of a lineon dipole around the subsystem. This operator measures the presence of fractonic excitations which live on the corners of membrane operators which are pierced by the wireframe at the black cross. b) The phase diagram of the wavefunction with different symmetry fractionalization. c) Non-local order parameter $O_f$ along the entire phase line in b). If the symmetry fractionalizes non-trivially on the fracton $f$, the order parameter will vanish.
    }
\end{figure}

The $2^3 \times 2^9$ $W$-matrix remains normalized for all values of the parameter $g$; thus the tensor network is isometric. This implies a simple contraction scheme along the direction from the incoming legs to the outgoing legs (see Appendix \ref{apx:IsoTNS} for a detailed discussion). A parent Hamiltonian can be constructed in an equivalent fashion as discussed in Section \ref{sec:Lineon} for the lineon wavefunction. Accordingly, the vertex stabilizers $A_{\nu, i}$ are modified; the deformed local operators $\tilde{A}_{\nu, i}(g)^\dagger \tilde{A}_{\nu, i}(g)$ are 54-local operators acting on the edges of eight cubes surrounding the vertex $\nu$.

We may again consider how a local cube tensor dressed with twelve physical legs on the edges transforms under application of the time-reversal symmetry $\mathcal{T}_Z = \left( \prod_{i = 1}^{12} Z_i \right) \mathcal{K}$. For positive values, $g > 0$, the symmetry fractionalization is trivial, as obtained from \eq{eq:PlumbingX}, we find
\begin{equation}
    \includegraphics[scale=0.75]{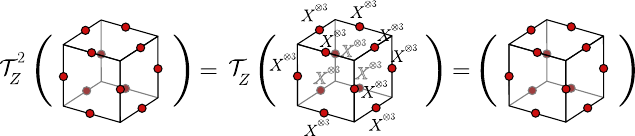},
    \label{eq:FractonTrivial}
\end{equation}
where $X^{\otimes 3}$ signifies an $X$ operator acting on each of the three virtual degrees of freedom which connect the considered tensor with the three tensors at the center of the cubes which share the edge in question. As the double application of the symmetry cancels, we find $\mathcal{T}_Z^2 = 1$ and no symmetry fractionalization happens in this parameter range.

\begin{figure*}
    \includegraphics[]{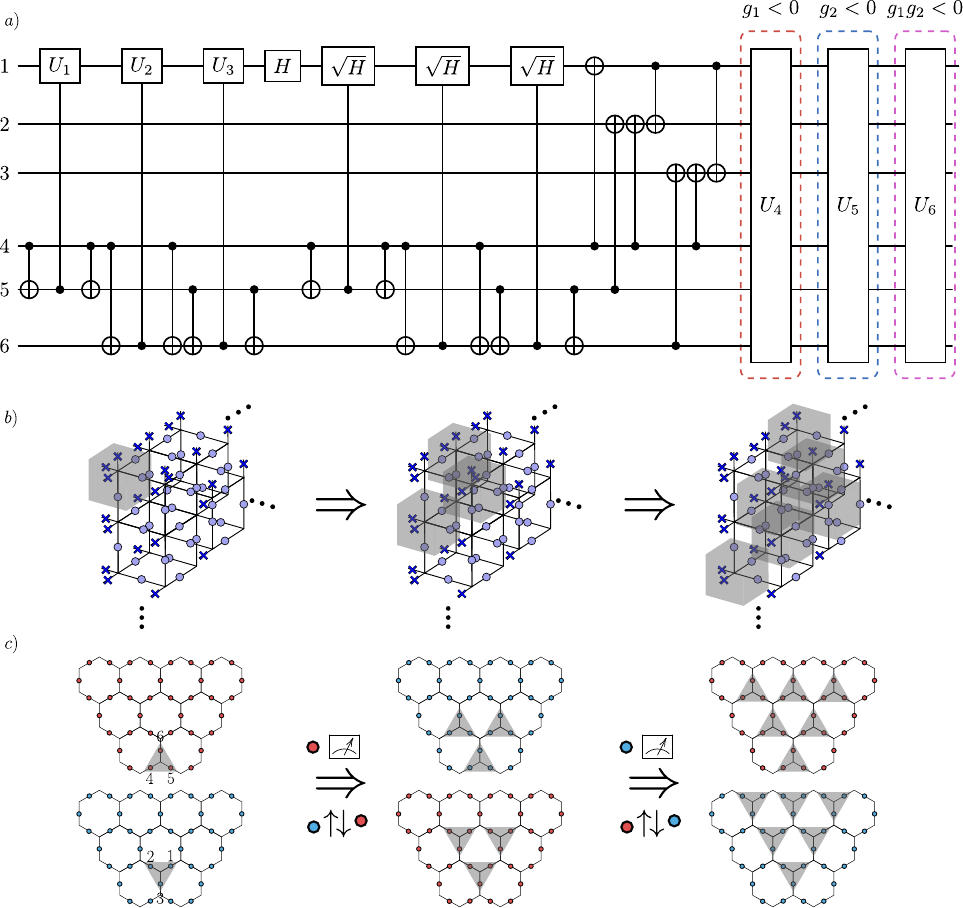}
    \caption{\label{fig:LineonCircuit}
        \textbf{Quantum circuit for lineon enrichment.} a) A parametrized six-qubit gate to sequentially generate the isoTNS defined by the $W$-matrix (\ref{eq:WSETLineon}). The six qubits are arranged around one vertex as shown in \fig{fig:Tensors}. The unitaries $U_i$ for $i = 1, \dots, 6$ are described in the main text and in Appendix~\ref{apx:Gates}. b) Initialization of the wavefunction. The state is initialized as $ \ket{\psi_0} \otimes \ket{00\cdots 0}$, where $\ket{\psi_0}$ is a boundary state (blue crosses on three boundaries of the three dimensional lattice). In each step, the circuit of a) (gray cube) is applied to a number of vertices, thereby sequentially preparing the state in the bulk (light blue circles). c) Holographic measurement-based construction. Using two honeycomb lattices and measurements, the state is created with a reduced number of qubits. At each step, a part of the boundary state $\ket{\psi_0}$ is implemented on the red lattice, then the circuit from a) is applied to certain qubits on both lattices, the red (blue) lattice corresponding to the outgoing (incoming) legs of the $W$-matrices at the vertices treated in the respective step. This mimics a step from the 3D preparation in b). Afterwards, the qubits on the red lattice are measured in the eigenbasis of operators $O_k$ and the two lattices are exchanged. After resetting the measured qubits and preparing qubits of the boundary state $\ket{\psi_0}$, the next step is implemented, allowing for sampling of correlation functions $\langle O_1 O_2 \cdots O_N \rangle$.
    }
\end{figure*}

For $g < 0$, the action of the time-reversal symmetry on the virtual legs becomes
\begin{equation}
    \includegraphics[scale=0.75]{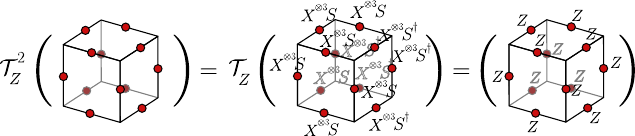},
    \label{eq:FractonEnriched}
\end{equation}
with phase gates $S^{(\dagger)}$ being present due to the effect of complex conjugation on the complex entries. In the same vein as in the last section, applying the symmetry to a subsystem can be reduced to an operator on the virtual legs only on the boundary. Here, due to the properties of the plumbing tensor $(\delta_X)^\sigma_{abcd}$, a $Z$ operator on a single virtual leg as present here is equivalent to an $X$ operator on the physical leg, see the second identity in \eq{eq:PlumbingX}. Therefore, in this phase the action of $\mathcal{T}_Z^2$ on a $L \times L \times L$ cubic subsystem is the same as a wireframe of $X$ operators around the cube. This is nothing but the braiding of a lineon dipole around the subsystem~\cite{shirley_fractional_2019}. In the presence of a fracton $f$ in the subsystem, the fractonic membrane operator is pierced by this wireframe, leading to interferometric detection of the fracton $f$ and implying $\mathcal{T}_Z^2 = -1$ (see \figc{fig:FractonDrawings}{a} for a graphical depiction). The parametrization therefore explores both possible fractionalization patterns on the fractons, as shown in the phase diagram of \figc{fig:FractonDrawings}{b}.

The definition of a non-local volumetric order parameter for fractons proceeds analogously to the lineon case; however, it entails certain numerical difficulties. As fractons are created in fours at the corners of a two-dimensional membrane of $Z$ operators, a state hosting a single fracton would have to be of infinite extent in at least two directions, which is numerically not feasible. Nevertheless, one may still prepare a wavefunction on a lattice with open boundary conditions in two directions and infinite extent in the third. A $Z$ membrane operator creates one fracton in the bulk, while the other three are absorbed by the boundaries. Evaluating the expectation value of the symmetry $\prod_i Z_i$ on a subsystem of this state $\ket{\Psi_f}$ hosting only the single fracton, this returns an order parameter which vanishes for sufficiently large $L$ when symmetry fractionalization is present due to a superselection rule. We write this non-local volumetric order parameter as
\begin{equation}
    O_f = \lim_{L \rightarrow \infty} \Big\vert\bra{\Psi_f} \prod_{\alpha \in S} Z_\alpha  \ket{\Psi_f}\Big\vert^{1/L^2},
    \label{eq:MOPFracton}
\end{equation}
where $S$ is the single-fracton subsystem.
In \figc{fig:FractonDrawings}{c}, we show the order parameter $O_f$ over the whole parameter range in a $2 \times 2 \times \infty$ system, indicating the number of cubes in each direction. The quantity clearly distinguishes the two phases with respect to the symmetry enrichment. We provide further details on the numerical approach in Appendix~\ref{apx:MOP}.

We now study $\langle X_i X_j \rangle$ correlations in this tensor network. Starting from a boundary state $\ket{\psi_0}$, we evaluate these correlations using stochastic Monte Carlo sampling with a probability distribution given by $\vert W_c\vert^2$, analogously to the preceding section. Each step updates the three qubits on the incoming legs taking the configuration of the nine qubits on the outgoing legs as input, moving in opposite direction to the contraction scheme. For $g \neq 0$, the bulk correlations are independent of the boundary conditions and show no signatures of algebraic decay or long-range order, as expected for a gapped phase. Away from the fixed point the correlation length is finite and diverges as the critical point at $g = 0$ is reached. At the critical point, it can be checked analytically that an $X$ basis GHZ state $\ket{\psi_0} = \frac{1}{\sqrt{2}} \left( \ket{++\cdots +} + \ket{--\cdots-} \right)$ on the boundary (where $\ket{\pm} =( \ket{0} \pm \ket{1})/\sqrt{2} $) also leads to a GHZ state in the bulk, featuring infinite-length correlations. This is suggestive of some gap closing at this point, as for a fully gapped state all boundary conditions would eventually lead to a generic bulk state with exponential correlations.

\section{Preparation on a quantum processor}\label{sec:Preparation}
Although it is well-established that one-dimensional matrix product states can be efficiently prepared using a sequential quantum circuit whose depth scales with system size $L$~\cite{schon_sequential_2005}, the situation is much more involved in higher dimensions. Typical classes of tensor network states such as Projected Entangled Pair States (PEPS) generally need quantum circuits of a depth that scales exponentially in total system size, i.e. the number of qubits, posing a severe restriction on their implementation~\cite{schuch_computational_2007}. There have therefore been continued efforts to establish subclasses of tensor networks with further restrictions which guarantee a more favorable scaling. Among those, isoTNS are particularly promising as it has been shown that they can be efficiently created using a sequential quantum circuit. The protocol is applied to a lattice with one open boundary in each spatial direction; on this boundary, some state $\ket{\psi_0}$ has already been prepared. The rest of the lattice is initialized as $\ket{00 \cdots 0}$. Starting from the boundary state $\ket{\psi_0}$, layers of mutually commuting local gates are applied; as the number of gates per layer increases with each step, the circuit runtime scales with the biggest linear size $L$ of the system~\cite{satzinger_realizing_2021,wei_sequential_2022,liu_simulating_2023}. A necessary condition for this scheme to be experimentally viable is the possibility of efficiently preparing the boundary state $\ket{\psi_0}$ in the first place. However, this is assured for a wide range of states, for instance when $\ket{\psi_0}$ itself can be expressed as an isoTNS (or, in the case of a one-dimensional boundary, as an MPS). In particular, for a gapped system the choice of the boundary state $\ket{\psi_0}$ does not affect any properties of the state sufficiently far away from the boundary. Therefore, one may choose a product state on the boundary, which can be prepared in a single step by simultaneous application of local unitaries. At a gapless point, different choices of boundary states $\ket{\psi_0}$ can be used to access different aspects of criticality.

In this picture, the isometry properties of isoTNS can be seen as reverting the unitary circuit: When contracting the physical legs of the wavefunction and its Hermitian conjugate, each layer of local unitaries cancels out with its inverse on the other layer, starting from the uppermost set of unitaries. This leaves only the trivial state $\ket{00 \cdots 0}$ on both sides, the local contraction of which naturally returns identities. The efficient contraction direction of the state is therefore opposite to the direction in which the layers of unitaries spread through the lattice. In principle, the local few-qubit gates can be obtained from the tensors of the isoTNS. Higher bond dimensions necessitate either gates with bigger local support or the presence of ancilla qubits to store the virtual information. In that regard, isoTNS expressed in terms of a $W$-matrix are particularly favorable, as the equivalence between the physical degrees of freedom and the virtual legs naturally leads to a circuit without ancillary qubits. Each qubit is acted on only once, whereby it becomes entangled with the preceding qubits in the circuit; afterwards, it may only serve as a control qubit in control gates acting on qubits in its vicinity.

\begin{figure*}
    \includegraphics[width=\textwidth]{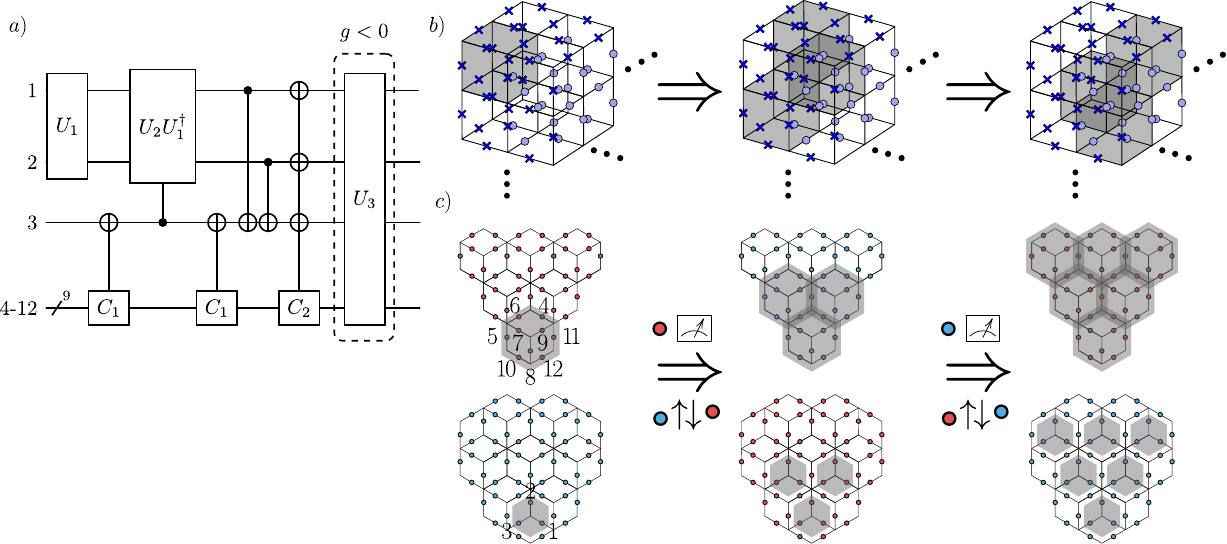}
    \caption{\label{fig:FractonCircuit}
        \textbf{Quantum circuit for fracton enrichment.} a) A parametrized twelve-qubit gate to sequentially generate the isoTNS defined by the $W$-matrix (\ref{eq:WSETFracton}). The twelve qubits are arranged around one vertex as show in \fig{fig:Tensors}. The unitaries $U_i$ for $i = 1, 2, 3$ and the control gates $C_i$ for $i = 1,2$ are described in the main text and in Appendix~\ref{apx:Gates}. This circuit prepares the state in the $X$ basis. It can be transformed to the $Z$ basis by applying Hadamard gates to every qubit after all qubits in the lattice have been prepared. b) Initialization of the wavefunction. The state is initialized as $ \ket{\psi_0} \otimes \ket{00\cdots 0}$, where $\ket{\psi_0}$ is a boundary state (blue crosses). In each step, the circuit from a) (gray cube) is applied to a number of vertices, thereby sequentially preparing the state in the bulk (light blue circles). c) Holographic measurement-based construction for fracton enrichment on two rhombille lattices. The structure of the holographic construction follows the one of lineon enrichment in \figc{fig:LineonCircuit}{c}.
    }
\end{figure*}

We apply this general framework to the parametrized wavefunctions introduced in this work. The circuits for the two classes of wavefunctions introduced so far are shown in Figs. \ref{fig:LineonCircuit} and \ref{fig:FractonCircuit}. Gates involving only a few qubits are presented directly in this section; a possible implementation of the more involved gates is given in Appendix \ref{apx:Gates}.

We present the six-qubit gate which creates the tensor described by the vertex matrix $W_\nu(g_1, g_2)$ in Eq.~\ref{eq:WSETLineon}; see \figc{fig:LineonCircuit}{a}. The controlled unitaries $CU_{i}$ with $i = 1,2,3$ at the start of the circuit are constructed as follows: We introduce two parametrized single-qubit unitaries $A(g)$ and $B(g)$ as
\begin{align}
    A(g) = & \frac{1}{\sqrt{1+ \vert g \vert}} 
        \begin{pNiceMatrix}[first-row,last-col]
        \ket{0} & \ket{1} &    \\
        1 & -\sqrt{\vert g \vert} & \; \ket{0} \\ 
        \sqrt{\vert g \vert} & 1 & \; \ket{1}
        \end{pNiceMatrix} \nonumber
    \\ B(g) = & A(f(g)), \qquad f(g) =  \frac{\left(\sqrt{1+ \vert g \vert} -1\right)^2}{\vert g \vert}
\end{align}
with $B^2(g) = A(g)$. 
Then, the three unitaries are given as

\begin{align}
    U_1 = B^\dagger(g_1) B(g_2) B(g_1g_2) \nonumber
    \\ U_2 = B(g_1) B^\dagger(g_2) B(g_1g_2) \nonumber
    \\ U_3 = B(g_1) B(g_2) B^\dagger(g_1g_2).
\end{align}
The Hadamard gate $H$ and its square root $\sqrt{H}$ are defined as usual:

\begin{align}
    H = & \frac{1}{\sqrt{2}} 
        \begin{pNiceMatrix}[first-row,last-col]
        \ket{0} & \ket{1} &    \\
        1 & 1 & \; \ket{0} \\ 
        1 & -1 & \; \ket{1}
        \end{pNiceMatrix} \nonumber
    \\     \sqrt{H} = & \frac{1-i}{2\sqrt{2}} 
        \begin{pNiceMatrix}[first-row,last-col]
        \ket{0} & \ket{1} &    \\
        1 + \sqrt{2}i & 1 & \; \ket{0} \\ 
        1 & -1 + \sqrt{2}i & \; \ket{1}
        \end{pNiceMatrix} 
\end{align}
with $(\sqrt{H})^2 = H$. 
The three final unitaries $U_i$ for $i = 4,5,6$ will be only applied if some of the parameters $g_{1/2}$ are negative and allocate complex coefficients to certain vertex configurations. They can be decomposed into controlled-phase gates acting on different sets of qubits; see Appendix \ref{apx:Gates} for an exact definition.
The state can then be prepared as shown in \figc{fig:LineonCircuit}{b}: Starting from an initial state $\ket{\psi_0} \otimes \ket{00\cdots 0}$, where $\ket{\psi_0}$ is the state on the boundary of dangling edges as shown in the figure, instances of this six-qubit gate are sequentially applied to diagonal layers of vertices, implying a circuit depth of $O(L)$ for a system of linear size $L$.

A similar approach is also applicable to the wavefunctions discussed for the symmetry enrichment of fractons. One can derive a circuit to create a cube tensor with the structure given by the cube matrix $W_c(g)$ in \eq{eq:WSETFracton}. This twelve-qubit gate is shown in \figc{fig:FractonCircuit}{a}. The unitary operators in the controlled gates of that circuit are
\begin{equation}
    U_1 = \frac{1}{\sqrt{1 + 3 \vert g \vert}} \; \begin{pNiceMatrix}[first-row,last-col]
 \ket{00} & \ket{01} & \ket{10} & \ket{11} &    \\
 1  & -\sqrt{\vert g \vert}   & -\sqrt{\vert g \vert}  & -\sqrt{\vert g \vert}  & \;\ket{00} \\
 \sqrt{\vert g \vert}   & 1  & -\sqrt{\vert g \vert} & \sqrt{\vert g \vert} & \;\ket{01} \\
\sqrt{\vert g \vert}   & \sqrt{\vert g \vert}   & 1 & -\sqrt{\vert g \vert}  & \;\ket{10} \\
\sqrt{\vert g \vert}   & -\sqrt{\vert g \vert}  & \sqrt{\vert g \vert}  & 1  & \;\ket{11} \\
\end{pNiceMatrix}
\label{eq:U1Fracton}
\end{equation}

and
\begin{equation}
    U_2 = \frac{1}{\sqrt{3 + \vert g \vert}} \; \begin{pNiceMatrix}[first-row,last-col]
 \ket{00} & \ket{01} & \ket{10} & \ket{11} &    \\
\sqrt{\vert g \vert}  & -1   & -1  & -1  & \;\ket{00} \\
1   & \sqrt{\vert g \vert}  & -1 & 1 & \;\ket{01} \\
1  & 1   & \sqrt{\vert g \vert} & -1  & \;\ket{10} \\
1   & -1  & 1  & \sqrt{\vert g \vert} & \;\ket{11} \\
\end{pNiceMatrix}.
\label{eq:U2Fracton}
\end{equation}
The precise construction of the controlled gates $C_k$ and the unitary $U_3$ is reported in Appendix~\ref{apx:Gates}.

In \figc{fig:FractonCircuit}{b}, we show the sequential circuit which results in the parametrized wavefunction: Starting again from a state with a prepared boundary and an empty bulk $\ket{\psi_0} \otimes \ket{00\cdots 0}$, where the boundary is slightly different compared to the lineon case, the full wavefunction is constructed with a linear-depth circuit. Each local circuit prepares a unit cell of three qubits from the input of the nine remaining qubits on the edges of the respective cube, reminiscent of the structure of the underlying $W$-matrix. The so-created state is in the $X$ basis and may be returned to the $Z$ basis by applying Hadamard gates to every qubit after the full circuit. It should be noted that each local gate has overlap on one qubit with other close gates in the same layer; nevertheless, this does not spoil the efficiency of the circuit. Both gates use this qubit only as a control qubit, i.e. they are diagonal in its local Hilbert space and therefore commute. Each layer can therefore be split up into three subsequent sublayers of mutually non-overlapping gates, introducing only a constant factor in the scaling.

While these approaches allow for the definition of circuits which are efficient in respect to circuit depth, preparing a three-dimensional system is still difficult on an actual quantum processor, as the number of qubits will quickly increase when considering a reasonably sized system. The three-dimensional structure can be avoided by measurements leading to a holographic construction of the state: Instead of preparing a $D$-dimensional wavefunction directly, one can also exploit the special causal structure of isoTNS to evaluate general correlation functions $\langle O_1 O_2 \cdots O_N \rangle$ using two $(D-1)$-dimensional arrays of qubits~\cite{raussendorf_one-way_2001, raussendorf_measurement-based_2003, foss-feig_holographic_2021}. This has already been successfully used to simulate the dynamics of a kicked Ising model on a trapped-ion quantum computer~\cite{chertkov_holographic_2022}; more recently, this idea has been proposed for two-dimensional states which allow an isoTNS representation~\cite{liu_simulating_2023}.

We illustrate the idea first for the lineon-enriched wavefunction; see \figc{fig:LineonCircuit}{c}. By applying the logic sketched in the last paragraph to the sequential circuit in \figc{fig:LineonCircuit}{b}, we can realize the holographic construction of the state, starting from some given boundary state $\ket{\psi_0}$. In this circuit one can sequentially evaluate expectation values $\langle O_1 O_2 \cdots O_N \rangle$ of tensor products of local operators $O_j$ using a pair of two-dimensional lattices. The protocol requires two initially decoupled honeycomb lattices with qubits living on each edge. For the first step, we initialize three qubits around a vertex on the first honeycomb lattice as prescribed by the boundary state $\ket{\psi_0}$ for the three qubits living on the topmost corner of the three-dimensional cubic lattice (see the first panel of \figc{fig:LineonCircuit}{b}), while the rest of the honeycomb lattice is in the trivial state $\ket{00 \cdots 0}$. The second lattice is equally in a trivial state $\ket{00 \cdots 0}$. 

The six-qubit quantum circuit from \figc{fig:LineonCircuit}{a} is then applied to the qubits as shown in the first column of \figc{fig:LineonCircuit}{c}, between the three prepared qubits on the first lattice and three qubits around a corner of the second lattice. Afterwards, each site $k$ on the first lattice is measured in the basis of the operator $O_k$ of interest, collapsing the site to the eigenbasis of the operator. This operator is the identity when the site is not measured. After the measurement, the qubits on the first lattice are all reset to $\ket{0}$. For the next step, the role of the two lattices is switched, with the next set of boundary qubits being implemented around the already treated qubits on the second lattice, thus realizing the second layer in the sequential circuit, with multiple separate six-qubit gates being applied to three qubits on each of the two lattices, followed by a measurement of the local operator $O_k$. Repeating this process with an increasing number of qubits and gates in each step, each layer of the sequential circuit is implemented successively between the two lattices, allowing direct sampling of the correlator $\langle O_1 O_2 \cdots O_N \rangle$.
The peculiar structure of the entanglement spread in the isoTNS, which follows from the causal structure induced by the contraction property, thus allows a significant reduction in the number of qubits needed to access properties of these non-trivial quantum states.

We also adapt this idea to the fracton enrichment quantum circuit, as shown in \figc{fig:FractonCircuit}{c}. The lattices in this case are rhombille lattices. Otherwise, the circuit follows a similar logic as the lineon case: For the first step, three qubits on the first lattice are initialized as given by the boundary state $\ket{\psi_0}$, which are then used to prepare nine qubits on the second lattice by means of the twelve-qubit gate in \figc{fig:FractonCircuit}{a}, the twelve qubits representing the topmost cube in \figc{fig:FractonCircuit}{b}; after measuring the first three qubits in the eigenbasis of the given operator $O_k$, the process is repeated for several steps to explore the correlations in the cubic lattice. 

Different relevant observables can be measured using this method. The symmetry fractionalization may be evaluated using a simple protocol of creating an excitation in the system, which is achieved by inserting an $X$ operator where we want the excitation to be localized right after the local quantum circuit has been applied to the site in question. Due to the open boundary conditions of our system, a single lineon or fracton (up to dressing by some excitations of the other type) can be created; even though they arise only in pairs or fours respectively, the other excitations condense on the boundary of the system, as they are pushed away by the sequential circuit. The relevant non-local volumetric order parameter is given by

\begin{equation}
        O = \lim_{A \rightarrow \infty} \Big\vert\bra{\Psi} \prod_{\alpha \in S} X_\alpha  \ket{\Psi}\Big\vert^{1/A},
\end{equation}
where $\ket{\Psi}$ is the wavefunction including the the defect, $S$ is a region surrounding the defect which is chosen so that the $X$ operators do not create further excitations, and $A$ is its surface. This quantity distinguishes between the different symmetry fractionalizations in the same way as the order parameters above. As it is the expectation value of a product of $X$ operators, it can be evaluated in the holographic setup.  

This protocol further allows one to explore the phase diagrams of the states experimentally. In addition to the correlation functions that can be efficiently evaluated classically (i.e. Z-correlations in the case of lineon enrichment and X-correlations in the case of fracton enrichment), other correlations of mixed operators are accessible on quantum processors that are exponentially hard to compute with classical resources.

\section{Outlook}\label{sec:Outlook}
We have presented plumbed isoTNS as a versatile tool to both construct new phases with fracton order and to describe phase transitions between them. In these states plumbing refers to the direct relation between virtual and physical degrees of freedom, which allows us to identify paths that leave the virtual symmetries associated to the fracton order invariant. Likewise, by the plumbing property one may identify modifications which associate a symmetry operation to a non-trivial operator on the virtual legs, which is equivalent to non-trivial symmetry fractionalization. It also guarantees an efficient circuit to prepare the state in linear time. We have illustrated these ideas on the X-Cube model, where we have introduced two parametrizations which host trivial and non-trivial fractionalization patterns on lineons and fractons, respectively.

Our results are relevant to current quantum processing platforms: Normally, studying a 3D quantum phase transition on a quantum processor would be a daunting task and would in general require a prohibitively high number of qubits. The families of quantum states in this work can be efficiently prepared and evaluated, which allows the study of non-trivial many-body properties in three dimensions. Furthermore, our holographic preparation scheme, which is based on the sequential circuit structure of isoTNS, can do with much fewer resources than a preparation of the full state. In fact, a pair of two-dimensional lattices and feedback suffices to study any point in our phase diagrams, including those with critical correlations. This significant reduction in required qubits puts first studies of three-dimensional states on quantum processors within reach.

The X-Cube model serves as an interesting illustration of our method for symmetry enrichment. However, the approach that we present is much more general: tensor networks where physical and virtual legs are equivalent, i.e. tensor networks with a plumbing structure, can be constructed for many stabilizer states; in Appendix \ref{apx:3DTC}, we consider the additional example of the pointlike excitation of the three-dimensional Toric Code. Other examples of stabilizer codes with anti-unitary symmetries are the Checkerboard model~\cite{shirley_foliated_2019} or Haah's code~\cite{haah_local_2011}; especially the latter case is intriguing as it is a type-II fracton model without any mobile excitation. All these models also feature $\mathbb{Z}_N$ generalizations where the symmetry group is enlarged to $\mathbb{Z}_N \times \mathbb{Z}_2^T$: the richer topological gauge group in these models implies that an anti-unitary symmetry necessarily permutes some excitations~\cite{essin_classifying_2013}. Likewise, internal symmetries which are consistent with the plumbing rules can also be treated. Finally, isoTNS can be used to derive simple wavefunctions where loop-like excitations feature non-trivial symmetry fractionalization, such as the plaquette excitations in the 3D Toric Code~\cite{chen_symmetry2016}.

The transitions discussed so far all occur between phases with the same topological or fracton order, but with different fractionalization patterns on some excitations. In addition, the $W$-matrix approach is amenable to a wide class of transitions where topological order changes as well: the path then connects to known stabilizer code fixed points. For example, one may connect two copies of a regular X-Cube model to a $\mathbb{Z}_4$ X-Cube by reducing on both sides the weight of the local configurations which violate the virtual symmetries of the other phase; the two paths meet at a critical point where both constraints are fulfilled. Furthermore, similarly to the 2D transition between the toric code and the double semion model, a more involved direction is to construct $W$-matrix paths with more general plumbing tensors between normal and twisted fracton phases, where the states exhibit distinct braiding statistics~\cite{song_fracton_2024}. Finally, recent work~\cite{lee_ZN2025} introduced generalized $\mathbb{Z}_N$ stabilizer codes whose ground state degeneracy and excitation mobility exhibit novel features: constructing $W$-matrices for their fixed points and for interconnecting paths is a promising strategy to improve our understanding of their properties. In summary, these ideas suggest a dense, interwoven network of exactly solvable paths between widely different phases. The analytical isoTNS construction we present here can be seen as a comparatively simple tool to construct and understand many topological and fracton phases in higher dimensions and, crucially, provides a systematic first step in the program to efficiently realize exotic 3D quantum ground states.

Moreover, the $W$-matrix lends itself to further investigations beyond the study of phase transitions. As isoTNS exhibit appealing properties, the form an attractive class of ansatz wavefunctions both for classical simulation and for quantum circuits. However, understanding the full expressibility of the isoTNS class remains an open question. While some understanding has been developed regarding possible correlations in isoTNS~\cite{haag_typical_2023, malz_computational_2024}, explicitly constructed isoTNS with critical correlations remain rare. The relation between stochastic circuits and plumbed isoTNS may help fill this gap: by adopting examples from the much better understood zoo of critical cellular automata, it seems possible to construct a variety of quantum states exhibiting algebraic correlations.

{\textbf{Acknowledgments:}} We thank Zhi-Yuan Wei for helpful discussions. We acknowledge support from the Deutsche Forschungsgemeinschaft (DFG, German Research Foundation) under Germany’s Excellence Strategy--EXC--2111--390814868, TRR 360 – 492547816 and DFG grants No. KN1254/1-2, KN1254/2-1, the European Research Council (ERC) under the European Union’s Horizon 2020 research and innovation programme (grant agreement No. 851161 and No. 771537), as well as the Munich Quantum Valley, which is supported by the Bavarian state government with funds from the Hightech Agenda Bayern Plus.

\textbf{Data and Code availability:} Numerical data and simulation codes are available on Zenodo~\cite{zenodo}.

\appendix

\section{Mapping to classical partition functions}\label{apx:ClassPart}
In this section, we review how tensor networks with a plumbing structure (locking of physical and virtual degreees of freedom) can be mapped to classical spin problems. As an example, we may consider a tensor network state of the form of \eq{eq:TVertex}, where a three-leg plumbing tensor connects $W$-matrices on the vertices $\nu$. Due to the equivalence between physical and virtual degrees of freedom, the squared norm $\langle \Psi\ket{\Psi}$ of this state can be expressed as a product over the matrices $\vert W_\nu\vert^2$, which are contracted along the edges.

We now consider a classical Hamiltonian $H = \sum_{\langle i,j \rangle} h(\sigma_i,\sigma_j)$ on the same lattice, where spins $\sigma_\nu$ on the vertices interact via nearest-neighbor two-body interactions. Its partition function $Z = \text{Tr}(e^{-\beta H})$ can also be expressed as the product of local weight matrices $R_{\nu} = e^{-\beta/2\sum_{\langle \nu,i \rangle} h(\sigma_\nu, \sigma_i)}$ at every vertex $\nu$. Identifying the two vertex matrices $R_\nu = \vert W_{\nu}\vert^2$ maps the quantum state to a classical spin system, the coupling strengths of which are determined by the local amplitudes of the state.

In the case of a TNS state where $W$-matrices in the middle of cubes are linked by plumbing tensors which couple four cubes sharing an edge, as in \eq{eq:TCube}, the identification proceeds analogously. Here, the classical spins live in the middle of a cube cell and are subject to four-spin interactions $h(\sigma_i,\sigma_j, \sigma_k,\sigma_l)$ between spins whose cubes are connected by a plumbing tensor in the quantum state. These two networks serve as examples of a general method to express the norm of a tensor network state with a plumbing structure as the partition function of a classical vertex problem.

\section{Three-dimensional isoTNS}\label{apx:IsoTNS}
In this appendix, we review some properties of isometric tensor network states (isoTNS). We start by introducing the most commonly discussed type, which is a two-dimensional isoTNS on a square lattice, before generalizing this to arbitrary dimensions for the simple case of a hypercubic geometry of the tensor network. For states with qubits living on the edges, certain isoTNS can be expressed in terms of a $W$-matrix, which contains information about the virtual degrees of freedom, and plumbing tensors which relate the edge degrees of freedom to the virtual legs. The isometry property is guaranteed by a specific normalization of the $W$-matrix. We then apply this construction to geometries beyond the hypercubic case, as is of interest for example in Section \ref{sec:Fracton}.

IsoTNS~\cite{zaletel_isometric_2020, haghshenas_conversion_2019} were originally introduced as a subclass of projected entangled-pair states (PEPS)~\cite{cirac_matrix_2021} in two dimensions, which are usually defined as the contraction of local rank-five tensors $T_{i_1i_2i_3i_4}^\sigma$, with four virtual legs of bond dimension $D$ and one physical leg with dimension $d$. Each virtual leg emanates in one direction, thereby allowing a PEPS network to span the entire two-dimensional space. The physical leg $\sigma$ describes one or more degrees of freedom on the lattice, depending on the unit cell and the distribution of the degrees of freedom. The wavefunction for a system with $N$ tensors is given as

\begin{equation}
        \ket{\Psi} = \sum_{\sigma_1, \cdots, \sigma_N} \text{tTR}\left( \left\{ T^{\sigma_1}, \cdots, T^{\sigma_N} \right\} \right) \ket{\sigma_1\cdots\sigma_N},.
    \label{eq:PEPSAppendix}
\end{equation}
where tTR is the tensor trace. 
While PEPS are a highly expressive class of wavefunctions which are believed to approximate the ground states of a large number of many-body Hamiltonians, they do not admit a canonical form in general, which renders their exact contraction an exponentially hard problem~\cite{schuch_computational_2007}. IsoTNS circumvent this issue by enforcing an additional isometry constraint on the tensors

\begin{equation}
    \sum_{\sigma, i_1, i_2} \left( (T)^\sigma_{i_1i_2i_3i_4} \right)^* (T)^\sigma_{i_1i_2i_3^\prime i_4^\prime} = \delta_{i_3i_3^\prime} \delta_{i_4i_4^\prime},
    \label{eq:isoTNS2D}
\end{equation}
where $\delta_{ab}$ is the Kronecker delta, $\delta_{ab} = 1$ for $a = b$ and zero otherwise. This property induces a directionality and allows for an efficient contraction of the tensor network in this direction.

While isoTNS are most commonly discussed for two-dimensional lattices, there is a straightforward generalization to higher dimensions. For an $n$-dimensional lattice model, one may define local tensors $(T)^\sigma_{i_1\cdots i_n  i_{n+1}\cdots i_{2n}}$ with $2n$ virtual legs $i_1, \dots, i_{2n}$, each pointing in one $n$-dimensional orthogonal direction (incoming and outgoing), and a physical leg $\sigma$. Its contraction describes a quantum state in this system. The isometry condition becomes

\begin{equation}
    \sum_{\sigma, i_1, \dots i_{n}} \left( (T)^\sigma_{i_1\cdots i_n  i_{n+1}\cdots i_{2n}} \right)^* (T)^\sigma_{i_1\cdots i_n i_{n+1}^\prime\cdots i_{2n}^\prime} = \delta_{i_{n+1}i_{n+1}^\prime} \cdots  \delta_{i_{2n}i_{2n}^\prime},
    \label{eq:isoTNSnD}
\end{equation}
where the virtual legs $i_j$ and $i_{j+n}$ point in opposite direction. This condition ensures the double-layer network $\langle\Psi\vert\Psi\rangle$ can be efficiently contracted in the same way as for the 2D case.

A transparent strategy to create isoTNS for certain states is offered by the $W$-matrix as introduced in the main text. For our purposes, we consider states on an $n$-dimensional hypercubic lattice which host a physical degree of freedom with dimension $d$ on every edge of the lattice. For the $n$ degrees of freedom $\sigma_1,\dots, \sigma_n$ in a unit cell, we decompose the local tensor as

\begin{equation}
    (T)^{\sigma_1\cdots\sigma_n}_{i_1\cdots i_n  i_{n+1}\cdots i_{2n}} = \prod_{j=1}^n \sum_{i_{n+j}'} \delta^{\sigma_j}_{i_{n+j}i^\prime_{n+j}} W_{(i_1\cdots i_n)(i^\prime_{n+1}\cdots i^\prime_{2n})},
\end{equation}
where $W_{(i_1\cdots i_n)(i_{n+1}\cdots i_{2n})}$ is a $d^n \times d^n$ matrix. The isometry condition (\ref{eq:isoTNSnD}) for this type of tensor is fulfilled if and only if

\begin{equation}
    \sum_{i_1\cdots i_n}\vert W_{(i_1\cdots i_n)(i_{n+1}\cdots i_{2n})} \vert^2 = 1 \; \forall i_{n+1}, \dots i_{2n},
\end{equation}
i.e., if its columns are normalized. The $W$-matrix naturally sorts the virtual legs with regard to the contraction direction. In the form given here, it locks the physical degrees of freedom to the associated virtual bond degrees of freedom. The network of vertex tensors $T_\nu$ defined by the prescription in \eq{eq:TVertex} with the $W$-matrix in \eq{eq:WSETLineon} falls in this category and is thus an example of a three-dimensional isoTNS. Other types of plumbing tensors may be considered as well: the modified plumbing tensor $(\delta_X)^\sigma_{ab}$ from the main text is a straightforward example. It associates a $\ket{0}$ ($\ket{1}$) state on the virtual legs to a $\ket{+} = (\ket{0} + \ket{1})/\sqrt{2}$ ($\ket{-} = (\ket{0} - \ket{1})/\sqrt{2}$) state on the physical leg. This amounts to a basis change into the $X$ basis on the physical leg. More distinct prescriptions for the plumbing tensors are possible as well: In Ref. \cite{liu_simulating_2023}, a double-line tensor network for the double-semion and toric code wavefunction is expressed in terms of a $W$-matrix and plumbing tensors which express the physical spins as domain walls of the virtual degrees of freedom.

We can define tensor networks on graphs with higher connectivity than the hypercube. Here, plumbing constructions akin to the more standard hybercubic tensor networks are also possible, albeit with plumbing tensors with more than two virtual legs. This naturally leads to an expression in terms of $W$-matrices. An example is given in the main text when considering the enrichment of fractons: there, we consider a twelve-leg tensor in the center of each cube, with each leg emanating towards one edge. One set of three legs sharing a corner are set as the incoming legs, while the other nine are outgoing. This allows us to re-organize the cube tensor in a $2^3 \times 2^9$ $W$-matrix as given in \eq{eq:WSETFracton}, which fulfills a normalization condition

\begin{equation}
    \sum_{i_1, i_2, i_3} \vert W_{(i_1i_2i_3)(i_4\dots i_{12})} \vert^2  = 1 \; \forall i_4, \dots i_{12}.
    \label{eq:NormCube}
\end{equation}
Each edge features a generalized plumbing tensor $(\delta_X)^\sigma_{abcd}$ as described in the main text, with one physical leg $\sigma$ and four virtual legs connecting it to the four adjacent cubes. It essentially sets the physical spins to the $\ket{+}$ ($\ket{-}$) state if all virtual degrees of freedom are in the $\ket{0}$ ($\ket{1}$) state and is zero otherwise. This structure allows for efficient contraction $\langle \Psi \vert \Psi \rangle$ of the wavefunction with its Hermitian conjugate as the normalization condition (\ref{eq:NormCube}) implies
\begin{equation}
    \begin{aligned}
    \sum_{i_1, i_2, i_3} &\left[ W_{(i_1i_2i_3)(i_4\dots i_{12})} W^*_{(i_1i_2i_3)(i_4^\prime\dots i_{12}^\prime)} \prod_{j=4}^{12}  \left( \sum_{\sigma_j} (\delta_X)^{\sigma_j}_{i_ja_jb_jc_j} (\delta_X)^{\sigma_j}_{i_j^\prime a_j^\prime b_j^\prime c_j^\prime} \right) \right] \\ =& \prod_{j=4}^{12}\delta_{a_jb_j} \delta_{a_jc_j}  \delta_{a_ja_j^\prime} \delta_{b_jb_j^\prime} \delta_{c_jc_j^\prime}.
    \end{aligned}
\end{equation}
This identity describes the local isometry condition: Contracting the $W$-matrix in a cube with its Hermitian conjugate $W^H$ on the incoming legs and the physical legs of the nine plumbing tensors on the outgoing legs between both layers reduces to identities, with the additional constraint that virtual legs coming from the same plumbing tensor have to agree. This property allows for a sequential contraction of the tensor network, similar to more common isoTNS.

The combination of the $W$-matrix construction and the isometry condition implies a number of favorable properties of the tensor networks. First of all, the matrix $\vert W\vert^2$ can be interpreted as a probability distribution of possible physical configurations on the incoming legs given some configuration on the outgoing legs. This naturally leads to the identification of a $d$-dimensional plumbed isoTNS with a stochastic process in $(d-1)$ dimensions, where the time direction takes the role of the contraction direction. In particular, operators diagonal in the computational basis can be evaluated from Monte Carlo sampling of this stochastic process, providing insight into the correlation structure of the state. Furthermore, it is known that isoTNS can be prepared in sequential circuits of a depth which scales with the linear system size $L$~\cite{soejima_isometric_2020, wei_sequential_2022}; the $W$-matrix provides a transparent strategy to define the local gate for this circuit. In particular, the equivalence of virtual and physical degrees of freedom allows for a significant simplification of this gate: each local degree of freedom is acted on only one time; after this, it will at most serve as a control qubit on qubits deeper in the bulk. Although there is no guarantee that this procedure represents an optimal implementation, it does represent a natural choice for the algorithm with a clear physical motivation, which also allows for holographic schemes as discussed in the main text.

\section{Non-local Volumetric Order Parameters}\label{apx:MOP}
In this appendix, we show how the non-local volumetric order parameters of the lineon and fracton enrichment in the X-cube model, as well as the electric excitation enrichment in the 3D toric code model, are calculated. These order parameters vanish when the symmetry fractionalizes non-trivially on the excitation in question, while they take a finite value for trivial fractionalization. When the obtained numerical values for the contracted tensor networks are zero up to machine precision, we set them to exactly zero before applying the power $1/L^2$ from the surface area between the two halves to avoid artificially amplified values in the non-trivial phase.

\subsection{Symmetry fractionalization on the lineons}
The non-local volumetric order parameter of the lineon enrichment is given by Eq.~\ref{eq:MOPLineon}. For $O_{\ell_z}$, we consider a system which is of finite length $L$ and periodic in the $x$ and $y$ directions and of infinite length $L_z \rightarrow \infty$ in the $z$ direction, and the subsystem $R$ on which the $X$ operators act is chosen to be the lower half of the system. Notice that the chosen subsystem $R$ must be filled by cube operators $\prod_{j \in c} X_j$, otherwise the operator $\prod_{\alpha\in R} X_{\alpha}$ will create lineon excitations such that the non-local volumetric order parameter is generically zero. One choice of the operator satisfying such a condition is:
\begin{equation}\label{eq:lineon_VOP_mid}
 \vcenter{\hbox{
  \includegraphics[width=6cm]{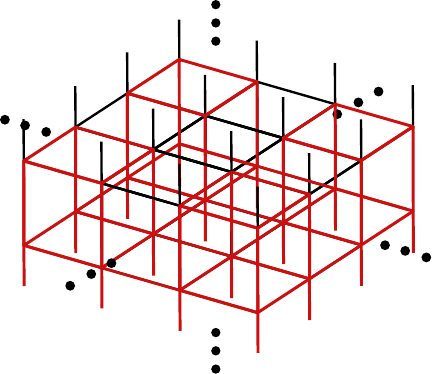}}},
\end{equation}
where red lines denote the edges on which the $X$ operators apply. The uppermost plane shown here is the boundary between the two semi-infinite halves, i.e. $X$ operators are applied on all qubits below, while no operators act on the qubits above. Due to the checkerboard structure some of the bonds in the plane separating the semi-infinite halves remain untouched. 

Since the isoTNS is unnormalized, in order to numerically evaluate the non-local volumetric order parameter, we need to express it as the ratio of two contracted tensor networks, i.e. as
\begin{equation}
        O_{\ell_z} = \lim_{L_z \rightarrow \infty} \left( \frac{\bra{\Psi_{\ell_z}} \prod_{\alpha \in R} X_\alpha  \ket{\Psi_{\ell_z}}}{\langle{\Psi_{\ell_z}} \vert{\Psi_{\ell_z}}\rangle} \right)^{1/L^2}.
        \label{eq:VOPNormalized}
\end{equation}
We can construct the norm $\langle{\Psi_{\ell_z}} \vert{\Psi_{\ell_z}}\rangle$ from the tensor network $\langle{\Psi} \vert{\Psi}\rangle$, where $\ket{\Psi}$ is given by Eq.~\ref{eq:TVertex} with the $W$-matrix in Eq.~\ref{eq:WSETLineon}. Although $\langle \Psi|\Psi\rangle$ is a double-layer (bra and ket layer) 3D tensor network, it can be reduced to a single-layer 3D tensor network as the isoTNS tensor has a ``plumbing" structure, and the contraction of the physical indices of two tensors in the bra and ket layers can be simplified to an entry-wise product of two tensors in the bra and ket layers without physical indices.
If we consider the reduced 3D tensor network $\langle \Psi|\Psi\rangle$, the transfer matrix $T_z$ in the $xy$ plane is a 2D tensor network operator with periodic boundary conditions. Importantly, the transfer matrix has the subsystem symmetry generated by $W_x=\prod_{y}Z_{x,y}$ and $W_y=\prod_{x}Z_{x,y}$: 
\begin{equation}
 \vcenter{\hbox{
    \includegraphics[width=8cm]{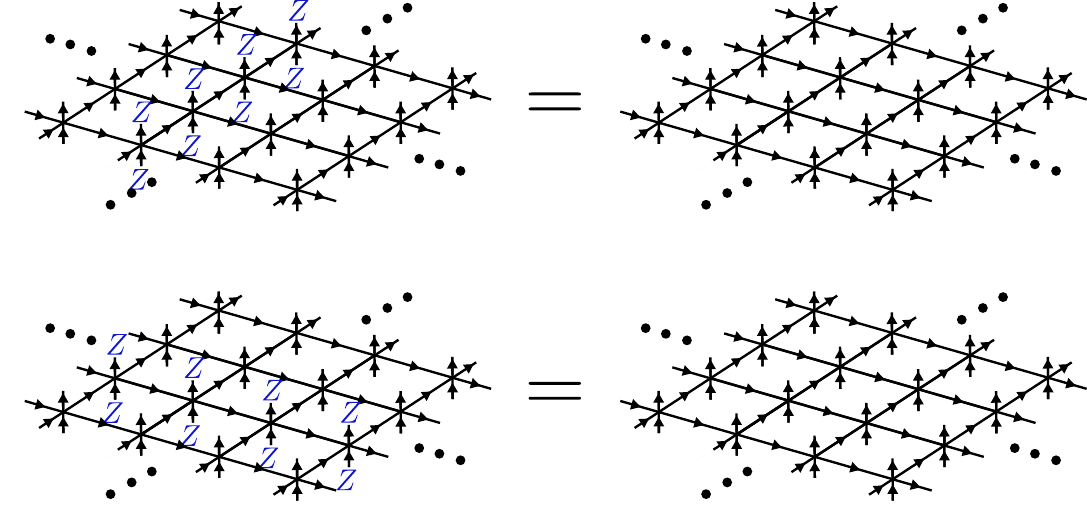}}} .
\end{equation}
For a transfer matrix with size $L_x\times L_y$, there are $L_x+L_y-1$ independent subsystem symmetry generators. The dominant eigenvalue of the transfer matrix is $2^{L_x+L_y-1}$-fold degenerate (up to an exponentially small finite-size splitting). We can thus construct a projector 
\begin{equation}
P_{x_0,y_0}=\prod_{x\neq x_0}\frac{1+W_x}{2} \prod_{y\neq y_0}\frac{1+W_y}{2} \frac{1-W_{x_0}}{2}\frac{1-W_{y_0}}{2},
\end{equation}
which commutes with the transfer matrix $T_z$ and projects to one specific dominant eigenvector when applied on the space spanned by the dominant eigenvectors. With the transfer matrix and the projector, for system size $L_z$ in $z$ direction we have

\begin{align}
    \langle \Psi|\Psi\rangle &= \bra{\tilde{V}_u}{T^{L_z}_z}\ket{\tilde{V}_d}, \nonumber
    \\ \langle \Psi_{\ell_z}|\Psi_{\ell_z}\rangle &= \bra{\tilde{V}_u}{P_{x_0,y_0}T^{L_z}_z}\ket{\tilde{V}_d},
\end{align} 
where $\bra{\tilde{V}_u}$ and $\ket{\tilde{V}_d}$ are the top and bottom vectors fixing the boundaries of the 3D tensor networks.
The projector $P_{x_0,y_0}$ corresponds to a lineon string along the $z$ direction with coordinates $(x_0,y_0)$ in the $xy$ plane, at the ends of which two lineons of infinite separation sit when $L_z\rightarrow+\infty$. Similarly, we can express the numerator as
\begin{equation}
    \langle \Psi_{\ell_z}|\prod_{\alpha\in R}X_{\alpha}|\Psi_{\ell_z}\rangle= \bra{\tilde{V}_u}P_{x_0,y_0}T^{L_z/2}_{z}T_{z,\mathrm{mid}}T^{L_z/2 -1}_{z, X})\ket{\tilde{V}_d},
\end{equation}
where $T_{z,X}$ is the transfer matrix of the reduced single layer 3D tensor network of $\bra{\Psi}\prod_{\alpha} X_{\alpha}\ket{\Psi}$ with $X$ operators present on all bonds and $T_{z,\mathrm{mid}}$ is the transfer matrix at the middle of $\langle \Psi_{\ell_z}|\prod_{\alpha\in R}X_{\alpha}|\Psi_{\ell_z}\rangle$. At the boundary of $R$, see Eq.~\ref{eq:lineon_VOP_mid} for which bonds feature $X$ operators due to the checkerboard structure.

When $L_z\rightarrow\infty$, the above relations can be simplified using the dominant eigenvectors of the transfer matrices. The volumetric order parameter in Eq.~\ref{eq:VOPNormalized} can be expressed as:
\begin{equation}
    O_{\ell_z}=\left( \frac{\langle V_u|P_{x_0,y_0} T_{z,\mathrm{mid}}|V_{d,X} \rangle}{\langle V_u|P_{x_0,y_0} T_z |V_d \rangle} \right)^{1/L^2}, 
\end{equation}
where $\bra{V_u}, \ket{V_d}$ are the top and bottom dominant eigenvectors of $T_zP_{x_0,y_0}$, and $\ket{V_{d,X}}$ is the bottom dominant eigenvector of $T_{z,X}P_{x_0,y_0}$. The order parameter for the symmetry fractionalization on the lineons in other directions can be derived similarly.

\subsection{Symmetry fractionalization on the fractons}
Unlike in the case of lineons, which are created in pairs, for a system with periodic boundary conditions we have to create at least four fractons. However, for a state hosting four fractons we cannot define a volumetric operator which covers only one fracton because such an operator cannot commute with all $B_c$ operators and therefore necessarily creates additional fractons when applied to the isoTNS. To avoid this subtlety, we instead consider a fracton isoTNS with open boundary conditions such that we can create a single fracton. For example, we may consider an isoTNS which is finite in the $x$ and $y$ directions and of infinite length in the $z$ direction. Notice that the open boundary conditions must be defined such that they do not break the global symmetry $\prod_{\alpha} Z_{\alpha}$. Since the 3D isoTNS is repeated in $z$ direction, we show the definition of the open boundary wavefunction using a repeated unit of the virtual degrees of freedom:
\begin{equation}
 \vcenter{\hbox{
  \includegraphics[width=5cm]{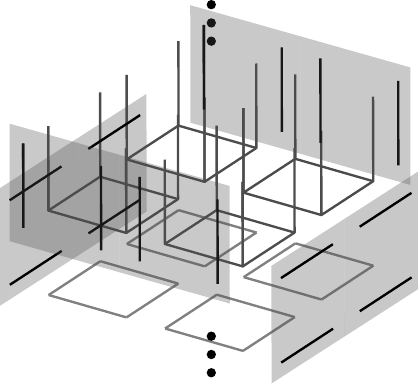}}} \;\;,
\end{equation}
where each line represents one virtual degree of freedom. The twelve virtual legs of each cube in the bulk are defined by the cube $W$-matrix given in Eq.~\ref{eq:WSETFracton}, while the four virtual legs on each shaded plane on the boundary are fixed to a GHZ state $\ket{\text{GHZ}} = (\ket{0000}+\ket{1111})/\sqrt{2}$. The physical degrees of freedom $\sigma$ on the edges are determined by plumbing tensors $(\delta_X)_{abcd}^\sigma$ where four virtual legs meet and by $(\delta_X)_{ab}^\sigma$ where two virtual legs meet.

Similar to the lineon enrichment case, the subsystem $S$ covering the single fracton must be filled by star operators such that the volume operator $\prod_{\alpha\in S}Z_{\alpha}$ does not create additional excitations. One choice of the volumetric operator satisfying such a condition is:
\begin{equation}
 \vcenter{\hbox{
    \includegraphics[width=4cm]{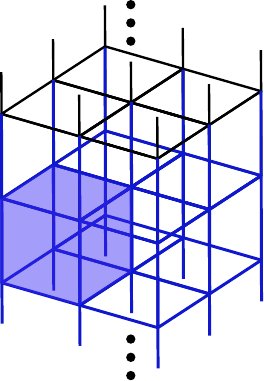}}} \; \;,
\end{equation}
where the blue lines indicate the edges where the $Z$ operators are applied. The cube hosting the single fracton excitation is indicated using light blue shading. The open boundary conditions allow for such a state, as the other excitations which would normally be present are absorbed by the boundary tensors. By reducing the double layer tensor networks $\bra{\Psi_f} \prod_{\alpha \in S} Z_\alpha  \ket{\Psi_f}$ and $\langle\Psi_f\vert\Psi_f\rangle$ to single layer tensor networks and taking the limit $L_z \rightarrow \infty$ using transfer matrix methods, we can evaluate the non-local volumetric order parameter $O_f$ as given in \eq{eq:MOPFracton} analogously to the lineon case.

\subsection{Symmetry fractionalization on the vertex excitations of the 3D toric code}
The calculation of the volumetric order parameter $O_e$ of the vertex charge enrichment in the 3D toric code (see Appendix~\ref{apx:3DTC}) proceeds along similar lines as the lineon enrichment in the X-cube model. One difference is that for the isoTNS of the 3D toric code, the projector to the electric excitation sector is $(1+Z^{\otimes L_xL_y})/2$, which replaces the projector $P_{x_0,y_0}$ of the lineon state. Moreover, the subsystem $R$ has to be filled with plaquette operators such that the volume operator $\prod_{\alpha\in R}X_{\alpha}$ itself does not create electric charge excitations. One choice of the operator satisfying such a condition is:
\begin{equation}
 \vcenter{\hbox{
  \includegraphics[width=6cm]{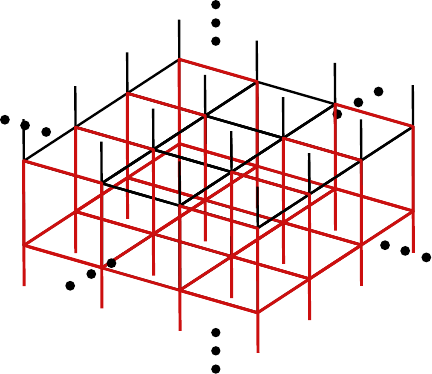}}},
\end{equation}
where the red lines imply the edges where the $X$ operators are applied. Otherwise, the numerical calculation is equivalent to the lineon case.

\section{Definition of many-qubits gates in the state preparation circuits}\label{apx:Gates}
We provide the definition for some more complicated gates we use in the circuits presented in Section \ref{sec:Preparation}. These gates can be decomposed into sequences of controlled unitaries. While the actual feasibility of an implementation depends on the hardware used, these decompositions may serve as a possible guideline for concrete realizations.

In the following, we will introduce several controlled unitaries $C^{(n)}U^{(m)}$ which act on $n$ control qubits and $m$ target qubits. The action of these gates is defined as
\begin{equation}
    C^{(n)}U^{(m)} \underset{n}{\ket{11\cdots1}} \;\underset{m}{\ket{\psi}} = \underset{n}{\ket{11\cdots1}} \; U\underset{m}{\ket{\psi}} \text{, trivially otherwise},
    \label{eq:CU}
\end{equation}
i.e. the unitary $U$ will be applied to the target qubits if and only if all control qubits are set to $\ket{1}$, otherwise the gate will act as the identity. All unitaries used in the following constructions are of the types $X, Z$ or the phase gate $S$ defined as

\begin{equation}
    S = \begin{pNiceMatrix}[first-row,last-col]
        \ket{0} & \ket{1} &    \\
        1 & 0 & \; \ket{0} \\ 
        0& i & \; \ket{1}.
        \end{pNiceMatrix}
\end{equation}

We first discuss the three unitaries $U_i$ for $i = 4,5,6$ at the end of the lineon circuit in \figc{fig:LineonCircuit}{a}. They are only applied if at least one of the parameters $g_{1/2}$ is negative, i.e. if we want to create a state featuring non-trivial symmetry fractionalization on the lineons. Each unitary consists of two $C^{(2)}S$ gates, which are already diagonal in the $Z$ basis. For $U_4$, the two $C^{(2)}S$ gates act on the triples of qubits named $1, 5, 6$ and $2, 3, 4$ in \figc{fig:LineonCircuit}{a}, respectively. The coefficients of two allowed qubit configurations become imaginary under application of this gate. Likewise, $U_5$ acts on the triples $2,4,6$ and $1,3,5$, while $U_6$ acts on $1,2,6$ and $3,4,5$. In all possible cases where $g_{1/2} < 0$, precisely two of these unitaries are present, leading to four local configurations with complex coefficients. The fully occupied configuration remains invariant, as all four $C^{(2)}S$ act non-trivially on it. Thus, the structure as given in \eq{eq:WSETLineon} is achieved.

The fracton circuit presented in \figc{fig:FractonCircuit}{a} features two controlled gates $C_1$ and $C_2$. They check the nine input qubits and apply $X$ gates to one or all of the output qubits depending on the input configuration. Therefore, they can be decomposed into $C^{(n)}X$ gates. The gate $C_1$ is a product of six $CX$ gates and 36 $C^{(2)}X$ gates acting on the third qubit, following the numeration given in the figure and the main text. The six $CX$ gates run over the set of qubits $P_2$, i.e. the qubits $7, \dots, 12$, while the $\frac{9\cdot 8}{2}=36 \; C^{(2)}X$ gates run over every possible pair of the nine input qubits $4, \dots,12$. The gate $C_2$ checks each input qubit separately and flips all output qubits once for every $\ket{1}$ state. It can therefore be decomposed into a sequence of nine $CX^{(3)}$ gates, which can be further decomposed into three $CX$ gates each, leading to a total of 27 two-qubit gates.

The final unitary $U_3$ in the fracton circuit is applied if and only if $g <0$ and allocates the complex phase factors in the non-trivial phase, thereby playing the same role as the unitaries in the lineon circuit discussed above. Due to the higher number of possible configurations, its effect is more complicated; we may decompose it into 30 $CS$ gates and 495 $C^{(3)}Z$ gates. The controlled-phase gates act on each pair in the two sets $P_1$ and $P_2$, respectively, leading to $2 \cdot \frac{6\cdot5}{2} = 30$ two-qubit gates, while the $C^{(3)}Z$ gates act on every possible set of four between the twelve qubits, leading to $\frac{12 \cdot 11 \cdot10 \cdot9}{4 \cdot 3 \cdot 2} = 495$ four-qubit gates. 

\section{Symmetry enrichment in the 3D toric code}\label{apx:3DTC}

The construction developed in the main text to enrich the excitations of the X-Cube model is in fact not limited to this specific model and can be applied more generally to other fixed points with a simple tensor network structure. In this section, we illustrate an analogous procedure to treat another archetypal three-dimensional model, the 3D toric code~\cite{kitaev_fault-tolerant_2003, he_entanglement_2018}. The associated Hamiltonian on a cubic lattice with qubits on each edge is
\begin{equation}
    H_{\text{3DTC}} =  \sum_f B_f + \sum_\nu A_\nu,
    \label{eq:3DTC}
\end{equation}
where the projectors $B_f = (1 - \prod_{j \in f} X_j)/2$ and $A_\nu = (1- \prod_{j \in \nu} Z_j)/2$ are defined from products of $X$ operators around the face $f$ and $Z$ operators around the vertex $\nu$. The individual terms of the Hamiltonian commute and thus are stabilizers. Its ground state is an equal-weight superposition of all loop configurations in the $Z$ basis. The 3D toric code state is a typical example of topological order in three dimensions. On a torus geometry, the ground state is eightfold degenerate as three Wilson loops each of $X$ and $Z$ operators can be defined. As a side note, it can be obtained from coupled layers in a similar fashion as the X-Cube model, using an $X$ coupling instead of a $Z$ coupling between qubits of different layers~\cite{ma_fracton_2017}. It features the same symmetries $\mathcal{T}_{X/Z}$ as the X-Cube model. The natural excitations of the system are the electric excitations $e$, which correspond to violations of a vertex stabilizer $A_\nu = -1$, and the magnetic excitations $m$, which correspond to violations of a plaquette stabilizer $B_f = -1$. Electric excitations are created at the endpoints of $X$ strings, whereas the loop-like magnetic excitations are created by applying $Z$ operators.
\begin{figure*}
    \includegraphics[width=\textwidth]{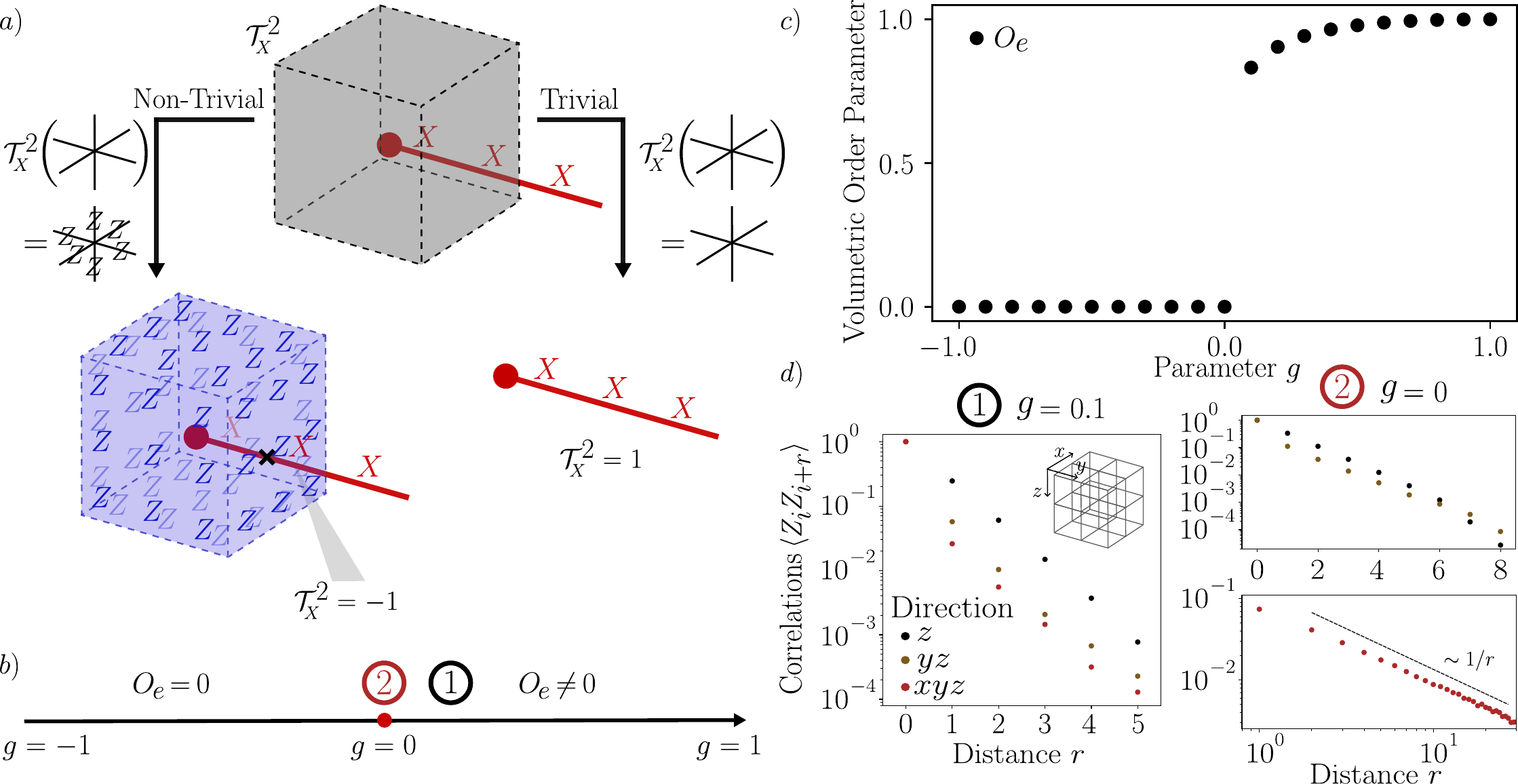}
    \caption{\label{fig:3DTC}
        \textbf{Enriching the electric excitation in the 3D toric code.} a) Possible fractionalization schemes in the phase diagram spanned by the $W$-matrix given in \eq{eq:W3DTCEnriched}. In the trivial phase, the anti-unitary symmetry $\mathcal{T}_X = (\prod_i X_i) \mathcal{K}$ on a subsystem squares to the identity; in the phase featuring non-trivial symmetry fractionalization, it squares to a cage of $Z$ operators on the boundary of the subsystem, which can detect the presence of electric excitations on the inside. Note the additional $Z$ operators compared to the X-Cube case shown in \fig{fig:LineonCubeOperator}, which are necessary due to the additional mobility of a 3D TC excitation compared to a lineon. b) The phase diagram of the wavefunction with different symmetry fractionalization. c) The non-local volumetric order parameter $O_e$ in the entire phase diagram of b). If the symmetry fractionalizes non-trivially on the electric excitation $e$, the order parameter will vanish. d) Correlation functions at two points in the phase diagram starting from a disordered boundary $\ket{\psi_0} = \ket{++ \cdots +}$, evaluated using Monte Carlo sampling techniques. Left panel: Away from the critical point, all correlations decay exponentially. Right panel: At the critical point $g = 0$, the $\langle Z_i Z_j \rangle$ correlations decay algebraically in the diagonal direction as $\sim 1/r$, as the state can be mapped to the propagation of hard-core particles in two spatial dimensions, interpreting the diagonal as the time direction.
    }
\end{figure*}

The bulk ground state of the 3D toric code also has a simple description as an isoTNS which can be readily expressed using the plumbing scheme featuring vertex tensors $(T_\nu)^{\sigma \rho \tau}_{i_1, \dots i_6}$ as given in \eq{eq:TVertex}. The $W$-matrix for this state is

\begin{widetext}
\begin{equation}
    W_\nu = \begin{pNiceMatrix}[first-row,last-col]
 \ket{000} & \ket{001} & \ket{010} & \ket{011} & \ket{100} & \ket{101} & \ket{110} & \ket{111} &    \\
 1/2  & 0   & 0  & 1/2  & 0 & 1/2  & 1/2 & 0  & \;\ket{000} \\
 0   & 1/2   & 1/2 & 0  & 1/2 & 0 & 0 & 1/2  & \;\ket{001} \\
0   & 1/2   & 1/2  & 0  & 1/2 & 0  & 0 & 1/2  & \;\ket{010} \\
1/2  & 0  & 0  & 1/2  & 0 & 1/2  & 1/2 & 0  & \;\ket{011} \\
0   & 1/2  & 1/2  & 0  & 1/2 & 0  & 0 & 1/2  & \;\ket{100} \\
1/2   & 0   & 0  & 1/2  & 0 & 1/2 & 1/2 & 0  &  \;\ket{101} \\
1/2  & 0 & 0 & 1/2  & 0 & 1/2 & 1/2 & 0  & \;\ket{110} \\
0  & 1/2  & 1/2  & 0  & 1/2 & 0  & 0 & 1/2  & \;\ket{111} \\
\end{pNiceMatrix}.
\label{eq:W3DTC}
\end{equation}
\end{widetext}
In this $W$-matrix the loop condensate character is expressed by the condition that only even-parity configurations at every vertex are allowed.

Starting from this exact representation of the 3D TC fixed point, it is possible to construct a path towards a fixed point with non-trivial symmetry fractionalization using the same strategy as in the main text. As we consider vertex tensors, the natural objects to enrich are the electric excitations under the time-reversal symmetry $\mathcal{T}_X$. In the SET phase, we want the physical symmetry to be equivalent to non-trivial operators on the virtual legs consisting of the $X$ operators which are plumbed from the physical leg and a phase gate $S^{(\dagger)}$, the square of which has the same effect as checking the vertex stabilizer $A_\nu$. This detects the presence of a vertex excitation due to the additional minus sign which indicates the half-integer $\mathcal{T}_X^2 = -1$ operation on the excitation. Following the same procedure as in the main text, this goal can be achieved by splitting the six qubits around the vertex into two sets of threes (for example, the qubits on incoming and outgoing legs, respectively), counting the number of possible pairs in each set and allocating complex phases if the parity of the sum of both numbers is odd. A full paramaterization of such a path depending on a single parameter $g \in [-1,1]$ is

\begin{widetext}
\begin{equation}
    W_\nu(g) = \begin{pNiceMatrix}[first-row,last-col]
 \ket{000} & \ket{001} & \ket{010} & \ket{011} & \ket{100} & \ket{101} & \ket{110} & \ket{111} &    \\
 \frac{1}{\sqrt{1+3\vert g \vert}}  & 0   & 0  & \sqrt{\frac{\text{sign}(g)\vert g \vert}{3+\vert g \vert}}  & 0 & \sqrt{\frac{\text{sign}(g)\vert g \vert}{3+\vert g \vert}}  & \sqrt{\frac{\text{sign}(g)\vert g \vert}{3+\vert g \vert}} & 0  & \;\ket{000} \\
 0   & \frac{1}{\sqrt{3+\vert g \vert}}   & \frac{1}{\sqrt{3+\vert g \vert}} & 0  & \frac{1}{\sqrt{3+\vert g \vert}} & 0 & 0 & \sqrt{\frac{\text{sign}(g)\vert g \vert}{1+3\vert g \vert}}  & \;\ket{001} \\
0   & \frac{1}{\sqrt{3+\vert g \vert}}   & \frac{1}{\sqrt{3+\vert g \vert}}  & 0  & \frac{1}{\sqrt{3+\vert g \vert}} & 0  & 0 & \sqrt{\frac{\text{sign}(g)\vert g \vert}{1+3\vert g \vert}}  & \;\ket{010} \\
\sqrt{\frac{\text{sign}(g)\vert g \vert}{1+3\vert g \vert}}  & 0  & 0  & \frac{1}{\sqrt{3+\vert g \vert}}  & 0 & \frac{1}{\sqrt{3+\vert g \vert}}  & \frac{1}{\sqrt{3+\vert g \vert}} & 0  & \;\ket{011} \\
0   & \frac{1}{\sqrt{3+\vert g \vert}}  & \frac{1}{\sqrt{3+\vert g \vert}}  & 0  & \frac{1}{\sqrt{3+\vert g \vert}} & 0  & 0 & \sqrt{\frac{\text{sign}(g)\vert g \vert}{1+3\vert g \vert}}  & \;\ket{100} \\
\sqrt{\frac{\text{sign}(g)\vert g \vert}{1+3\vert g \vert}}   & 0   & 0  & \frac{1}{\sqrt{3+\vert g \vert}}  & 0 & \frac{1}{\sqrt{3+\vert g \vert}} & \frac{1}{\sqrt{3+\vert g \vert}} & 0  &  \;\ket{101} \\
\sqrt{\frac{\text{sign}(g)\vert g \vert}{1+3\vert g \vert}}  & 0 & 0 & \frac{1}{\sqrt{3+\vert g \vert}}  & 0 & \frac{1}{\sqrt{3+\vert g \vert}} & \frac{1}{\sqrt{3+\vert g \vert}} & 0  & \;\ket{110} \\
0  & \sqrt{\frac{\text{sign}(g)\vert g \vert}{3+\vert g \vert}}  & 1\sqrt{\frac{\text{sign}(g)\vert g \vert}{3+\vert g \vert}}  & 0  & \sqrt{\frac{\text{sign}(g)\vert g \vert}{3+\vert g \vert}} & 0  & 0 & \frac{1}{\sqrt{1+ 3\vert g \vert}}  & \;\ket{111} \\
\end{pNiceMatrix}.
\label{eq:W3DTCEnriched}
\end{equation}
\end{widetext}
The virtual symmetry protecting the closed loop constraint is respected along the entire path.

From this expression, the action of the local symmetry operation $\mathcal{T}_X = \left (\prod_{i=1}^6 X_i \right) \mathcal{K}$ can be checked explicitly. For $g > 0$, its acts trivially on the tensor, without any symmetry fractionalization:
\begin{equation}
    \includegraphics[scale=0.75]{LineonSymmetryTrivial.pdf},
    \label{eq:3DTCTrivial}
\end{equation}
while for $g < 0$, a non-trivial virtual operator is present:
\begin{equation}
    \includegraphics[scale=0.75]{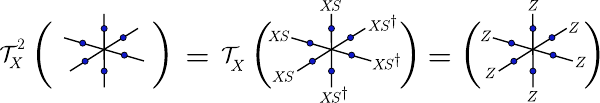},
    \label{eq:3DTCEnriched}
\end{equation}
which is equivalent to a vertex stabilizer. Applying the symmetry to a subsystem twice leads to a cage operator which measures the parity of $e$ excitations in this subsystem. This implies that the electric excitation exhibits Kramer's degeneracy in this phase, $\mathcal{T}_X^2 = -1$. For the 3D toric code, a minimally entangled state $\ket{\Psi_e}$ corresponding to the $e$ excitation can be defined without the intricacies associated with fracton phases, as the number of Wilson loop operators does not increase with system size. Defining a non-local volumetric order parameter $O_e$ in the same way as for the lineon exictations in \eq{eq:MOPLineon}, we use this quantity to distinguish the different enrichment schemes, with non-trivial fractionalization implying $O_e = 0$.

The critical point $g = 0$ between the two phases is partly amenable to an analytical treatment. While away from this point, the correlations $\langle Z_i Z_j \rangle$ decay exponentially in all directions, as can be checked numerically by mapping the state to a stochastic process as in the main text. At the critical point a $U(1)$ conservation law in this circuit emerges: If we initialize the state from a product state $\ket{\psi_0}$ in the $Z$ basis on three boundaries of the lattice which meet at a corner, the number of $\ket{1}$ states will be conserved on each slice orthogonal to the diagonal line emanating from this corner. Here, the process of sequentially building up the wavefunction from the boundary is therefore equivalent to the dynamical proliferation of hard-core particles in a two-dimensional system, with the diagonal being the time direction; the total wavefunction is a superposition of all possible outcomes. Similar to the decoupled planes at the lineon critical points, the evolving configurations thus obey a diffusive process: along the diagonal direction, the correlations are critical obeying a power-law $\langle Z_i Z_{i+r} \rangle \sim 1/r$ in the case of a finite density of $\ket{1}$ states, while other directions exhibit exponential decay.

The results of the 3D toric code are summarized in \fig{fig:3DTC}, including numerical data on the volumetric order parameter and the correlation functions.

\bibliography{references}

\providecommand{\noopsort}[1]{}\providecommand{\singleletter}[1]{#1}%
\begin{thebibliography}{86}%
\makeatletter
\providecommand \@ifxundefined [1]{%
 \@ifx{#1\undefined}
}%
\providecommand \@ifnum [1]{%
 \ifnum #1\expandafter \@firstoftwo
 \else \expandafter \@secondoftwo
 \fi
}%
\providecommand \@ifx [1]{%
 \ifx #1\expandafter \@firstoftwo
 \else \expandafter \@secondoftwo
 \fi
}%
\providecommand \natexlab [1]{#1}%
\providecommand \enquote  [1]{``#1''}%
\providecommand \bibnamefont  [1]{#1}%
\providecommand \bibfnamefont [1]{#1}%
\providecommand \citenamefont [1]{#1}%
\providecommand \href@noop [0]{\@secondoftwo}%
\providecommand \href [0]{\begingroup \@sanitize@url \@href}%
\providecommand \@href[1]{\@@startlink{#1}\@@href}%
\providecommand \@@href[1]{\endgroup#1\@@endlink}%
\providecommand \@sanitize@url [0]{\catcode `\\12\catcode `\$12\catcode `\&12\catcode `\#12\catcode `\^12\catcode `\_12\catcode `\%12\relax}%
\providecommand \@@startlink[1]{}%
\providecommand \@@endlink[0]{}%
\providecommand \url  [0]{\begingroup\@sanitize@url \@url }%
\providecommand \@url [1]{\endgroup\@href {#1}{\urlprefix }}%
\providecommand \urlprefix  [0]{URL }%
\providecommand \Eprint [0]{\href }%
\providecommand \doibase [0]{https://doi.org/}%
\providecommand \selectlanguage [0]{\@gobble}%
\providecommand \bibinfo  [0]{\@secondoftwo}%
\providecommand \bibfield  [0]{\@secondoftwo}%
\providecommand \translation [1]{[#1]}%
\providecommand \BibitemOpen [0]{}%
\providecommand \bibitemStop [0]{}%
\providecommand \bibitemNoStop [0]{.\EOS\space}%
\providecommand \EOS [0]{\spacefactor3000\relax}%
\providecommand \BibitemShut  [1]{\csname bibitem#1\endcsname}%
\let\auto@bib@innerbib\@empty
\bibitem [{\citenamefont {Dijkgraaf}\ and\ \citenamefont {Witten}(1990)}]{dijkgraaf_topological_1990}%
  \BibitemOpen
  \bibfield  {author} {\bibinfo {author} {\bibfnamefont {R.}~\bibnamefont {Dijkgraaf}}\ and\ \bibinfo {author} {\bibfnamefont {E.}~\bibnamefont {Witten}},\ }\bibfield  {title} {\bibinfo {title} {Topological gauge theories and group cohomology},\ }\href {https://doi.org/10.1007/BF02096988} {\bibfield  {journal} {\bibinfo  {journal} {Commun. Math. Phys}\ }\textbf {\bibinfo {volume} {129}},\ \bibinfo {pages} {393} (\bibinfo {year} {1990})}\BibitemShut {NoStop}%
\bibitem [{\citenamefont {Wen}(1990)}]{wen_topological_1990}%
  \BibitemOpen
  \bibfield  {author} {\bibinfo {author} {\bibfnamefont {X.~G.}\ \bibnamefont {Wen}},\ }\bibfield  {title} {\bibinfo {title} {Topological orders in rigid states},\ }\href {https://doi.org/10.1142/S0217979290000139} {\bibfield  {journal} {\bibinfo  {journal} {International Journal of Modern Physics B}\ }\textbf {\bibinfo {volume} {04}},\ \bibinfo {pages} {239} (\bibinfo {year} {1990})}\BibitemShut {NoStop}%
\bibitem [{\citenamefont {Wen}(2017)}]{wen_zoo_2017}%
  \BibitemOpen
  \bibfield  {author} {\bibinfo {author} {\bibfnamefont {X.-G.}\ \bibnamefont {Wen}},\ }\bibfield  {title} {\bibinfo {title} {Colloquium: Zoo of quantum-topological phases of matter},\ }\href {https://doi.org/10.1103/RevModPhys.89.041004} {\bibfield  {journal} {\bibinfo  {journal} {Rev. Mod. Phys.}\ }\textbf {\bibinfo {volume} {89}},\ \bibinfo {pages} {041004} (\bibinfo {year} {2017})}\BibitemShut {NoStop}%
\bibitem [{\citenamefont {Leinaas}\ and\ \citenamefont {Myrheim}(1977)}]{LeinaasMyrheim}%
  \BibitemOpen
  \bibfield  {author} {\bibinfo {author} {\bibfnamefont {J.~M.}\ \bibnamefont {Leinaas}}\ and\ \bibinfo {author} {\bibfnamefont {J.}~\bibnamefont {Myrheim}},\ }\bibfield  {title} {\bibinfo {title} {On the theory of identical particles},\ }\href {https://doi.org/https://doi.org/10.1007/BF02727953} {\bibfield  {journal} {\bibinfo  {journal} {Il Nuovo Cimento B}\ }\textbf {\bibinfo {volume} {37}},\ \bibinfo {pages} {1} (\bibinfo {year} {1977})}\BibitemShut {NoStop}%
\bibitem [{\citenamefont {Wilczek}(1982)}]{Wilczek1982}%
  \BibitemOpen
  \bibfield  {author} {\bibinfo {author} {\bibfnamefont {F.}~\bibnamefont {Wilczek}},\ }\bibfield  {title} {\bibinfo {title} {Quantum mechanics of fractional-spin particles},\ }\href {https://doi.org/10.1103/PhysRevLett.49.957} {\bibfield  {journal} {\bibinfo  {journal} {Phys. Rev. Lett.}\ }\textbf {\bibinfo {volume} {49}},\ \bibinfo {pages} {957} (\bibinfo {year} {1982})}\BibitemShut {NoStop}%
\bibitem [{\citenamefont {Kitaev}(2003)}]{kitaev_fault-tolerant_2003}%
  \BibitemOpen
  \bibfield  {author} {\bibinfo {author} {\bibfnamefont {A.~Y.}\ \bibnamefont {Kitaev}},\ }\bibfield  {title} {\bibinfo {title} {Fault-tolerant quantum computation by anyons},\ }\href {https://doi.org/10.1016/S0003-4916(02)00018-0} {\bibfield  {journal} {\bibinfo  {journal} {Annals of Physics}\ }\textbf {\bibinfo {volume} {303}},\ \bibinfo {pages} {2} (\bibinfo {year} {2003})}\BibitemShut {NoStop}%
\bibitem [{\citenamefont {Essin}\ and\ \citenamefont {Hermele}(2013)}]{essin_classifying_2013}%
  \BibitemOpen
  \bibfield  {author} {\bibinfo {author} {\bibfnamefont {A.~M.}\ \bibnamefont {Essin}}\ and\ \bibinfo {author} {\bibfnamefont {M.}~\bibnamefont {Hermele}},\ }\bibfield  {title} {\bibinfo {title} {Classifying fractionalization: {Symmetry} classification of gapped $\mathbb{Z}_2$ spin liquids in two dimensions},\ }\href {https://doi.org/10.1103/PhysRevB.87.104406} {\bibfield  {journal} {\bibinfo  {journal} {Phys. Rev. B}\ }\textbf {\bibinfo {volume} {87}},\ \bibinfo {pages} {104406} (\bibinfo {year} {2013})}\BibitemShut {NoStop}%
\bibitem [{\citenamefont {Chen}(2017)}]{chen_symmetry_2017}%
  \BibitemOpen
  \bibfield  {author} {\bibinfo {author} {\bibfnamefont {X.}~\bibnamefont {Chen}},\ }\bibfield  {title} {\bibinfo {title} {Symmetry fractionalization in two dimensional topological phases},\ }\href {https://doi.org/10.1016/j.revip.2017.02.002} {\bibfield  {journal} {\bibinfo  {journal} {Reviews in Physics}\ }\textbf {\bibinfo {volume} {2}},\ \bibinfo {pages} {3} (\bibinfo {year} {2017})}\BibitemShut {NoStop}%
\bibitem [{\citenamefont {Pollmann}\ \emph {et~al.}(2010)\citenamefont {Pollmann}, \citenamefont {Turner}, \citenamefont {Berg},\ and\ \citenamefont {Oshikawa}}]{pollmann_entanglement_2010}%
  \BibitemOpen
  \bibfield  {author} {\bibinfo {author} {\bibfnamefont {F.}~\bibnamefont {Pollmann}}, \bibinfo {author} {\bibfnamefont {A.~M.}\ \bibnamefont {Turner}}, \bibinfo {author} {\bibfnamefont {E.}~\bibnamefont {Berg}},\ and\ \bibinfo {author} {\bibfnamefont {M.}~\bibnamefont {Oshikawa}},\ }\bibfield  {title} {\bibinfo {title} {Entanglement spectrum of a topological phase in one dimension},\ }\href {https://doi.org/10.1103/PhysRevB.81.064439} {\bibfield  {journal} {\bibinfo  {journal} {Phys. Rev. B}\ }\textbf {\bibinfo {volume} {81}},\ \bibinfo {pages} {064439} (\bibinfo {year} {2010})}\BibitemShut {NoStop}%
\bibitem [{\citenamefont {Tsui}\ \emph {et~al.}(1982)\citenamefont {Tsui}, \citenamefont {Stormer},\ and\ \citenamefont {Gossard}}]{tsui_two-dimensional_1982}%
  \BibitemOpen
  \bibfield  {author} {\bibinfo {author} {\bibfnamefont {D.~C.}\ \bibnamefont {Tsui}}, \bibinfo {author} {\bibfnamefont {H.~L.}\ \bibnamefont {Stormer}},\ and\ \bibinfo {author} {\bibfnamefont {A.~C.}\ \bibnamefont {Gossard}},\ }\bibfield  {title} {\bibinfo {title} {Two-{Dimensional} {Magnetotransport} in the {Extreme} {Quantum} {Limit}},\ }\href {https://doi.org/10.1103/PhysRevLett.48.1559} {\bibfield  {journal} {\bibinfo  {journal} {Phys. Rev. Lett.}\ }\textbf {\bibinfo {volume} {48}},\ \bibinfo {pages} {1559} (\bibinfo {year} {1982})}\BibitemShut {NoStop}%
\bibitem [{\citenamefont {Barkeshli}\ \emph {et~al.}(2019)\citenamefont {Barkeshli}, \citenamefont {Bonderson}, \citenamefont {Cheng},\ and\ \citenamefont {Wang}}]{barkeshli_symmetry_2019}%
  \BibitemOpen
  \bibfield  {author} {\bibinfo {author} {\bibfnamefont {M.}~\bibnamefont {Barkeshli}}, \bibinfo {author} {\bibfnamefont {P.}~\bibnamefont {Bonderson}}, \bibinfo {author} {\bibfnamefont {M.}~\bibnamefont {Cheng}},\ and\ \bibinfo {author} {\bibfnamefont {Z.}~\bibnamefont {Wang}},\ }\bibfield  {title} {\bibinfo {title} {Symmetry fractionalization, defects, and gauging of topological phases},\ }\href {https://doi.org/10.1103/PhysRevB.100.115147} {\bibfield  {journal} {\bibinfo  {journal} {Phys. Rev. B}\ }\textbf {\bibinfo {volume} {100}},\ \bibinfo {pages} {115147} (\bibinfo {year} {2019})}\BibitemShut {NoStop}%
\bibitem [{\citenamefont {Mesaros}\ and\ \citenamefont {Ran}(2013)}]{mesaros_classification_2013}%
  \BibitemOpen
  \bibfield  {author} {\bibinfo {author} {\bibfnamefont {A.}~\bibnamefont {Mesaros}}\ and\ \bibinfo {author} {\bibfnamefont {Y.}~\bibnamefont {Ran}},\ }\bibfield  {title} {\bibinfo {title} {Classification of symmetry enriched topological phases with exactly solvable models},\ }\href {https://doi.org/10.1103/PhysRevB.87.155115} {\bibfield  {journal} {\bibinfo  {journal} {Phys. Rev. B}\ }\textbf {\bibinfo {volume} {87}},\ \bibinfo {pages} {155115} (\bibinfo {year} {2013})}\BibitemShut {NoStop}%
\bibitem [{\citenamefont {Swingle}(2014)}]{swingle_interplay_2014}%
  \BibitemOpen
  \bibfield  {author} {\bibinfo {author} {\bibfnamefont {B.}~\bibnamefont {Swingle}},\ }\bibfield  {title} {\bibinfo {title} {Interplay between short- and long-range entanglement in symmetry-protected phases},\ }\href {https://doi.org/10.1103/PhysRevB.90.035451} {\bibfield  {journal} {\bibinfo  {journal} {Phys. Rev. B}\ }\textbf {\bibinfo {volume} {90}},\ \bibinfo {pages} {035451} (\bibinfo {year} {2014})}\BibitemShut {NoStop}%
\bibitem [{\citenamefont {Levin}\ and\ \citenamefont {Gu}(2012)}]{levin_braiding_2012}%
  \BibitemOpen
  \bibfield  {author} {\bibinfo {author} {\bibfnamefont {M.}~\bibnamefont {Levin}}\ and\ \bibinfo {author} {\bibfnamefont {Z.-C.}\ \bibnamefont {Gu}},\ }\bibfield  {title} {\bibinfo {title} {Braiding statistics approach to symmetry-protected topological phases},\ }\href {https://doi.org/10.1103/PhysRevB.86.115109} {\bibfield  {journal} {\bibinfo  {journal} {Phys. Rev. B}\ }\textbf {\bibinfo {volume} {86}},\ \bibinfo {pages} {115109} (\bibinfo {year} {2012})}\BibitemShut {NoStop}%
\bibitem [{\citenamefont {Haller}\ \emph {et~al.}(2023)\citenamefont {Haller}, \citenamefont {Xu}, \citenamefont {Liu},\ and\ \citenamefont {Pollmann}}]{haller_quantum_2023}%
  \BibitemOpen
  \bibfield  {author} {\bibinfo {author} {\bibfnamefont {L.}~\bibnamefont {Haller}}, \bibinfo {author} {\bibfnamefont {W.-T.}\ \bibnamefont {Xu}}, \bibinfo {author} {\bibfnamefont {Y.-J.}\ \bibnamefont {Liu}},\ and\ \bibinfo {author} {\bibfnamefont {F.}~\bibnamefont {Pollmann}},\ }\bibfield  {title} {\bibinfo {title} {Quantum phase transition between symmetry enriched topological phases in tensor-network states},\ }\href {https://doi.org/10.1103/PhysRevResearch.5.043078} {\bibfield  {journal} {\bibinfo  {journal} {Phys. Rev. Res.}\ }\textbf {\bibinfo {volume} {5}},\ \bibinfo {pages} {043078} (\bibinfo {year} {2023})}\BibitemShut {NoStop}%
\bibitem [{\citenamefont {Ye}\ and\ \citenamefont {Zou}(2024)}]{ye_classification_2024}%
  \BibitemOpen
  \bibfield  {author} {\bibinfo {author} {\bibfnamefont {W.}~\bibnamefont {Ye}}\ and\ \bibinfo {author} {\bibfnamefont {L.}~\bibnamefont {Zou}},\ }\bibfield  {title} {\bibinfo {title} {Classification of {Symmetry}-{Enriched} {Topological} {Quantum} {Spin} {Liquids}},\ }\href {https://doi.org/10.1103/PhysRevX.14.021053} {\bibfield  {journal} {\bibinfo  {journal} {Phys. Rev. X}\ }\textbf {\bibinfo {volume} {14}},\ \bibinfo {pages} {021053} (\bibinfo {year} {2024})}\BibitemShut {NoStop}%
\bibitem [{\citenamefont {Chamon}(2005)}]{chamon_quantum_2005}%
  \BibitemOpen
  \bibfield  {author} {\bibinfo {author} {\bibfnamefont {C.}~\bibnamefont {Chamon}},\ }\bibfield  {title} {\bibinfo {title} {Quantum {Glassiness} in {Strongly} {Correlated} {Clean} {Systems}: {An} {Example} of {Topological} {Overprotection}},\ }\href {https://doi.org/10.1103/PhysRevLett.94.040402} {\bibfield  {journal} {\bibinfo  {journal} {Phys. Rev. Lett.}\ }\textbf {\bibinfo {volume} {94}},\ \bibinfo {pages} {040402} (\bibinfo {year} {2005})}\BibitemShut {NoStop}%
\bibitem [{\citenamefont {Haah}(2011)}]{haah_local_2011}%
  \BibitemOpen
  \bibfield  {author} {\bibinfo {author} {\bibfnamefont {J.}~\bibnamefont {Haah}},\ }\bibfield  {title} {\bibinfo {title} {Local stabilizer codes in three dimensions without string logical operators},\ }\href {https://doi.org/10.1103/PhysRevA.83.042330} {\bibfield  {journal} {\bibinfo  {journal} {Phys. Rev. A}\ }\textbf {\bibinfo {volume} {83}},\ \bibinfo {pages} {042330} (\bibinfo {year} {2011})}\BibitemShut {NoStop}%
\bibitem [{\citenamefont {Vijay}\ \emph {et~al.}(2015)\citenamefont {Vijay}, \citenamefont {Haah},\ and\ \citenamefont {Fu}}]{vijay_new_2015}%
  \BibitemOpen
  \bibfield  {author} {\bibinfo {author} {\bibfnamefont {S.}~\bibnamefont {Vijay}}, \bibinfo {author} {\bibfnamefont {J.}~\bibnamefont {Haah}},\ and\ \bibinfo {author} {\bibfnamefont {L.}~\bibnamefont {Fu}},\ }\bibfield  {title} {\bibinfo {title} {A new kind of topological quantum order: {A} dimensional hierarchy of quasiparticles built from stationary excitations},\ }\href {https://doi.org/10.1103/PhysRevB.92.235136} {\bibfield  {journal} {\bibinfo  {journal} {Phys. Rev. B}\ }\textbf {\bibinfo {volume} {92}},\ \bibinfo {pages} {235136} (\bibinfo {year} {2015})}\BibitemShut {NoStop}%
\bibitem [{\citenamefont {Vijay}\ \emph {et~al.}(2016)\citenamefont {Vijay}, \citenamefont {Haah},\ and\ \citenamefont {Fu}}]{vijay_fracton_2016}%
  \BibitemOpen
  \bibfield  {author} {\bibinfo {author} {\bibfnamefont {S.}~\bibnamefont {Vijay}}, \bibinfo {author} {\bibfnamefont {J.}~\bibnamefont {Haah}},\ and\ \bibinfo {author} {\bibfnamefont {L.}~\bibnamefont {Fu}},\ }\bibfield  {title} {\bibinfo {title} {Fracton topological order, generalized lattice gauge theory, and duality},\ }\href {https://doi.org/10.1103/PhysRevB.94.235157} {\bibfield  {journal} {\bibinfo  {journal} {Phys. Rev. B}\ }\textbf {\bibinfo {volume} {94}},\ \bibinfo {pages} {235157} (\bibinfo {year} {2016})}\BibitemShut {NoStop}%
\bibitem [{\citenamefont {Ma}\ \emph {et~al.}(2018)\citenamefont {Ma}, \citenamefont {Schmitz}, \citenamefont {Parameswaran}, \citenamefont {Hermele},\ and\ \citenamefont {Nandkishore}}]{ma_topological_2018}%
  \BibitemOpen
  \bibfield  {author} {\bibinfo {author} {\bibfnamefont {H.}~\bibnamefont {Ma}}, \bibinfo {author} {\bibfnamefont {A.~T.}\ \bibnamefont {Schmitz}}, \bibinfo {author} {\bibfnamefont {S.~A.}\ \bibnamefont {Parameswaran}}, \bibinfo {author} {\bibfnamefont {M.}~\bibnamefont {Hermele}},\ and\ \bibinfo {author} {\bibfnamefont {R.~M.}\ \bibnamefont {Nandkishore}},\ }\bibfield  {title} {\bibinfo {title} {Topological entanglement entropy of fracton stabilizer codes},\ }\href {https://doi.org/10.1103/PhysRevB.97.125101} {\bibfield  {journal} {\bibinfo  {journal} {Phys. Rev. B}\ }\textbf {\bibinfo {volume} {97}},\ \bibinfo {pages} {125101} (\bibinfo {year} {2018})}\BibitemShut {NoStop}%
\bibitem [{\citenamefont {Pai}\ and\ \citenamefont {Hermele}(2019)}]{pai_fracton_2019}%
  \BibitemOpen
  \bibfield  {author} {\bibinfo {author} {\bibfnamefont {S.}~\bibnamefont {Pai}}\ and\ \bibinfo {author} {\bibfnamefont {M.}~\bibnamefont {Hermele}},\ }\bibfield  {title} {\bibinfo {title} {Fracton fusion and statistics},\ }\href {https://doi.org/10.1103/PhysRevB.100.195136} {\bibfield  {journal} {\bibinfo  {journal} {Phys. Rev. B}\ }\textbf {\bibinfo {volume} {100}},\ \bibinfo {pages} {195136} (\bibinfo {year} {2019})}\BibitemShut {NoStop}%
\bibitem [{\citenamefont {Pretko}\ \emph {et~al.}(2020)\citenamefont {Pretko}, \citenamefont {Chen},\ and\ \citenamefont {You}}]{pretko_fracton_2020}%
  \BibitemOpen
  \bibfield  {author} {\bibinfo {author} {\bibfnamefont {M.}~\bibnamefont {Pretko}}, \bibinfo {author} {\bibfnamefont {X.}~\bibnamefont {Chen}},\ and\ \bibinfo {author} {\bibfnamefont {Y.}~\bibnamefont {You}},\ }\bibfield  {title} {\bibinfo {title} {Fracton phases of matter},\ }\href {https://doi.org/10.1142/S0217751X20300033} {\bibfield  {journal} {\bibinfo  {journal} {International Journal of Modern Physics A}\ }\textbf {\bibinfo {volume} {35}},\ \bibinfo {pages} {2030003} (\bibinfo {year} {2020})}\BibitemShut {NoStop}%
\bibitem [{\citenamefont {Zhu}\ \emph {et~al.}(2023)\citenamefont {Zhu}, \citenamefont {Chen}, \citenamefont {Ye},\ and\ \citenamefont {Trebst}}]{zhu_topological_2023}%
  \BibitemOpen
  \bibfield  {author} {\bibinfo {author} {\bibfnamefont {G.-Y.}\ \bibnamefont {Zhu}}, \bibinfo {author} {\bibfnamefont {J.-Y.}\ \bibnamefont {Chen}}, \bibinfo {author} {\bibfnamefont {P.}~\bibnamefont {Ye}},\ and\ \bibinfo {author} {\bibfnamefont {S.}~\bibnamefont {Trebst}},\ }\bibfield  {title} {\bibinfo {title} {Topological {Fracton} {Quantum} {Phase} {Transitions} by {Tuning} {Exact} {Tensor} {Network} {States}},\ }\href {https://doi.org/10.1103/PhysRevLett.130.216704} {\bibfield  {journal} {\bibinfo  {journal} {Phys. Rev. Lett.}\ }\textbf {\bibinfo {volume} {130}},\ \bibinfo {pages} {216704} (\bibinfo {year} {2023})}\BibitemShut {NoStop}%
\bibitem [{\citenamefont {Song}\ \emph {et~al.}(2024)\citenamefont {Song}, \citenamefont {Tantivasadakarn}, \citenamefont {Shirley},\ and\ \citenamefont {Hermele}}]{song_fracton_2024}%
  \BibitemOpen
  \bibfield  {author} {\bibinfo {author} {\bibfnamefont {H.}~\bibnamefont {Song}}, \bibinfo {author} {\bibfnamefont {N.}~\bibnamefont {Tantivasadakarn}}, \bibinfo {author} {\bibfnamefont {W.}~\bibnamefont {Shirley}},\ and\ \bibinfo {author} {\bibfnamefont {M.}~\bibnamefont {Hermele}},\ }\bibfield  {title} {\bibinfo {title} {Fracton {Self}-{Statistics}},\ }\href {https://doi.org/10.1103/PhysRevLett.132.016604} {\bibfield  {journal} {\bibinfo  {journal} {Phys. Rev. Lett.}\ }\textbf {\bibinfo {volume} {132}},\ \bibinfo {pages} {016604} (\bibinfo {year} {2024})}\BibitemShut {NoStop}%
\bibitem [{\citenamefont {Bravyi}\ and\ \citenamefont {Haah}(2013)}]{bravyi_quantum_2013}%
  \BibitemOpen
  \bibfield  {author} {\bibinfo {author} {\bibfnamefont {S.}~\bibnamefont {Bravyi}}\ and\ \bibinfo {author} {\bibfnamefont {J.}~\bibnamefont {Haah}},\ }\bibfield  {title} {\bibinfo {title} {Quantum self-correction in the 3d cubic code model},\ }\href {https://doi.org/10.1103/PhysRevLett.111.200501} {\bibfield  {journal} {\bibinfo  {journal} {Phys. Rev. Lett.}\ }\textbf {\bibinfo {volume} {111}},\ \bibinfo {pages} {200501} (\bibinfo {year} {2013})}\BibitemShut {NoStop}%
\bibitem [{\citenamefont {Prem}\ \emph {et~al.}(2017)\citenamefont {Prem}, \citenamefont {Haah},\ and\ \citenamefont {Nandkishore}}]{prem_glassy_2017}%
  \BibitemOpen
  \bibfield  {author} {\bibinfo {author} {\bibfnamefont {A.}~\bibnamefont {Prem}}, \bibinfo {author} {\bibfnamefont {J.}~\bibnamefont {Haah}},\ and\ \bibinfo {author} {\bibfnamefont {R.}~\bibnamefont {Nandkishore}},\ }\bibfield  {title} {\bibinfo {title} {Glassy quantum dynamics in translation invariant fracton models},\ }\href {https://doi.org/10.1103/PhysRevB.95.155133} {\bibfield  {journal} {\bibinfo  {journal} {Phys. Rev. B}\ }\textbf {\bibinfo {volume} {95}},\ \bibinfo {pages} {155133} (\bibinfo {year} {2017})}\BibitemShut {NoStop}%
\bibitem [{\citenamefont {Aasen}\ \emph {et~al.}(2020)\citenamefont {Aasen}, \citenamefont {Bulmash}, \citenamefont {Prem}, \citenamefont {Slagle},\ and\ \citenamefont {Williamson}}]{aasen_topological2020}%
  \BibitemOpen
  \bibfield  {author} {\bibinfo {author} {\bibfnamefont {D.}~\bibnamefont {Aasen}}, \bibinfo {author} {\bibfnamefont {D.}~\bibnamefont {Bulmash}}, \bibinfo {author} {\bibfnamefont {A.}~\bibnamefont {Prem}}, \bibinfo {author} {\bibfnamefont {K.}~\bibnamefont {Slagle}},\ and\ \bibinfo {author} {\bibfnamefont {D.~J.}\ \bibnamefont {Williamson}},\ }\bibfield  {title} {\bibinfo {title} {Topological defect networks for fractons of all types},\ }\href {https://doi.org/10.1103/PhysRevResearch.2.043165} {\bibfield  {journal} {\bibinfo  {journal} {Phys. Rev. Res.}\ }\textbf {\bibinfo {volume} {2}},\ \bibinfo {pages} {043165} (\bibinfo {year} {2020})}\BibitemShut {NoStop}%
\bibitem [{\citenamefont {Song}\ \emph {et~al.}(2023)\citenamefont {Song}, \citenamefont {Dua}, \citenamefont {Shirley},\ and\ \citenamefont {Williamson}}]{song_topological2023}%
  \BibitemOpen
  \bibfield  {author} {\bibinfo {author} {\bibfnamefont {Z.}~\bibnamefont {Song}}, \bibinfo {author} {\bibfnamefont {A.}~\bibnamefont {Dua}}, \bibinfo {author} {\bibfnamefont {W.}~\bibnamefont {Shirley}},\ and\ \bibinfo {author} {\bibfnamefont {D.~J.}\ \bibnamefont {Williamson}},\ }\bibfield  {title} {\bibinfo {title} {Topological defect network representations of fracton stabilizer codes},\ }\href {https://doi.org/10.1103/PRXQuantum.4.010304} {\bibfield  {journal} {\bibinfo  {journal} {PRX Quantum}\ }\textbf {\bibinfo {volume} {4}},\ \bibinfo {pages} {010304} (\bibinfo {year} {2023})}\BibitemShut {NoStop}%
\bibitem [{\citenamefont {Ma}\ \emph {et~al.}(2017)\citenamefont {Ma}, \citenamefont {Lake}, \citenamefont {Chen},\ and\ \citenamefont {Hermele}}]{ma_fracton_2017}%
  \BibitemOpen
  \bibfield  {author} {\bibinfo {author} {\bibfnamefont {H.}~\bibnamefont {Ma}}, \bibinfo {author} {\bibfnamefont {E.}~\bibnamefont {Lake}}, \bibinfo {author} {\bibfnamefont {X.}~\bibnamefont {Chen}},\ and\ \bibinfo {author} {\bibfnamefont {M.}~\bibnamefont {Hermele}},\ }\bibfield  {title} {\bibinfo {title} {Fracton topological order via coupled layers},\ }\href {https://doi.org/10.1103/PhysRevB.95.245126} {\bibfield  {journal} {\bibinfo  {journal} {Phys. Rev. B}\ }\textbf {\bibinfo {volume} {95}},\ \bibinfo {pages} {245126} (\bibinfo {year} {2017})}\BibitemShut {NoStop}%
\bibitem [{\citenamefont {Prem}\ \emph {et~al.}(2019)\citenamefont {Prem}, \citenamefont {Huang}, \citenamefont {Song},\ and\ \citenamefont {Hermele}}]{prem_cage-net_2019}%
  \BibitemOpen
  \bibfield  {author} {\bibinfo {author} {\bibfnamefont {A.}~\bibnamefont {Prem}}, \bibinfo {author} {\bibfnamefont {S.-J.}\ \bibnamefont {Huang}}, \bibinfo {author} {\bibfnamefont {H.}~\bibnamefont {Song}},\ and\ \bibinfo {author} {\bibfnamefont {M.}~\bibnamefont {Hermele}},\ }\bibfield  {title} {\bibinfo {title} {Cage-{Net} {Fracton} {Models}},\ }\href {https://doi.org/10.1103/PhysRevX.9.021010} {\bibfield  {journal} {\bibinfo  {journal} {Phys. Rev. X}\ }\textbf {\bibinfo {volume} {9}},\ \bibinfo {pages} {021010} (\bibinfo {year} {2019})}\BibitemShut {NoStop}%
\bibitem [{\citenamefont {Shirley}\ \emph {et~al.}(2018)\citenamefont {Shirley}, \citenamefont {Slagle}, \citenamefont {Wang},\ and\ \citenamefont {Chen}}]{shirley_fracton_2018}%
  \BibitemOpen
  \bibfield  {author} {\bibinfo {author} {\bibfnamefont {W.}~\bibnamefont {Shirley}}, \bibinfo {author} {\bibfnamefont {K.}~\bibnamefont {Slagle}}, \bibinfo {author} {\bibfnamefont {Z.}~\bibnamefont {Wang}},\ and\ \bibinfo {author} {\bibfnamefont {X.}~\bibnamefont {Chen}},\ }\bibfield  {title} {\bibinfo {title} {Fracton {Models} on {General} {Three}-{Dimensional} {Manifolds}},\ }\href {https://doi.org/10.1103/PhysRevX.8.031051} {\bibfield  {journal} {\bibinfo  {journal} {Phys. Rev. X}\ }\textbf {\bibinfo {volume} {8}},\ \bibinfo {pages} {031051} (\bibinfo {year} {2018})}\BibitemShut {NoStop}%
\bibitem [{\citenamefont {Shirley}\ \emph {et~al.}(2019{\natexlab{a}})\citenamefont {Shirley}, \citenamefont {Slagle},\ and\ \citenamefont {Chen}}]{shirley_foliated_2019}%
  \BibitemOpen
  \bibfield  {author} {\bibinfo {author} {\bibfnamefont {W.}~\bibnamefont {Shirley}}, \bibinfo {author} {\bibfnamefont {K.}~\bibnamefont {Slagle}},\ and\ \bibinfo {author} {\bibfnamefont {X.}~\bibnamefont {Chen}},\ }\bibfield  {title} {\bibinfo {title} {Foliated fracton order in the checkerboard model},\ }\href {https://doi.org/10.1103/PhysRevB.99.115123} {\bibfield  {journal} {\bibinfo  {journal} {Phys. Rev. B}\ }\textbf {\bibinfo {volume} {99}},\ \bibinfo {pages} {115123} (\bibinfo {year} {2019}{\natexlab{a}})}\BibitemShut {NoStop}%
\bibitem [{\citenamefont {Slagle}\ \emph {et~al.}(2019)\citenamefont {Slagle}, \citenamefont {Aasen},\ and\ \citenamefont {Williamson}}]{slagle_foliated_2019}%
  \BibitemOpen
  \bibfield  {author} {\bibinfo {author} {\bibfnamefont {K.}~\bibnamefont {Slagle}}, \bibinfo {author} {\bibfnamefont {D.}~\bibnamefont {Aasen}},\ and\ \bibinfo {author} {\bibfnamefont {D.}~\bibnamefont {Williamson}},\ }\bibfield  {title} {\bibinfo {title} {Foliated field theory and string-membrane-net condensation picture of fracton order},\ }\href {https://doi.org/10.21468/SciPostPhys.6.4.043} {\bibfield  {journal} {\bibinfo  {journal} {SciPost Physics}\ }\textbf {\bibinfo {volume} {6}},\ \bibinfo {pages} {043} (\bibinfo {year} {2019})}\BibitemShut {NoStop}%
\bibitem [{\citenamefont {Shirley}\ \emph {et~al.}(2020)\citenamefont {Shirley}, \citenamefont {Slagle},\ and\ \citenamefont {Chen}}]{shirley_twisted_2020}%
  \BibitemOpen
  \bibfield  {author} {\bibinfo {author} {\bibfnamefont {W.}~\bibnamefont {Shirley}}, \bibinfo {author} {\bibfnamefont {K.}~\bibnamefont {Slagle}},\ and\ \bibinfo {author} {\bibfnamefont {X.}~\bibnamefont {Chen}},\ }\bibfield  {title} {\bibinfo {title} {Twisted foliated fracton phases},\ }\href {https://doi.org/10.1103/PhysRevB.102.115103} {\bibfield  {journal} {\bibinfo  {journal} {Phys. Rev. B}\ }\textbf {\bibinfo {volume} {102}},\ \bibinfo {pages} {115103} (\bibinfo {year} {2020})}\BibitemShut {NoStop}%
\bibitem [{\citenamefont {You}\ \emph {et~al.}(2020)\citenamefont {You}, \citenamefont {Devakul}, \citenamefont {Burnell},\ and\ \citenamefont {Sondhi}}]{you_symmetric_2020}%
  \BibitemOpen
  \bibfield  {author} {\bibinfo {author} {\bibfnamefont {Y.}~\bibnamefont {You}}, \bibinfo {author} {\bibfnamefont {T.}~\bibnamefont {Devakul}}, \bibinfo {author} {\bibfnamefont {F.~J.}\ \bibnamefont {Burnell}},\ and\ \bibinfo {author} {\bibfnamefont {S.~L.}\ \bibnamefont {Sondhi}},\ }\bibfield  {title} {\bibinfo {title} {Symmetric fracton matter: {Twisted} and enriched},\ }\href {https://doi.org/10.1016/j.aop.2020.168140} {\bibfield  {journal} {\bibinfo  {journal} {Annals of Physics}\ }\textbf {\bibinfo {volume} {416}},\ \bibinfo {pages} {168140} (\bibinfo {year} {2020})}\BibitemShut {NoStop}%
\bibitem [{\citenamefont {Tantivasadakarn}\ and\ \citenamefont {Vijay}(2020)}]{tantivasadakarn_searching_2020}%
  \BibitemOpen
  \bibfield  {author} {\bibinfo {author} {\bibfnamefont {N.}~\bibnamefont {Tantivasadakarn}}\ and\ \bibinfo {author} {\bibfnamefont {S.}~\bibnamefont {Vijay}},\ }\bibfield  {title} {\bibinfo {title} {Searching for fracton orders via symmetry defect condensation},\ }\href {https://doi.org/10.1103/PhysRevB.101.165143} {\bibfield  {journal} {\bibinfo  {journal} {Phys. Rev. B}\ }\textbf {\bibinfo {volume} {101}},\ \bibinfo {pages} {165143} (\bibinfo {year} {2020})}\BibitemShut {NoStop}%
\bibitem [{\citenamefont {Rayhaun}\ and\ \citenamefont {Williamson}(2023)}]{rayhaun_higher2023}%
  \BibitemOpen
  \bibfield  {author} {\bibinfo {author} {\bibfnamefont {B.~C.}\ \bibnamefont {Rayhaun}}\ and\ \bibinfo {author} {\bibfnamefont {D.~J.}\ \bibnamefont {Williamson}},\ }\bibfield  {title} {\bibinfo {title} {{Higher-form subsystem symmetry breaking: Subdimensional criticality and fracton phase transitions}},\ }\href {https://doi.org/10.21468/SciPostPhys.15.1.017} {\bibfield  {journal} {\bibinfo  {journal} {SciPost Phys.}\ }\textbf {\bibinfo {volume} {15}},\ \bibinfo {pages} {017} (\bibinfo {year} {2023})}\BibitemShut {NoStop}%
\bibitem [{\citenamefont {Williamson}\ \emph {et~al.}(2016)\citenamefont {Williamson}, \citenamefont {Bultinck}, \citenamefont {Mari\"en}, \citenamefont {\ifmmode \mbox{\c{S}}\else \c{S}\fi{}ahino\ifmmode~\breve{g}\else \u{g}\fi{}lu}, \citenamefont {Haegeman},\ and\ \citenamefont {Verstraete}}]{williamson_matrix2016}%
  \BibitemOpen
  \bibfield  {author} {\bibinfo {author} {\bibfnamefont {D.~J.}\ \bibnamefont {Williamson}}, \bibinfo {author} {\bibfnamefont {N.}~\bibnamefont {Bultinck}}, \bibinfo {author} {\bibfnamefont {M.}~\bibnamefont {Mari\"en}}, \bibinfo {author} {\bibfnamefont {M.~B.}\ \bibnamefont {\ifmmode \mbox{\c{S}}\else \c{S}\fi{}ahino\ifmmode~\breve{g}\else \u{g}\fi{}lu}}, \bibinfo {author} {\bibfnamefont {J.}~\bibnamefont {Haegeman}},\ and\ \bibinfo {author} {\bibfnamefont {F.}~\bibnamefont {Verstraete}},\ }\bibfield  {title} {\bibinfo {title} {Matrix product operators for symmetry-protected topological phases: Gauging and edge theories},\ }\href {https://doi.org/10.1103/PhysRevB.94.205150} {\bibfield  {journal} {\bibinfo  {journal} {Phys. Rev. B}\ }\textbf {\bibinfo {volume} {94}},\ \bibinfo {pages} {205150} (\bibinfo {year} {2016})}\BibitemShut {NoStop}%
\bibitem [{\citenamefont {Shukla}\ \emph {et~al.}(2018)\citenamefont {Shukla}, \citenamefont {\ifmmode \mbox{\c{S}}\else \c{S}\fi{}ahino\ifmmode~\breve{g}\else \u{g}\fi{}lu}, \citenamefont {Pollmann},\ and\ \citenamefont {Chen}}]{shukla_boson2018}%
  \BibitemOpen
  \bibfield  {author} {\bibinfo {author} {\bibfnamefont {S.~K.}\ \bibnamefont {Shukla}}, \bibinfo {author} {\bibfnamefont {M.~B.}\ \bibnamefont {\ifmmode \mbox{\c{S}}\else \c{S}\fi{}ahino\ifmmode~\breve{g}\else \u{g}\fi{}lu}}, \bibinfo {author} {\bibfnamefont {F.}~\bibnamefont {Pollmann}},\ and\ \bibinfo {author} {\bibfnamefont {X.}~\bibnamefont {Chen}},\ }\bibfield  {title} {\bibinfo {title} {Boson condensation and instability in the tensor network representation of string-net states},\ }\href {https://doi.org/10.1103/PhysRevB.98.125112} {\bibfield  {journal} {\bibinfo  {journal} {Phys. Rev. B}\ }\textbf {\bibinfo {volume} {98}},\ \bibinfo {pages} {125112} (\bibinfo {year} {2018})}\BibitemShut {NoStop}%
\bibitem [{\citenamefont {Liu}\ \emph {et~al.}(2024)\citenamefont {Liu}, \citenamefont {Shtengel},\ and\ \citenamefont {Pollmann}}]{liu_simulating_2023}%
  \BibitemOpen
  \bibfield  {author} {\bibinfo {author} {\bibfnamefont {Y.-J.}\ \bibnamefont {Liu}}, \bibinfo {author} {\bibfnamefont {K.}~\bibnamefont {Shtengel}},\ and\ \bibinfo {author} {\bibfnamefont {F.}~\bibnamefont {Pollmann}},\ }\bibfield  {title} {\bibinfo {title} {Simulating two-dimensional topological quantum phase transitions on a digital quantum computer},\ }\href {https://doi.org/10.1103/PhysRevResearch.6.043256} {\bibfield  {journal} {\bibinfo  {journal} {Phys. Rev. Res.}\ }\textbf {\bibinfo {volume} {6}},\ \bibinfo {pages} {043256} (\bibinfo {year} {2024})}\BibitemShut {NoStop}%
\bibitem [{\citenamefont {Yu}\ \emph {et~al.}(2024)\citenamefont {Yu}, \citenamefont {Cirac}, \citenamefont {Kos},\ and\ \citenamefont {Styliaris}}]{yu_dual2024}%
  \BibitemOpen
  \bibfield  {author} {\bibinfo {author} {\bibfnamefont {X.-H.}\ \bibnamefont {Yu}}, \bibinfo {author} {\bibfnamefont {J.~I.}\ \bibnamefont {Cirac}}, \bibinfo {author} {\bibfnamefont {P.}~\bibnamefont {Kos}},\ and\ \bibinfo {author} {\bibfnamefont {G.}~\bibnamefont {Styliaris}},\ }\bibfield  {title} {\bibinfo {title} {Dual-isometric projected entangled pair states},\ }\href {https://doi.org/10.1103/PhysRevLett.133.190401} {\bibfield  {journal} {\bibinfo  {journal} {Phys. Rev. Lett.}\ }\textbf {\bibinfo {volume} {133}},\ \bibinfo {pages} {190401} (\bibinfo {year} {2024})}\BibitemShut {NoStop}%
\bibitem [{\citenamefont {Verstraete}\ \emph {et~al.}(2006)\citenamefont {Verstraete}, \citenamefont {Wolf}, \citenamefont {Perez-Garcia},\ and\ \citenamefont {Cirac}}]{verstraete_criticality_2006}%
  \BibitemOpen
  \bibfield  {author} {\bibinfo {author} {\bibfnamefont {F.}~\bibnamefont {Verstraete}}, \bibinfo {author} {\bibfnamefont {M.~M.}\ \bibnamefont {Wolf}}, \bibinfo {author} {\bibfnamefont {D.}~\bibnamefont {Perez-Garcia}},\ and\ \bibinfo {author} {\bibfnamefont {J.~I.}\ \bibnamefont {Cirac}},\ }\bibfield  {title} {\bibinfo {title} {Criticality, the area law, and the computational power of projected entangled pair states},\ }\href {https://doi.org/10.1103/PhysRevLett.96.220601} {\bibfield  {journal} {\bibinfo  {journal} {Phys. Rev. Lett.}\ }\textbf {\bibinfo {volume} {96}},\ \bibinfo {pages} {220601} (\bibinfo {year} {2006})}\BibitemShut {NoStop}%
\bibitem [{\citenamefont {Cirac}\ \emph {et~al.}(2021)\citenamefont {Cirac}, \citenamefont {Pérez-García}, \citenamefont {Schuch},\ and\ \citenamefont {Verstraete}}]{cirac_matrix_2021}%
  \BibitemOpen
  \bibfield  {author} {\bibinfo {author} {\bibfnamefont {J.~I.}\ \bibnamefont {Cirac}}, \bibinfo {author} {\bibfnamefont {D.}~\bibnamefont {Pérez-García}}, \bibinfo {author} {\bibfnamefont {N.}~\bibnamefont {Schuch}},\ and\ \bibinfo {author} {\bibfnamefont {F.}~\bibnamefont {Verstraete}},\ }\bibfield  {title} {\bibinfo {title} {Matrix product states and projected entangled pair states: {Concepts}, symmetries, theorems},\ }\href {https://link.aps.org/doi/10.1103/RevModPhys.93.045003} {\bibfield  {journal} {\bibinfo  {journal} {Rev. of Mod. Phys.}\ }\textbf {\bibinfo {volume} {93}},\ \bibinfo {pages} {045003} (\bibinfo {year} {2021})}\BibitemShut {NoStop}%
\bibitem [{\citenamefont {Zaletel}\ and\ \citenamefont {Pollmann}(2020)}]{zaletel_isometric_2020}%
  \BibitemOpen
  \bibfield  {author} {\bibinfo {author} {\bibfnamefont {M.~P.}\ \bibnamefont {Zaletel}}\ and\ \bibinfo {author} {\bibfnamefont {F.}~\bibnamefont {Pollmann}},\ }\bibfield  {title} {\bibinfo {title} {Isometric {Tensor} {Network} {States} in {Two} {Dimensions}},\ }\href {https://doi.org/10.1103/PhysRevLett.124.037201} {\bibfield  {journal} {\bibinfo  {journal} {Phys. Rev. Lett.}\ }\textbf {\bibinfo {volume} {124}},\ \bibinfo {pages} {037201} (\bibinfo {year} {2020})}\BibitemShut {NoStop}%
\bibitem [{\citenamefont {Haghshenas}\ \emph {et~al.}(2019)\citenamefont {Haghshenas}, \citenamefont {O'Rourke},\ and\ \citenamefont {Chan}}]{haghshenas_conversion_2019}%
  \BibitemOpen
  \bibfield  {author} {\bibinfo {author} {\bibfnamefont {R.}~\bibnamefont {Haghshenas}}, \bibinfo {author} {\bibfnamefont {M.~J.}\ \bibnamefont {O'Rourke}},\ and\ \bibinfo {author} {\bibfnamefont {G.~K.-L.}\ \bibnamefont {Chan}},\ }\bibfield  {title} {\bibinfo {title} {Conversion of projected entangled pair states into a canonical form},\ }\href {https://doi.org/10.1103/PhysRevB.100.054404} {\bibfield  {journal} {\bibinfo  {journal} {Phys. Rev. B}\ }\textbf {\bibinfo {volume} {100}},\ \bibinfo {pages} {054404} (\bibinfo {year} {2019})}\BibitemShut {NoStop}%
\bibitem [{\citenamefont {Kadow}\ \emph {et~al.}(2023)\citenamefont {Kadow}, \citenamefont {Pollmann},\ and\ \citenamefont {Knap}}]{Kadow2023}%
  \BibitemOpen
  \bibfield  {author} {\bibinfo {author} {\bibfnamefont {W.}~\bibnamefont {Kadow}}, \bibinfo {author} {\bibfnamefont {F.}~\bibnamefont {Pollmann}},\ and\ \bibinfo {author} {\bibfnamefont {M.}~\bibnamefont {Knap}},\ }\bibfield  {title} {\bibinfo {title} {Isometric tensor network representations of two-dimensional thermal states},\ }\href {https://doi.org/10.1103/physrevb.107.205106} {\bibfield  {journal} {\bibinfo  {journal} {Physical Review B}\ }\textbf {\bibinfo {volume} {107}},\ \bibinfo {pages} {205106} (\bibinfo {year} {2023})}\BibitemShut {NoStop}%
\bibitem [{\citenamefont {Malz}\ and\ \citenamefont {Trivedi}(2025)}]{malz_computational_2024}%
  \BibitemOpen
  \bibfield  {author} {\bibinfo {author} {\bibfnamefont {D.}~\bibnamefont {Malz}}\ and\ \bibinfo {author} {\bibfnamefont {R.}~\bibnamefont {Trivedi}},\ }\bibfield  {title} {\bibinfo {title} {Computational complexity of isometric tensor-network states},\ }\href {https://doi.org/10.1103/PRXQuantum.6.020310} {\bibfield  {journal} {\bibinfo  {journal} {PRX Quantum}\ }\textbf {\bibinfo {volume} {6}},\ \bibinfo {pages} {020310} (\bibinfo {year} {2025})}\BibitemShut {NoStop}%
\bibitem [{\citenamefont {Soejima}\ \emph {et~al.}(2020)\citenamefont {Soejima}, \citenamefont {Siva}, \citenamefont {Bultinck}, \citenamefont {Chatterjee}, \citenamefont {Pollmann},\ and\ \citenamefont {Zaletel}}]{soejima_isometric_2020}%
  \BibitemOpen
  \bibfield  {author} {\bibinfo {author} {\bibfnamefont {T.}~\bibnamefont {Soejima}}, \bibinfo {author} {\bibfnamefont {K.}~\bibnamefont {Siva}}, \bibinfo {author} {\bibfnamefont {N.}~\bibnamefont {Bultinck}}, \bibinfo {author} {\bibfnamefont {S.}~\bibnamefont {Chatterjee}}, \bibinfo {author} {\bibfnamefont {F.}~\bibnamefont {Pollmann}},\ and\ \bibinfo {author} {\bibfnamefont {M.~P.}\ \bibnamefont {Zaletel}},\ }\bibfield  {title} {\bibinfo {title} {Isometric tensor network representation of string-net liquids},\ }\href {https://doi.org/10.1103/PhysRevB.101.085117} {\bibfield  {journal} {\bibinfo  {journal} {Phys. Rev. B}\ }\textbf {\bibinfo {volume} {101}},\ \bibinfo {pages} {085117} (\bibinfo {year} {2020})}\BibitemShut {NoStop}%
\bibitem [{\citenamefont {Schön}\ \emph {et~al.}(2005)\citenamefont {Schön}, \citenamefont {Solano}, \citenamefont {Verstraete}, \citenamefont {Cirac},\ and\ \citenamefont {Wolf}}]{schon_sequential_2005}%
  \BibitemOpen
  \bibfield  {author} {\bibinfo {author} {\bibfnamefont {C.}~\bibnamefont {Schön}}, \bibinfo {author} {\bibfnamefont {E.}~\bibnamefont {Solano}}, \bibinfo {author} {\bibfnamefont {F.}~\bibnamefont {Verstraete}}, \bibinfo {author} {\bibfnamefont {J.~I.}\ \bibnamefont {Cirac}},\ and\ \bibinfo {author} {\bibfnamefont {M.~M.}\ \bibnamefont {Wolf}},\ }\bibfield  {title} {\bibinfo {title} {Sequential {Generation} of {Entangled} {Multiqubit} {States}},\ }\href {https://doi.org/10.1103/PhysRevLett.95.110503} {\bibfield  {journal} {\bibinfo  {journal} {Phys. Rev. Lett.}\ }\textbf {\bibinfo {volume} {95}},\ \bibinfo {pages} {110503} (\bibinfo {year} {2005})}\BibitemShut {NoStop}%
\bibitem [{\citenamefont {Wolf}\ \emph {et~al.}(2006)\citenamefont {Wolf}, \citenamefont {Ortiz}, \citenamefont {Verstraete},\ and\ \citenamefont {Cirac}}]{wolf_quantum_2006}%
  \BibitemOpen
  \bibfield  {author} {\bibinfo {author} {\bibfnamefont {M.~M.}\ \bibnamefont {Wolf}}, \bibinfo {author} {\bibfnamefont {G.}~\bibnamefont {Ortiz}}, \bibinfo {author} {\bibfnamefont {F.}~\bibnamefont {Verstraete}},\ and\ \bibinfo {author} {\bibfnamefont {J.~I.}\ \bibnamefont {Cirac}},\ }\bibfield  {title} {\bibinfo {title} {Quantum {Phase} {Transitions} in {Matrix} {Product} {Systems}},\ }\href {https://doi.org/10.1103/PhysRevLett.97.110403} {\bibfield  {journal} {\bibinfo  {journal} {Phys. Rev. Lett.}\ }\textbf {\bibinfo {volume} {97}},\ \bibinfo {pages} {110403} (\bibinfo {year} {2006})}\BibitemShut {NoStop}%
\bibitem [{\citenamefont {Jones}\ \emph {et~al.}(2021)\citenamefont {Jones}, \citenamefont {Bibo}, \citenamefont {Jobst}, \citenamefont {Pollmann}, \citenamefont {Smith},\ and\ \citenamefont {Verresen}}]{jones_skeleton_2021}%
  \BibitemOpen
  \bibfield  {author} {\bibinfo {author} {\bibfnamefont {N.~G.}\ \bibnamefont {Jones}}, \bibinfo {author} {\bibfnamefont {J.}~\bibnamefont {Bibo}}, \bibinfo {author} {\bibfnamefont {B.}~\bibnamefont {Jobst}}, \bibinfo {author} {\bibfnamefont {F.}~\bibnamefont {Pollmann}}, \bibinfo {author} {\bibfnamefont {A.}~\bibnamefont {Smith}},\ and\ \bibinfo {author} {\bibfnamefont {R.}~\bibnamefont {Verresen}},\ }\bibfield  {title} {\bibinfo {title} {Skeleton of matrix-product-state-solvable models connecting topological phases of matter},\ }\href {https://doi.org/10.1103/PhysRevResearch.3.033265} {\bibfield  {journal} {\bibinfo  {journal} {Phys. Rev. Res.}\ }\textbf {\bibinfo {volume} {3}},\ \bibinfo {pages} {033265} (\bibinfo {year} {2021})}\BibitemShut {NoStop}%
\bibitem [{\citenamefont {Smith}\ \emph {et~al.}(2022)\citenamefont {Smith}, \citenamefont {Jobst}, \citenamefont {Green},\ and\ \citenamefont {Pollmann}}]{smith_crossing_2022}%
  \BibitemOpen
  \bibfield  {author} {\bibinfo {author} {\bibfnamefont {A.}~\bibnamefont {Smith}}, \bibinfo {author} {\bibfnamefont {B.}~\bibnamefont {Jobst}}, \bibinfo {author} {\bibfnamefont {A.~G.}\ \bibnamefont {Green}},\ and\ \bibinfo {author} {\bibfnamefont {F.}~\bibnamefont {Pollmann}},\ }\bibfield  {title} {\bibinfo {title} {Crossing a topological phase transition with a quantum computer},\ }\href {https://doi.org/10.1103/PhysRevResearch.4.L022020} {\bibfield  {journal} {\bibinfo  {journal} {Phys. Rev. Res.}\ }\textbf {\bibinfo {volume} {4}},\ \bibinfo {pages} {L022020} (\bibinfo {year} {2022})}\BibitemShut {NoStop}%
\bibitem [{\citenamefont {Wei}\ \emph {et~al.}(2022)\citenamefont {Wei}, \citenamefont {Malz},\ and\ \citenamefont {Cirac}}]{wei_sequential_2022}%
  \BibitemOpen
  \bibfield  {author} {\bibinfo {author} {\bibfnamefont {Z.-Y.}\ \bibnamefont {Wei}}, \bibinfo {author} {\bibfnamefont {D.}~\bibnamefont {Malz}},\ and\ \bibinfo {author} {\bibfnamefont {J.~I.}\ \bibnamefont {Cirac}},\ }\bibfield  {title} {\bibinfo {title} {Sequential {Generation} of {Projected} {Entangled}-{Pair} {States}},\ }\href {https://doi.org/10.1103/PhysRevLett.128.010607} {\bibfield  {journal} {\bibinfo  {journal} {Phys. Rev. Lett.}\ }\textbf {\bibinfo {volume} {128}},\ \bibinfo {pages} {010607} (\bibinfo {year} {2022})}\BibitemShut {NoStop}%
\bibitem [{\citenamefont {Chen}\ \emph {et~al.}(2024{\natexlab{a}})\citenamefont {Chen}, \citenamefont {Dua}, \citenamefont {Hermele}, \citenamefont {Stephen}, \citenamefont {Tantivasadakarn}, \citenamefont {Vanhove},\ and\ \citenamefont {Zhao}}]{chen_sequential_2024}%
  \BibitemOpen
  \bibfield  {author} {\bibinfo {author} {\bibfnamefont {X.}~\bibnamefont {Chen}}, \bibinfo {author} {\bibfnamefont {A.}~\bibnamefont {Dua}}, \bibinfo {author} {\bibfnamefont {M.}~\bibnamefont {Hermele}}, \bibinfo {author} {\bibfnamefont {D.~T.}\ \bibnamefont {Stephen}}, \bibinfo {author} {\bibfnamefont {N.}~\bibnamefont {Tantivasadakarn}}, \bibinfo {author} {\bibfnamefont {R.}~\bibnamefont {Vanhove}},\ and\ \bibinfo {author} {\bibfnamefont {J.-Y.}\ \bibnamefont {Zhao}},\ }\bibfield  {title} {\bibinfo {title} {Sequential quantum circuits as maps between gapped phases},\ }\href {https://doi.org/10.1103/PhysRevB.109.075116} {\bibfield  {journal} {\bibinfo  {journal} {Phys. Rev. B}\ }\textbf {\bibinfo {volume} {109}},\ \bibinfo {pages} {075116} (\bibinfo {year} {2024}{\natexlab{a}})}\BibitemShut {NoStop}%
\bibitem [{\citenamefont {Satzinger}\ \emph {et~al.}(2021)\citenamefont {Satzinger}, \citenamefont {Liu}, \citenamefont {Smith}, \citenamefont {Knapp}, \citenamefont {Newman}, \citenamefont {Jones}, \citenamefont {Chen}, \citenamefont {Quintana}, \citenamefont {Mi}, \citenamefont {Dunsworth} \emph {et~al.}}]{satzinger_realizing_2021}%
  \BibitemOpen
  \bibfield  {author} {\bibinfo {author} {\bibfnamefont {K.~J.}\ \bibnamefont {Satzinger}}, \bibinfo {author} {\bibfnamefont {Y.-J.}\ \bibnamefont {Liu}}, \bibinfo {author} {\bibfnamefont {A.}~\bibnamefont {Smith}}, \bibinfo {author} {\bibfnamefont {C.}~\bibnamefont {Knapp}}, \bibinfo {author} {\bibfnamefont {M.}~\bibnamefont {Newman}}, \bibinfo {author} {\bibfnamefont {C.}~\bibnamefont {Jones}}, \bibinfo {author} {\bibfnamefont {Z.}~\bibnamefont {Chen}}, \bibinfo {author} {\bibfnamefont {C.}~\bibnamefont {Quintana}}, \bibinfo {author} {\bibfnamefont {X.}~\bibnamefont {Mi}}, \bibinfo {author} {\bibfnamefont {A.}~\bibnamefont {Dunsworth}}, \emph {et~al.},\ }\bibfield  {title} {\bibinfo {title} {Realizing topologically ordered states on a quantum processor},\ }\href {https://doi.org/10.1126/science.abi8378} {\bibfield  {journal} {\bibinfo  {journal} {Science}\ }\textbf {\bibinfo {volume} {374}},\ \bibinfo {pages} {1237} (\bibinfo {year} {2021})}\BibitemShut {NoStop}%
\bibitem [{\citenamefont {Liu}\ \emph {et~al.}(2022)\citenamefont {Liu}, \citenamefont {Shtengel}, \citenamefont {Smith},\ and\ \citenamefont {Pollmann}}]{liu_methods_2022}%
  \BibitemOpen
  \bibfield  {author} {\bibinfo {author} {\bibfnamefont {Y.-J.}\ \bibnamefont {Liu}}, \bibinfo {author} {\bibfnamefont {K.}~\bibnamefont {Shtengel}}, \bibinfo {author} {\bibfnamefont {A.}~\bibnamefont {Smith}},\ and\ \bibinfo {author} {\bibfnamefont {F.}~\bibnamefont {Pollmann}},\ }\bibfield  {title} {\bibinfo {title} {Methods for {Simulating} {String}-{Net} {States} and {Anyons} on a {Digital} {Quantum} {Computer}},\ }\href {https://doi.org/10.1103/PRXQuantum.3.040315} {\bibfield  {journal} {\bibinfo  {journal} {PRX Quantum}\ }\textbf {\bibinfo {volume} {3}},\ \bibinfo {pages} {040315} (\bibinfo {year} {2022})}\BibitemShut {NoStop}%
\bibitem [{\citenamefont {Cochran}\ \emph {et~al.}(2025)\citenamefont {Cochran}, \citenamefont {Jobst}, \citenamefont {Rosenberg}, \citenamefont {Lensky}, \citenamefont {Gyawali}, \citenamefont {Eassa}, \citenamefont {Will}, \citenamefont {Szasz}, \citenamefont {Abanin}, \citenamefont {Acharya} \emph {et~al.}}]{cochran_visualizing_2024}%
  \BibitemOpen
  \bibfield  {author} {\bibinfo {author} {\bibfnamefont {T.~A.}\ \bibnamefont {Cochran}}, \bibinfo {author} {\bibfnamefont {B.}~\bibnamefont {Jobst}}, \bibinfo {author} {\bibfnamefont {E.}~\bibnamefont {Rosenberg}}, \bibinfo {author} {\bibfnamefont {Y.~D.}\ \bibnamefont {Lensky}}, \bibinfo {author} {\bibfnamefont {G.}~\bibnamefont {Gyawali}}, \bibinfo {author} {\bibfnamefont {N.}~\bibnamefont {Eassa}}, \bibinfo {author} {\bibfnamefont {M.}~\bibnamefont {Will}}, \bibinfo {author} {\bibfnamefont {A.}~\bibnamefont {Szasz}}, \bibinfo {author} {\bibfnamefont {D.}~\bibnamefont {Abanin}}, \bibinfo {author} {\bibfnamefont {R.}~\bibnamefont {Acharya}}, \emph {et~al.},\ }\bibfield  {title} {\bibinfo {title} {Visualizing {Dynamics} of {Charges} and {Strings} in (2+1){D} {Lattice} {Gauge} {Theories}},\ }\href {https://doi.org/10.1038/s41586-025-08999-9} {\bibfield  {journal} {\bibinfo  {journal} {Nature}\ }\textbf {\bibinfo {volume} {642}},\ \bibinfo {pages} {315} (\bibinfo {year} {2025})}\BibitemShut {NoStop}%
\bibitem [{\citenamefont {Semeghini}\ \emph {et~al.}(2021)\citenamefont {Semeghini}, \citenamefont {Levine}, \citenamefont {Keesling}, \citenamefont {Ebadi}, \citenamefont {Wang}, \citenamefont {Bluvstein}, \citenamefont {Verresen}, \citenamefont {Pichler}, \citenamefont {Kalinowski}, \citenamefont {Samajdar}, \citenamefont {Omran}, \citenamefont {Sachdev}, \citenamefont {Vishwanath}, \citenamefont {Greiner}, \citenamefont {Vuletić},\ and\ \citenamefont {Lukin}}]{semeghini_probing_2021}%
  \BibitemOpen
  \bibfield  {author} {\bibinfo {author} {\bibfnamefont {G.}~\bibnamefont {Semeghini}}, \bibinfo {author} {\bibfnamefont {H.}~\bibnamefont {Levine}}, \bibinfo {author} {\bibfnamefont {A.}~\bibnamefont {Keesling}}, \bibinfo {author} {\bibfnamefont {S.}~\bibnamefont {Ebadi}}, \bibinfo {author} {\bibfnamefont {T.~T.}\ \bibnamefont {Wang}}, \bibinfo {author} {\bibfnamefont {D.}~\bibnamefont {Bluvstein}}, \bibinfo {author} {\bibfnamefont {R.}~\bibnamefont {Verresen}}, \bibinfo {author} {\bibfnamefont {H.}~\bibnamefont {Pichler}}, \bibinfo {author} {\bibfnamefont {M.}~\bibnamefont {Kalinowski}}, \bibinfo {author} {\bibfnamefont {R.}~\bibnamefont {Samajdar}}, \bibinfo {author} {\bibfnamefont {A.}~\bibnamefont {Omran}}, \bibinfo {author} {\bibfnamefont {S.}~\bibnamefont {Sachdev}}, \bibinfo {author} {\bibfnamefont {A.}~\bibnamefont {Vishwanath}}, \bibinfo {author} {\bibfnamefont {M.}~\bibnamefont {Greiner}}, \bibinfo {author} {\bibfnamefont {V.}~\bibnamefont {Vuletić}},\ and\ \bibinfo {author} {\bibfnamefont
  {M.~D.}\ \bibnamefont {Lukin}},\ }\bibfield  {title} {\bibinfo {title} {Probing topological spin liquids on a programmable quantum simulator},\ }\href {https://doi.org/10.1126/science.abi8794} {\bibfield  {journal} {\bibinfo  {journal} {Science}\ }\textbf {\bibinfo {volume} {374}},\ \bibinfo {pages} {1242} (\bibinfo {year} {2021})}\BibitemShut {NoStop}%
\bibitem [{\citenamefont {Verresen}\ \emph {et~al.}(2021)\citenamefont {Verresen}, \citenamefont {Lukin},\ and\ \citenamefont {Vishwanath}}]{verresen_prediction_2021}%
  \BibitemOpen
  \bibfield  {author} {\bibinfo {author} {\bibfnamefont {R.}~\bibnamefont {Verresen}}, \bibinfo {author} {\bibfnamefont {M.~D.}\ \bibnamefont {Lukin}},\ and\ \bibinfo {author} {\bibfnamefont {A.}~\bibnamefont {Vishwanath}},\ }\bibfield  {title} {\bibinfo {title} {Prediction of {Toric} {Code} {Topological} {Order} from {Rydberg} {Blockade}},\ }\href {https://doi.org/10.1103/PhysRevX.11.031005} {\bibfield  {journal} {\bibinfo  {journal} {Phys. Rev. X}\ }\textbf {\bibinfo {volume} {11}},\ \bibinfo {pages} {031005} (\bibinfo {year} {2021})}\BibitemShut {NoStop}%
\bibitem [{\citenamefont {Iqbal}\ \emph {et~al.}(2024)\citenamefont {Iqbal}, \citenamefont {Tantivasadakarn}, \citenamefont {Verresen}, \citenamefont {Campbell}, \citenamefont {Dreiling}, \citenamefont {Figgatt}, \citenamefont {Gaebler}, \citenamefont {Johansen}, \citenamefont {Mills}, \citenamefont {Moses}, \citenamefont {Pino}, \citenamefont {Ransford}, \citenamefont {Rowe}, \citenamefont {Siegfried}, \citenamefont {Stutz}, \citenamefont {Foss-Feig}, \citenamefont {Vishwanath},\ and\ \citenamefont {Dreyer}}]{iqbal_non-abelian_2024}%
  \BibitemOpen
  \bibfield  {author} {\bibinfo {author} {\bibfnamefont {M.}~\bibnamefont {Iqbal}}, \bibinfo {author} {\bibfnamefont {N.}~\bibnamefont {Tantivasadakarn}}, \bibinfo {author} {\bibfnamefont {R.}~\bibnamefont {Verresen}}, \bibinfo {author} {\bibfnamefont {S.~L.}\ \bibnamefont {Campbell}}, \bibinfo {author} {\bibfnamefont {J.~M.}\ \bibnamefont {Dreiling}}, \bibinfo {author} {\bibfnamefont {C.}~\bibnamefont {Figgatt}}, \bibinfo {author} {\bibfnamefont {J.~P.}\ \bibnamefont {Gaebler}}, \bibinfo {author} {\bibfnamefont {J.}~\bibnamefont {Johansen}}, \bibinfo {author} {\bibfnamefont {M.}~\bibnamefont {Mills}}, \bibinfo {author} {\bibfnamefont {S.~A.}\ \bibnamefont {Moses}}, \bibinfo {author} {\bibfnamefont {J.~M.}\ \bibnamefont {Pino}}, \bibinfo {author} {\bibfnamefont {A.}~\bibnamefont {Ransford}}, \bibinfo {author} {\bibfnamefont {M.}~\bibnamefont {Rowe}}, \bibinfo {author} {\bibfnamefont {P.}~\bibnamefont {Siegfried}}, \bibinfo {author} {\bibfnamefont {R.~P.}\ \bibnamefont {Stutz}}, \bibinfo {author}
  {\bibfnamefont {M.}~\bibnamefont {Foss-Feig}}, \bibinfo {author} {\bibfnamefont {A.}~\bibnamefont {Vishwanath}},\ and\ \bibinfo {author} {\bibfnamefont {H.}~\bibnamefont {Dreyer}},\ }\bibfield  {title} {\bibinfo {title} {Non-{Abelian} topological order and anyons on a trapped-ion processor},\ }\href {https://doi.org/10.1038/s41586-023-06934-4} {\bibfield  {journal} {\bibinfo  {journal} {Nature}\ }\textbf {\bibinfo {volume} {626}},\ \bibinfo {pages} {505} (\bibinfo {year} {2024})}\BibitemShut {NoStop}%
\bibitem [{\citenamefont {Tantivasadakarn}\ \emph {et~al.}(2023)\citenamefont {Tantivasadakarn}, \citenamefont {Verresen},\ and\ \citenamefont {Vishwanath}}]{tantivasadakarn_shortest_2023}%
  \BibitemOpen
  \bibfield  {author} {\bibinfo {author} {\bibfnamefont {N.}~\bibnamefont {Tantivasadakarn}}, \bibinfo {author} {\bibfnamefont {R.}~\bibnamefont {Verresen}},\ and\ \bibinfo {author} {\bibfnamefont {A.}~\bibnamefont {Vishwanath}},\ }\bibfield  {title} {\bibinfo {title} {Shortest {Route} to {Non}-{Abelian} {Topological} {Order} on a {Quantum} {Processor}},\ }\href {https://doi.org/10.1103/PhysRevLett.131.060405} {\bibfield  {journal} {\bibinfo  {journal} {Phys. Rev. Lett.}\ }\textbf {\bibinfo {volume} {131}},\ \bibinfo {pages} {060405} (\bibinfo {year} {2023})}\BibitemShut {NoStop}%
\bibitem [{\citenamefont {Nevidomskyy}\ \emph {et~al.}(2024)\citenamefont {Nevidomskyy}, \citenamefont {Bernien},\ and\ \citenamefont {Canright}}]{nevidomskyy_realizing_2024}%
  \BibitemOpen
  \bibfield  {author} {\bibinfo {author} {\bibfnamefont {A.~H.}\ \bibnamefont {Nevidomskyy}}, \bibinfo {author} {\bibfnamefont {H.}~\bibnamefont {Bernien}},\ and\ \bibinfo {author} {\bibfnamefont {A.}~\bibnamefont {Canright}},\ }\href@noop {} {\bibinfo {title} {Realizing fracton order from long-range quantum entanglement in programmable rydberg atom arrays}} (\bibinfo {year} {2024}),\ \Eprint {https://arxiv.org/abs/2407.05885} {arXiv:2407.05885 [quant-ph]} \BibitemShut {NoStop}%
\bibitem [{\citenamefont {Chen}\ \emph {et~al.}(2024{\natexlab{b}})\citenamefont {Chen}, \citenamefont {Yan},\ and\ \citenamefont {Cui}}]{chen_quantum_2024}%
  \BibitemOpen
  \bibfield  {author} {\bibinfo {author} {\bibfnamefont {P.}~\bibnamefont {Chen}}, \bibinfo {author} {\bibfnamefont {B.}~\bibnamefont {Yan}},\ and\ \bibinfo {author} {\bibfnamefont {S.~X.}\ \bibnamefont {Cui}},\ }\bibfield  {title} {\bibinfo {title} {Quantum circuits for toric code and {X}-cube fracton model},\ }\href {https://doi.org/10.22331/q-2024-03-13-1276} {\bibfield  {journal} {\bibinfo  {journal} {{Quantum}}\ }\textbf {\bibinfo {volume} {8}},\ \bibinfo {pages} {1276} (\bibinfo {year} {2024}{\natexlab{b}})}\BibitemShut {NoStop}%
\bibitem [{\citenamefont {Slagle}\ and\ \citenamefont {Kim}(2018)}]{slagle_x-cube_2018}%
  \BibitemOpen
  \bibfield  {author} {\bibinfo {author} {\bibfnamefont {K.}~\bibnamefont {Slagle}}\ and\ \bibinfo {author} {\bibfnamefont {Y.~B.}\ \bibnamefont {Kim}},\ }\bibfield  {title} {\bibinfo {title} {X-cube model on generic lattices: {Fracton} phases and geometric order},\ }\href {https://doi.org/10.1103/PhysRevB.97.165106} {\bibfield  {journal} {\bibinfo  {journal} {Phys. Rev. B}\ }\textbf {\bibinfo {volume} {97}},\ \bibinfo {pages} {165106} (\bibinfo {year} {2018})}\BibitemShut {NoStop}%
\bibitem [{\citenamefont {Levin}\ and\ \citenamefont {Wen}(2005)}]{levin_string-net_2005}%
  \BibitemOpen
  \bibfield  {author} {\bibinfo {author} {\bibfnamefont {M.~A.}\ \bibnamefont {Levin}}\ and\ \bibinfo {author} {\bibfnamefont {X.-G.}\ \bibnamefont {Wen}},\ }\bibfield  {title} {\bibinfo {title} {String-net condensation: {A} physical mechanism for topological phases},\ }\href {https://doi.org/10.1103/PhysRevB.71.045110} {\bibfield  {journal} {\bibinfo  {journal} {Phys. Rev. B}\ }\textbf {\bibinfo {volume} {71}},\ \bibinfo {pages} {045110} (\bibinfo {year} {2005})}\BibitemShut {NoStop}%
\bibitem [{\citenamefont {Lin}\ \emph {et~al.}(2021)\citenamefont {Lin}, \citenamefont {Levin},\ and\ \citenamefont {Burnell}}]{lin_generalized_2021}%
  \BibitemOpen
  \bibfield  {author} {\bibinfo {author} {\bibfnamefont {C.-H.}\ \bibnamefont {Lin}}, \bibinfo {author} {\bibfnamefont {M.}~\bibnamefont {Levin}},\ and\ \bibinfo {author} {\bibfnamefont {F.~J.}\ \bibnamefont {Burnell}},\ }\bibfield  {title} {\bibinfo {title} {Generalized string-net models: {A} thorough exposition},\ }\href {https://doi.org/10.1103/PhysRevB.103.195155} {\bibfield  {journal} {\bibinfo  {journal} {Phys. Rev. B}\ }\textbf {\bibinfo {volume} {103}},\ \bibinfo {pages} {195155} (\bibinfo {year} {2021})}\BibitemShut {NoStop}%
\bibitem [{\citenamefont {He}\ \emph {et~al.}(2018)\citenamefont {He}, \citenamefont {Zheng}, \citenamefont {Bernevig},\ and\ \citenamefont {Regnault}}]{he_entanglement_2018}%
  \BibitemOpen
  \bibfield  {author} {\bibinfo {author} {\bibfnamefont {H.}~\bibnamefont {He}}, \bibinfo {author} {\bibfnamefont {Y.}~\bibnamefont {Zheng}}, \bibinfo {author} {\bibfnamefont {B.~A.}\ \bibnamefont {Bernevig}},\ and\ \bibinfo {author} {\bibfnamefont {N.}~\bibnamefont {Regnault}},\ }\bibfield  {title} {\bibinfo {title} {Entanglement entropy from tensor network states for stabilizer codes},\ }\href {https://doi.org/10.1103/PhysRevB.97.125102} {\bibfield  {journal} {\bibinfo  {journal} {Phys. Rev. B}\ }\textbf {\bibinfo {volume} {97}},\ \bibinfo {pages} {125102} (\bibinfo {year} {2018})}\BibitemShut {NoStop}%
\bibitem [{\citenamefont {Shirley}\ \emph {et~al.}(2019{\natexlab{b}})\citenamefont {Shirley}, \citenamefont {Slagle},\ and\ \citenamefont {Chen}}]{shirley_fractional_2019}%
  \BibitemOpen
  \bibfield  {author} {\bibinfo {author} {\bibfnamefont {W.}~\bibnamefont {Shirley}}, \bibinfo {author} {\bibfnamefont {K.}~\bibnamefont {Slagle}},\ and\ \bibinfo {author} {\bibfnamefont {X.}~\bibnamefont {Chen}},\ }\bibfield  {title} {\bibinfo {title} {Fractional excitations in foliated fracton phases},\ }\href {https://doi.org/10.1016/j.aop.2019.167922} {\bibfield  {journal} {\bibinfo  {journal} {Annals of Physics}\ }\textbf {\bibinfo {volume} {410}},\ \bibinfo {pages} {167922} (\bibinfo {year} {2019}{\natexlab{b}})}\BibitemShut {NoStop}%
\bibitem [{\citenamefont {Pollmann}\ and\ \citenamefont {Turner}(2012)}]{pollmann_detection_2012}%
  \BibitemOpen
  \bibfield  {author} {\bibinfo {author} {\bibfnamefont {F.}~\bibnamefont {Pollmann}}\ and\ \bibinfo {author} {\bibfnamefont {A.~M.}\ \bibnamefont {Turner}},\ }\bibfield  {title} {\bibinfo {title} {Detection of symmetry-protected topological phases in one dimension},\ }\href {https://doi.org/10.1103/PhysRevB.86.125441} {\bibfield  {journal} {\bibinfo  {journal} {Phys. Rev. B}\ }\textbf {\bibinfo {volume} {86}},\ \bibinfo {pages} {125441} (\bibinfo {year} {2012})}\BibitemShut {NoStop}%
\bibitem [{\citenamefont {Zaletel}(2014)}]{zaletel_detecting_2014}%
  \BibitemOpen
  \bibfield  {author} {\bibinfo {author} {\bibfnamefont {M.~P.}\ \bibnamefont {Zaletel}},\ }\bibfield  {title} {\bibinfo {title} {Detecting two-dimensional symmetry-protected topological order in a ground-state wave function},\ }\href {https://doi.org/10.1103/PhysRevB.90.235113} {\bibfield  {journal} {\bibinfo  {journal} {Phys. Rev. B}\ }\textbf {\bibinfo {volume} {90}},\ \bibinfo {pages} {235113} (\bibinfo {year} {2014})}\BibitemShut {NoStop}%
\bibitem [{\citenamefont {Huang}\ \emph {et~al.}(2014)\citenamefont {Huang}, \citenamefont {Chen},\ and\ \citenamefont {Pollmann}}]{huang_detection_2014}%
  \BibitemOpen
  \bibfield  {author} {\bibinfo {author} {\bibfnamefont {C.-Y.}\ \bibnamefont {Huang}}, \bibinfo {author} {\bibfnamefont {X.}~\bibnamefont {Chen}},\ and\ \bibinfo {author} {\bibfnamefont {F.}~\bibnamefont {Pollmann}},\ }\bibfield  {title} {\bibinfo {title} {Detection of symmetry-enriched topological phases},\ }\href {https://doi.org/10.1103/PhysRevB.90.045142} {\bibfield  {journal} {\bibinfo  {journal} {Phys. Rev. B}\ }\textbf {\bibinfo {volume} {90}},\ \bibinfo {pages} {045142} (\bibinfo {year} {2014})}\BibitemShut {NoStop}%
\bibitem [{\citenamefont {Zhang}\ \emph {et~al.}(2012)\citenamefont {Zhang}, \citenamefont {Grover}, \citenamefont {Turner}, \citenamefont {Oshikawa},\ and\ \citenamefont {Vishwanath}}]{zhang_quasiparticle_2012}%
  \BibitemOpen
  \bibfield  {author} {\bibinfo {author} {\bibfnamefont {Y.}~\bibnamefont {Zhang}}, \bibinfo {author} {\bibfnamefont {T.}~\bibnamefont {Grover}}, \bibinfo {author} {\bibfnamefont {A.}~\bibnamefont {Turner}}, \bibinfo {author} {\bibfnamefont {M.}~\bibnamefont {Oshikawa}},\ and\ \bibinfo {author} {\bibfnamefont {A.}~\bibnamefont {Vishwanath}},\ }\bibfield  {title} {\bibinfo {title} {Quasiparticle statistics and braiding from ground-state entanglement},\ }\href {https://doi.org/10.1103/PhysRevB.85.235151} {\bibfield  {journal} {\bibinfo  {journal} {Phys. Rev. B}\ }\textbf {\bibinfo {volume} {85}},\ \bibinfo {pages} {235151} (\bibinfo {year} {2012})}\BibitemShut {NoStop}%
\bibitem [{\citenamefont {Slagle}\ and\ \citenamefont {Kim}(2017)}]{slagle_quantum_2017}%
  \BibitemOpen
  \bibfield  {author} {\bibinfo {author} {\bibfnamefont {K.}~\bibnamefont {Slagle}}\ and\ \bibinfo {author} {\bibfnamefont {Y.~B.}\ \bibnamefont {Kim}},\ }\bibfield  {title} {\bibinfo {title} {Quantum field theory of x-cube fracton topological order and robust degeneracy from geometry},\ }\href {https://doi.org/10.1103/PhysRevB.96.195139} {\bibfield  {journal} {\bibinfo  {journal} {Phys. Rev. B}\ }\textbf {\bibinfo {volume} {96}},\ \bibinfo {pages} {195139} (\bibinfo {year} {2017})}\BibitemShut {NoStop}%
\bibitem [{\citenamefont {Baxter}(1985)}]{baxter_exactly_1985}%
  \BibitemOpen
  \bibfield  {author} {\bibinfo {author} {\bibfnamefont {R.~J.}\ \bibnamefont {Baxter}},\ }\bibfield  {title} {\bibinfo {title} {Exactly {Solved} {Models} in {Statistical} {Mechanics}},\ }in\ \href {https://doi.org/10.1142/9789814415255_0002} {\emph {\bibinfo {booktitle} {Integrable {Systems} in {Statistical} {Mechanics}}}},\ \bibinfo {series} {Series on {Advances} in {Statistical} {Mechanics}}, Vol.\ \bibinfo {volume} {Volume 1}\ (\bibinfo  {publisher} {WORLD SCIENTIFIC},\ \bibinfo {year} {1985})\ pp.\ \bibinfo {pages} {5--63}\BibitemShut {NoStop}%
\bibitem [{\citenamefont {Friedman}\ \emph {et~al.}(2019)\citenamefont {Friedman}, \citenamefont {Chan}, \citenamefont {De~Luca},\ and\ \citenamefont {Chalker}}]{friedman_spectral_2019}%
  \BibitemOpen
  \bibfield  {author} {\bibinfo {author} {\bibfnamefont {A.~J.}\ \bibnamefont {Friedman}}, \bibinfo {author} {\bibfnamefont {A.}~\bibnamefont {Chan}}, \bibinfo {author} {\bibfnamefont {A.}~\bibnamefont {De~Luca}},\ and\ \bibinfo {author} {\bibfnamefont {J.}~\bibnamefont {Chalker}},\ }\bibfield  {title} {\bibinfo {title} {Spectral {Statistics} and {Many}-{Body} {Quantum} {Chaos} with {Conserved} {Charge}},\ }\href {https://doi.org/10.1103/PhysRevLett.123.210603} {\bibfield  {journal} {\bibinfo  {journal} {Phys. Rev. Lett.}\ }\textbf {\bibinfo {volume} {123}},\ \bibinfo {pages} {210603} (\bibinfo {year} {2019})}\BibitemShut {NoStop}%
\bibitem [{\citenamefont {Kos}\ \emph {et~al.}(2021)\citenamefont {Kos}, \citenamefont {Bertini},\ and\ \citenamefont {Prosen}}]{kos_chaos_2021}%
  \BibitemOpen
  \bibfield  {author} {\bibinfo {author} {\bibfnamefont {P.}~\bibnamefont {Kos}}, \bibinfo {author} {\bibfnamefont {B.}~\bibnamefont {Bertini}},\ and\ \bibinfo {author} {\bibfnamefont {T.~c.~v.}\ \bibnamefont {Prosen}},\ }\bibfield  {title} {\bibinfo {title} {Chaos and ergodicity in extended quantum systems with noisy driving},\ }\href {https://doi.org/10.1103/PhysRevLett.126.190601} {\bibfield  {journal} {\bibinfo  {journal} {Phys. Rev. Lett.}\ }\textbf {\bibinfo {volume} {126}},\ \bibinfo {pages} {190601} (\bibinfo {year} {2021})}\BibitemShut {NoStop}%
\bibitem [{\citenamefont {Schuch}\ \emph {et~al.}(2007)\citenamefont {Schuch}, \citenamefont {Wolf}, \citenamefont {Verstraete},\ and\ \citenamefont {Cirac}}]{schuch_computational_2007}%
  \BibitemOpen
  \bibfield  {author} {\bibinfo {author} {\bibfnamefont {N.}~\bibnamefont {Schuch}}, \bibinfo {author} {\bibfnamefont {M.~M.}\ \bibnamefont {Wolf}}, \bibinfo {author} {\bibfnamefont {F.}~\bibnamefont {Verstraete}},\ and\ \bibinfo {author} {\bibfnamefont {J.~I.}\ \bibnamefont {Cirac}},\ }\bibfield  {title} {\bibinfo {title} {Computational {Complexity} of {Projected} {Entangled} {Pair} {States}},\ }\href {https://doi.org/10.1103/PhysRevLett.98.140506} {\bibfield  {journal} {\bibinfo  {journal} {Phys. Rev. Lett.}\ }\textbf {\bibinfo {volume} {98}},\ \bibinfo {pages} {140506} (\bibinfo {year} {2007})}\BibitemShut {NoStop}%
\bibitem [{\citenamefont {Raussendorf}\ and\ \citenamefont {Briegel}(2001)}]{raussendorf_one-way_2001}%
  \BibitemOpen
  \bibfield  {author} {\bibinfo {author} {\bibfnamefont {R.}~\bibnamefont {Raussendorf}}\ and\ \bibinfo {author} {\bibfnamefont {H.~J.}\ \bibnamefont {Briegel}},\ }\bibfield  {title} {\bibinfo {title} {A one-way quantum computer},\ }\href {https://doi.org/10.1103/PhysRevLett.86.5188} {\bibfield  {journal} {\bibinfo  {journal} {Phys. Rev. Lett.}\ }\textbf {\bibinfo {volume} {86}},\ \bibinfo {pages} {5188} (\bibinfo {year} {2001})}\BibitemShut {NoStop}%
\bibitem [{\citenamefont {Raussendorf}\ \emph {et~al.}(2003)\citenamefont {Raussendorf}, \citenamefont {Browne},\ and\ \citenamefont {Briegel}}]{raussendorf_measurement-based_2003}%
  \BibitemOpen
  \bibfield  {author} {\bibinfo {author} {\bibfnamefont {R.}~\bibnamefont {Raussendorf}}, \bibinfo {author} {\bibfnamefont {D.~E.}\ \bibnamefont {Browne}},\ and\ \bibinfo {author} {\bibfnamefont {H.~J.}\ \bibnamefont {Briegel}},\ }\bibfield  {title} {\bibinfo {title} {Measurement-based quantum computation on cluster states},\ }\href {https://doi.org/10.1103/PhysRevA.68.022312} {\bibfield  {journal} {\bibinfo  {journal} {Phys. Rev. A}\ }\textbf {\bibinfo {volume} {68}},\ \bibinfo {pages} {022312} (\bibinfo {year} {2003})}\BibitemShut {NoStop}%
\bibitem [{\citenamefont {Foss-Feig}\ \emph {et~al.}(2021)\citenamefont {Foss-Feig}, \citenamefont {Hayes}, \citenamefont {Dreiling}, \citenamefont {Figgatt}, \citenamefont {Gaebler}, \citenamefont {Moses}, \citenamefont {Pino},\ and\ \citenamefont {Potter}}]{foss-feig_holographic_2021}%
  \BibitemOpen
  \bibfield  {author} {\bibinfo {author} {\bibfnamefont {M.}~\bibnamefont {Foss-Feig}}, \bibinfo {author} {\bibfnamefont {D.}~\bibnamefont {Hayes}}, \bibinfo {author} {\bibfnamefont {J.~M.}\ \bibnamefont {Dreiling}}, \bibinfo {author} {\bibfnamefont {C.}~\bibnamefont {Figgatt}}, \bibinfo {author} {\bibfnamefont {J.~P.}\ \bibnamefont {Gaebler}}, \bibinfo {author} {\bibfnamefont {S.~A.}\ \bibnamefont {Moses}}, \bibinfo {author} {\bibfnamefont {J.~M.}\ \bibnamefont {Pino}},\ and\ \bibinfo {author} {\bibfnamefont {A.~C.}\ \bibnamefont {Potter}},\ }\bibfield  {title} {\bibinfo {title} {Holographic quantum algorithms for simulating correlated spin systems},\ }\href {https://doi.org/10.1103/PhysRevResearch.3.033002} {\bibfield  {journal} {\bibinfo  {journal} {Phys. Rev. Res.}\ }\textbf {\bibinfo {volume} {3}},\ \bibinfo {pages} {033002} (\bibinfo {year} {2021})}\BibitemShut {NoStop}%
\bibitem [{\citenamefont {Chertkov}\ \emph {et~al.}(2022)\citenamefont {Chertkov}, \citenamefont {Bohnet}, \citenamefont {Francois}, \citenamefont {Gaebler}, \citenamefont {Gresh}, \citenamefont {Hankin}, \citenamefont {Lee}, \citenamefont {Hayes}, \citenamefont {Neyenhuis}, \citenamefont {Stutz}, \citenamefont {Potter},\ and\ \citenamefont {Foss-Feig}}]{chertkov_holographic_2022}%
  \BibitemOpen
  \bibfield  {author} {\bibinfo {author} {\bibfnamefont {E.}~\bibnamefont {Chertkov}}, \bibinfo {author} {\bibfnamefont {J.}~\bibnamefont {Bohnet}}, \bibinfo {author} {\bibfnamefont {D.}~\bibnamefont {Francois}}, \bibinfo {author} {\bibfnamefont {J.}~\bibnamefont {Gaebler}}, \bibinfo {author} {\bibfnamefont {D.}~\bibnamefont {Gresh}}, \bibinfo {author} {\bibfnamefont {A.}~\bibnamefont {Hankin}}, \bibinfo {author} {\bibfnamefont {K.}~\bibnamefont {Lee}}, \bibinfo {author} {\bibfnamefont {D.}~\bibnamefont {Hayes}}, \bibinfo {author} {\bibfnamefont {B.}~\bibnamefont {Neyenhuis}}, \bibinfo {author} {\bibfnamefont {R.}~\bibnamefont {Stutz}}, \bibinfo {author} {\bibfnamefont {A.~C.}\ \bibnamefont {Potter}},\ and\ \bibinfo {author} {\bibfnamefont {M.}~\bibnamefont {Foss-Feig}},\ }\bibfield  {title} {\bibinfo {title} {Holographic dynamics simulations with a trapped-ion quantum computer},\ }\href {https://doi.org/10.1038/s41567-022-01689-7} {\bibfield  {journal} {\bibinfo  {journal} {Nature Physics}\ }\textbf
  {\bibinfo {volume} {18}},\ \bibinfo {pages} {1074} (\bibinfo {year} {2022})}\BibitemShut {NoStop}%
\bibitem [{\citenamefont {Chen}\ and\ \citenamefont {Hermele}(2016)}]{chen_symmetry2016}%
  \BibitemOpen
  \bibfield  {author} {\bibinfo {author} {\bibfnamefont {X.}~\bibnamefont {Chen}}\ and\ \bibinfo {author} {\bibfnamefont {M.}~\bibnamefont {Hermele}},\ }\bibfield  {title} {\bibinfo {title} {Symmetry fractionalization and anomaly detection in three-dimensional topological phases},\ }\href {https://doi.org/10.1103/PhysRevB.94.195120} {\bibfield  {journal} {\bibinfo  {journal} {Phys. Rev. B}\ }\textbf {\bibinfo {volume} {94}},\ \bibinfo {pages} {195120} (\bibinfo {year} {2016})}\BibitemShut {NoStop}%
\bibitem [{\citenamefont {Lee}\ \emph {et~al.}(2025)\citenamefont {Lee}, \citenamefont {Hu}, \citenamefont {Cho},\ and\ \citenamefont {Watanabe}}]{lee_ZN2025}%
  \BibitemOpen
  \bibfield  {author} {\bibinfo {author} {\bibfnamefont {C.}~\bibnamefont {Lee}}, \bibinfo {author} {\bibfnamefont {Y.}~\bibnamefont {Hu}}, \bibinfo {author} {\bibfnamefont {G.~Y.}\ \bibnamefont {Cho}},\ and\ \bibinfo {author} {\bibfnamefont {H.}~\bibnamefont {Watanabe}},\ }\href@noop {} {\bibinfo {title} {$\mathbb{Z}_n$ generalizations of three-dimensional stabilizer codes}} (\bibinfo {year} {2025}),\ \Eprint {https://arxiv.org/abs/2504.09847} {arXiv:2504.09847 [cond-mat.str-el]} \BibitemShut {NoStop}%
\bibitem [{\citenamefont {Haag}\ \emph {et~al.}(2023)\citenamefont {Haag}, \citenamefont {Baccari},\ and\ \citenamefont {Styliaris}}]{haag_typical_2023}%
  \BibitemOpen
  \bibfield  {author} {\bibinfo {author} {\bibfnamefont {D.}~\bibnamefont {Haag}}, \bibinfo {author} {\bibfnamefont {F.}~\bibnamefont {Baccari}},\ and\ \bibinfo {author} {\bibfnamefont {G.}~\bibnamefont {Styliaris}},\ }\bibfield  {title} {\bibinfo {title} {Typical {Correlation} {Length} of {Sequentially} {Generated} {Tensor} {Network} {States}},\ }\href {https://doi.org/10.1103/PRXQuantum.4.030330} {\bibfield  {journal} {\bibinfo  {journal} {PRX Quantum}\ }\textbf {\bibinfo {volume} {4}},\ \bibinfo {pages} {030330} (\bibinfo {year} {2023})}\BibitemShut {NoStop}%
\bibitem [{zen()}]{zenodo}%
  \BibitemOpen
  \href@noop {} {\bibinfo {title} {Simulation codes are available at \href{https://doi.org/10.5281/zenodo.14719699}{10.5281/zenodo.14719699}}}\BibitemShut {NoStop}%
\end{thebibliography}%
\newpage

\end{document}